\documentstyle[aps,pra,eqsecnum,graphics,epsf]{revtex}

\input psfig.sty
\def\CC{{\rm\kern.24em \vrule width.04em height1.46ex depth-.07ex
\kern-.30em C}}

\begin{document}
\title{Theory of Decoherence-Free Fault-Tolerant Universal Quantum Computation}
\author{J. Kempe$^{1,3,4}$, D. Bacon$^{1,2}$, D.A. Lidar$^{1}$ and
K.B. Whaley$^{1}$}
\address{
$^{1}$Department of Chemistry, University of California, Berkeley\\
$^{2}$Department of Physics, University of California, Berkeley\\
$^{3}$Department of Mathematics, University of California, Berkeley\\
$^{4}$\'{E}cole Nationale Superieure des T\'{e}l\'{e}communications,
Paris, France\\
}
\date{\today}
\maketitle

\begin{abstract}
Universal quantum computation on decoherence-free subspaces and subsystems
(DFSs) is examined with particular emphasis on using only physically
relevant interactions. A necessary and sufficient condition for the
existence of decoherence-free (noiseless) subsystems in the Markovian
regime is derived 
here for the first time. A stabilizer formalism for DFSs is then developed
which allows for the explicit understanding of these in their dual role as
quantum error correcting codes. Conditions for the existence of Hamiltonians
whose induced evolution always preserves a DFS are derived within this
stabilizer formalism. Two possible collective decoherence mechanisms arising
from permutation symmetries of the system-bath coupling are examined within
this framework. It is shown that in both cases universal quantum computation
which {\em always} preserves the DFS ({\em natural fault-tolerant
computation}) can be performed using only two-body interactions. This is in
marked contrast to standard error correcting codes, where all known
constructions using one or two-body interactions must leave the codespace
during the on-time of the fault-tolerant gates. A further consequence of our
universality construction is that a single exchange Hamiltonian can be used
to perform universal quantum computation on an encoded space whose
asymptotic coding efficiency is unity. The exchange Hamiltonian, which is
naturally present in many quantum systems, is thus {\em asymptotically
universal}.
\end{abstract}

\section{Introduction}

The discovery that information encoded over quantum systems can exhibit
strange and wonderful computational \cite{Shor:94a,Grover:97a} and
information theoretic \cite{Bennett:93a,Bennett:98a} properties has led to
an explosion of interest in understanding and exploiting the ``quantumness''
of nature. For the use of quantum information to progress beyond mere
theoretical constructs into the realm of testable and useful implementations
and experiments, it is essential to develop techniques for preserving
quantum coherences. In particular, the coupling of a quantum system to its
environment leads to a process known as decoherence, in which encoded
quantum information is lost to the environment. In order to remedy this
problem active quantum error correction codes (QECCs) \cite
{Glossary,Shor:95a,Steane:96a,Knill:97b,Gottesman:97a} have been developed, by
analogy with classical error correction. These codes encode quantum
information over an entangled set of codewords, the structure of which
serves to preserve the quantum information, when used in conjunction with a
frequently recurring error correcting procedure. It has been shown that when
the rate of decoherence is below a certain threshold, fault tolerant quantum
information manipulation is possible\cite
{Aharonov:96a,Knill:98a,Preskill:98a}. Since it is believed that there are
no systems for which the decoherence mechanism entirely vanishes, QECCs will
be essential if quantum information manipulation is to become practical.

An alternative approach has been proposed and developed recently, in which
the central motivation is the desire to reduce the effect of a specific
decoherence mechanism. This is the decoherence-free subspace (DFS) approach
(also referred to as ``error avoiding'', or ``noiseless'' quantum
codes) \cite
{Palma:96,Duan:97b,Duan:98a,Zanardi:97a,Zanardi:97b,Zanardi:97c,Zanardi:98a,Zanardi:98b,Lidar:98a,Lidar:99a,Lidar:99b,Bacon:99a,Lidar:99c,Durdevich:00}.
In contrast to the active mode of QECC, DFS theory can be viewed as
providing a passive approach, where a specific symmetry of the system-bath
coupling is employed in order to seek out a quiet corner of the system's
Hilbert space which does not experience decoherence. Information encoded
here over a subspace of (usually entangled) system states is robust against
a specific form of decoherence. We shall refer to this as the ``DFS
supporting decoherence mechanism''. When this is the dominant form of
decoherence in the physical system, there are major gains to be had by
operating in the DFS. Previous work has shown that collective decoherence of
the type experienced in condensed phase systems at low temperatures can be
successfully eliminated in this way \cite{Zanardi:98c,Zanardi:99a}. Further
research showed that DFSs are robust to perturbing error processes \cite
{Lidar:98a,Bacon:99a}, and are thus ideally suited for concatenation in a
QECC \cite{Lidar:99a}.

The motivating goal behind the DFS approach is to {\em use symmetry first}.
Thus, one first identifies a DFS for the major sources of decoherence, via
the symmetry of the interaction with the environment. One then proceeds to
use the DFS states as a basis for a QECC which can deal with additional
perturbing error processes. In order for this scheme to be credible, DFSs
must support the ability to perform universal quantum computation on the
encoded states. Towards this end, certain existential results \cite
{Zanardi:99c} have been derived showing that in principle universal quantum
computation can be performed on any DFS. Constructive results for a set of
universal quantum gates on a particular class DFSs were subsequently
constructed in \cite{Lidar:99b} using known QECC constructions. However,
these gates were constructed in such a way that during the operation of the
gate, states within a DFS are taken outside of this subspace. Thus these
gates would necessarily need to operate on a timescale faster than the DFS
supporting decoherence mechanism, in order to be applied efficiently to a
concatenated DFS-QECC scheme.\footnote{Note that QECC fault-tolerant
gates are also required to operate faster than the decoherence time of the
main error process.} Similarly, a universal computation result on DFSs
for atoms
in cavities was recently presented by Beige {\it et al.} in
\cite{Beige:99,Beige:00}. It assumes that  
the interaction driving a system out of the DFS is much weaker than
the coupling of non DF-states to the environment. It is then possible
to make use of an environment-induced quantum Zeno effect. 
In order to make use of the robustness condition {\em
without} resorting to gates which can be made faster than the main DFS
supporting decoherence mechanism, one would prefer to explicitly construct a
set of Hamiltonians which can be used to perform universal quantum
computation, but which {\em never allow states in the DFS to leak out of the
DFS}. Imperfections in these gates may be dealt with by the concatenation
technique of \cite{Lidar:99a} (see also \cite{Lidar:99c}).

In addition, one would, from a practical standpoint, like to use
Hamiltonians which involve at most two-body interactions (under the
assumption that any three-body interactions will be weak and not useful for
operations which must compete with the decoherence rate). In \cite{Bacon:99b}
such Hamiltonians were used for the important decoherence mechanism known as
``collective decoherence'', on a system of $4$ physical qubits. In
collective decoherence the bath cannot distinguish between individual system
qubits, and thus couples in a collective manner to the qubits. The
corresponding two-body Hamiltonians used to implement universal quantum
computation are those that preserve the collective symmetry: the {\em
exchange interaction} between pairs of qubits. The first and main purpose of
this paper is therefore to extend the constructive results obtained in \cite
{Bacon:99b} to other forms of collective decoherence and to larger DFSs. Two
different forms of collective decoherence are considered here, and
constructive results are obtained for these on DFSs of arbitrary numbers of
qubits. These results have implications that extend far beyond the problem
of dealing with collective decoherence. Since they imply that the exchange
interaction {\em by itself} is sufficient to implement universal quantum
computation {\em on a subspace}, it follows that using encoded (rather than
physical) qubits can be advantageous when resources for physical operations
are limited. After all, the standard results for universal quantum
computation employ either arbitrary single-qubit operations in addition to a
non-trivial two-qubit gate (e.g., a controlled-{\sc NOT}), or at least
two non-commuting two-qubit Hamiltonians \cite
{Deutsch:95,Barenco:1,DiVincenzo:95a,Lloyd:95a,Sleator:95}. These issues
will be explored in a separate publication.

Previous work established that DFSs correspond to the degenerate component
of a QECC \cite{Lidar:99a,Duan:98d}. A second purpose of this work is
to present new 
results on a recently discovered generalization of DFSs, which has been
termed ``noiseless subsystems'', and arises from a theory of QECC for
general decoherence mechanisms \cite{Knill:99a,Viola:00a}. In line
with our previously 
established terminology \cite{Lidar:98a} we will refer to these as
``decoherence-free subsystems'', where we take the term ``decoherence'' to
mean both dephasing ($T_{2}$) and dissipation ($T_{1}$). Essentially, the
generalization corresponds to allowing for information to be encoded into
states transforming according to {\em arbitrary}-dimensional irreducible
representations (irreps) of the decoherence-operators' algebra, instead of
just one-dimensional irreps as in the decoherence-free subspace case (we
will present precise definitions later in this paper). These results all
arise from a basic theorem on algebras that are closed under the Hermitian
conjugation operation (``$\dagger $-closed algebras''), and thereby unify
the role of symmetry in both decoherence-free subspaces and quantum error
correction. In this paper we extend the decoherence-free subsystem concept
to situations governed by essentially non-$\dagger $-closed evolution. Such
situations arise from non-Hermitian terms in the system-bath interaction,
which may occur, e.g., in generalized master equation and conditional
Hamiltonian representations of open quantum dynamics \cite{Carmichael:93}.
In particular, we derive an if and only if (iff) condition for the existence
of decoherence-free subsystems with dynamics governed by a semigroup master
equation. This is important because it is well known in decoherence-free
subspace theory that such non-$\dagger $-closed evolution can support
different DFSs than in the $\dagger $-closed case. A similar result is now
shown here to hold for the decoherence-free subsystems.

Existential results for universal quantum computation on decoherence-free
subsystems also exist \cite{Zanardi:99d}. The universal quantum computation
results we obtain in this paper extend beyond decoherence-free subspaces: we
show how to achieve {\em constructive} universal quantum computation on the
decoherence-free subsystems supported under collective decoherence. This
most significant achievement of our paper settles the question of universal
quantum computation under collective decoherence using realistic
Hamiltonians.

Another aim of this paper is to elucidate the close link between DFS and
QECC. In \cite{Lidar:99a,Duan:98d} it was shown that DFSs are in fact
{\em maximally 
degenerate} QECCs. This result was derived from the general condition for a
code to be a QECC \cite{Knill:97b}. A very fruitful approach towards QECC
has been the stabilizer formalism developed in \cite{Gottesman:97a} which
led to the theory of universal fault-tolerant computation on QECCs \cite
{Gottesman:97}. In \cite{Lidar:99b} we considered DFSs as abelian stabilizer
codes. Here we generalize the stabilizer-framework to {\em non-abelian}
stabilizers, and show that in general DFSs are stabilizer-codes that protect
against errors in the stabilizer itself. This perspective allows in return
to view QECCs as DFSs against a certain kind of errors, and establishes a
kind of {\em duality} of QECCs and DFSs.

The paper is structured as follows: In Section \ref{Chapter1} we review
decoherence-free subsystems and place them into the context of the Markovian
master equation. For decoherence-free subspaces this has been done in \cite
{Lidar:98a,Bacon:99a}. These earlier results are therefore generalized here
to subsystems. In Section \ref{stabichapter} we introduce a generalized
stabilizer-formalism for DFS, and connect to the theory of stabilizers on
QECC developed in \cite{Gottesman:97a}. This allows us to treat DFS and QECC
within the same framework. It also sheds some light on the {\em duality}
between DFS and QECC, in particular on the performance of a DFS viewed as a
QECC and {\it vice versa}. In Section \ref{universalchapter} we deal with
universal computation on DFS within both the stabilizer-framework and the
representation-theoretic approach. We derive fault-tolerance properties of
the universal operations. In particular, we show how to obtain operations
that keep the states {\em within} a DFS during the {\em entire
switching-time } of a gate. Further we define the allowed compositions of
operations and review results on the length of gate sequences in terms of
the desired accuracy of the target gates. In Section \ref{collectivechapter}
we introduce the model of collective decoherence. Section \ref{weakchapter}
explicitly deals with the abelian case of {\em weak collective decoherence}
in which system-bath interaction coupling involves only a single system
operator. Stabilizer and error-correcting properties are developed for this
case, and it is shown how universal computation can be achieved. The same is
done for the non-abelian and more general case of {\em strong collective
decoherence} in Section \ref{strongchapter}. For both weak and strong collective decoherence we show how to fault-tolerantly encode into and read out of the respective DFSs. Finally, we analyze in Section
\ref{concatchapter} how to concatenate DFSs and QECCs to make them more robust
against perturbing errors (as proposed in \cite{Lidar:99a}) and show how the
universality results can be applied to achieve fault-tolerant universal
computation on these powerful concatenated codes. We conclude in Section \ref
{concludechapter}. Derivations and proofs of a more technical nature are presented in the Appendix.

\section{Overview of Decoherence-Free Subspaces and Subsystems}
\label{Chapter1}

\subsection{Decoherence-Free Subspaces}

Consider the dynamics of a system $S$ (the quantum computer) coupled to a
bath $B$ via the Hamiltonian
\begin{equation}
{\bf H}={\bf H}_{S}\otimes {\bf I}_{B}+{\bf I}_{S}\otimes {\bf H}_{B}+{\bf H}
_{I},
\end{equation}
where ${\bf H}_{S}$ (${\bf H}_{B}$) [the system (bath) Hamiltonian] acts on
the system (bath) Hilbert space ${\cal H}_{S}$ (${\cal H}_{B}$), ${\bf I}
_{S} $ (${\bf I}_{B}$) is the identity operator on the system (bath) Hilbert
space, and ${\bf H}_{I}$, which acts on both the system and bath Hilbert
spaces ${\cal H}_{S}\otimes {\cal H}_{B}$, is the interaction Hamiltonian
containing all the nontrivial couplings between system and bath. In general $
{\bf H}_{I}$ can be written as a sum of operators which act separately on
the system (${\bf S}_{\alpha }$'s) and on the bath (${\bf B}_{\alpha }$'s):
\begin{equation}
{\bf H}_{I}=\sum_{\alpha }{\bf S}_{\alpha }\otimes {\bf B}_{\alpha }.
\label{eq:HI}
\end{equation}
In the absence of an interaction Hamiltonian (${\bf H}_{I}=0$), the
evolution of the system and the bath are separately unitary: ${\bf U}
(t)=\exp [-i{\bf H}t]=\exp [-i{\bf H}_{S}t]\otimes \exp [-i{\bf H}_{B}t]$
(we set $\hbar =1$ throughout). Information that has been encoded (mapped)
into states of the system Hilbert space remains encoded in the system
Hilbert space if ${\bf H}_{I}=0$. However in the case when the interaction
Hamiltonian contains nontrivial couplings between the system and the bath,
information that has been encoded over the system Hilbert space does not
remain encoded over solely the system Hilbert space but spreads out instead
into the combined system and bath Hilbert space as the time evolution
proceeds. Such leakage of quantum information from the system to the bath is
the origin of the {\em decoherence} process in quantum mechanics.

Let $\tilde{{\cal H}}_{S}$ be a subspace of the system Hilbert space with a
basis $|\tilde{\imath}\rangle $. The evolution of such a subspace will be
unitary \cite{Zanardi:97a,Lidar:99a} if and only if (i)
\begin{equation}
{\bf S}_{\alpha }|\tilde{\imath}\rangle =c_{\alpha }|\tilde{\imath}\rangle
,\quad c_{\alpha }\in \CC  \label{eq:dfsS}
\end{equation}
for all $|\tilde{\imath}\rangle \in \tilde{{\cal H}}_{S}$ and for all ${\bf
S }_{\alpha }$, (ii) ${\bf H}_{S}$ does not mix states within the subspace
with states that are outside of the subspace ($\langle j^{\prime }|{\bf H}
_{S}|\tilde{\imath}\rangle =0$ for all $|\tilde{\imath}\rangle $ in the
subspace and all $|j^{\prime }\rangle $ outside of the subspace: ${\bf H}
_{S}=\tilde{{\bf H}}_{S}\oplus {\bf H}_{S}^{\prime }$) and (iii) system and
bath are initially decoupled ${\bf \rho }(0)={\bf \rho }_{S}(0)\otimes {\bf
\rho }_{B}(0)$. We call a subspace of the system's Hilbert space which
fulfills these requirements a {\em decoherence-free subspace}{\it \ }(DFS).

The above formulation of DFSs in terms of a larger closed system is exact.
It is extremely useful for finding DFSs, providing often the most direct
route via simple examination of the system components of the interaction
Hamiltonian. In practical situations, however, the closed-system formulation
of DFSs is often too strict. This is because the closed-system formulation
incorporates the possibility that information which is put into the bath
will back-react on the system and cause a recurrence. Such interactions will
always occur in the closed-system formulation (due to the the
Hamiltonian being Hermitian). However, in many practical situations the likelihood of such
an event is extremely small. Thus, for example, an excited atom which is in
a ``cold'' bath will radiate a photon and decohere but the bath will not in
turn excite the atom back to its excited state, except via the (extremely
long) recurrence time of the emission process. In these situations a more
appropriate way to describe the evolution of the system is via a quantum
dynamical semigroup master equation \cite{Lindblad:76a,Alicki:87a}. By
assuming that (i) the evolution of system density matrix is a one-parameter
semigroup, (ii) the system density matrix retains the properties of a
density matrix including ``complete positivity'', and (iii) the system and
bath density matrices are initially decoupled, Lindblad \cite{Lindblad:76a}
has shown that the most general evolution of the system density matrix ${\bf
\rho }_{S}(t)$ is governed by the master equation
\begin{eqnarray}
{\frac{d{\bf \rho }_{S}(t)}{dt}} &=&-i[{\bf H}_{S},{\bf \rho }_{S}(t)]+{\tt
L }_{D}[{\bf \rho }_{S}(t)]  \nonumber \\
{\tt L}_{D}[{\bf \rho }_{S}(t)] &=&{\frac{1}{2}}\sum_{\alpha ,\beta
=1}^{M}a_{\alpha \beta }\left( [{\bf F}_{\alpha },{\bf \rho }_{S}(t){\bf F}
_{\beta }^{\dagger }]+[{\bf F}_{\alpha }{\bf \rho }_{S}(t),{\bf F}_{\beta
}^{\dagger }]\right)  \label{eq:mastereq}
\end{eqnarray}
where ${\bf H}_{S}$ is the system Hamiltonian, the operators ${\bf F}
_{\alpha }$ constitute a basis for the $M$-dimensional space of all bounded
operators acting on ${\cal H}_{S}$, and $a_{\alpha \beta }$ are the elements
of a positive semi-definite Hermitian matrix. As above, let $\tilde{{\cal H}}
_{S}$ be a subspace of the system Hilbert space ${\cal H}_{S}$ with a basis $
|\tilde{\imath}\rangle $. The evolution over such a subspace is then unitary
\cite{Lidar:98a} iff
\begin{equation}
{\bf F}_{\alpha }|\tilde{\imath}\rangle =c_{\alpha }|\tilde{\imath}\rangle
,\quad c_{\alpha }\in \CC  \label{eq:dfsM}
\end{equation}
for all $|\tilde{\imath}\rangle $ and for all ${\bf F}_{\alpha }$. While
this condition appears to be identical to Eq~(\ref{eq:dfsS}), there is an
important difference between the ${\bf S}_{\alpha }$'s and the ${\bf F}
_{\alpha }$'s which makes these two decoherence-freeness conditions
different. In the Hamiltonian formulation of DFSs, the Hamiltonian is
Hermitian. Thus the expansion for the interaction Hamiltonian Eq.~(\ref
{eq:HI}) can always be written such that the ${\bf S}_{\alpha }$ are also
Hermitian. On the other hand, the ${\bf F}_{\alpha }$'s in the master
equation, Eq.~(\ref{eq:mastereq}), need only be bounded operators acting on $
{\cal H}_{S}$ and thus the ${\bf F}_{\alpha }$'s need not be Hermitian.
Because of this difference, Eq.~(\ref{eq:dfsM}) allows for a broader range
of subspaces than Eq.~(\ref{eq:dfsS}). For example consider the situation
where there are only two nonzero terms in a master equation for a two-level
system, corresponding to ${\bf F}_{1}=\sigma _{-}$ and ${\bf F}_{2}=\sigma
_{z}$ where $\sigma _{-}=|0\rangle \langle 1|$ and $\sigma _{z}=|0\rangle
\langle 0|-|1\rangle \langle 1|$ (e.g., cooling with phase damping). In this
case there is a DFS corresponding to the single state $|0\rangle $. In the
Hamiltonian formulation, inclusion of ${\bf S}_{1}=\sigma _{-}$ in the
interaction Hamiltonian expansion Eq.~(\ref{eq:HI}) would necessitate a
second term in the Hamiltonian with ${\bf S}_{2}=\sigma _{-}^{\dagger }$,
along with the ${\bf S}_{z}=\sigma _{z}$ as above. For this set of
operators, however, Eq.~(\ref{eq:dfsS}) allows for no DFS.

\subsection{Decoherence-Free Subsystems}

If one desires to encode quantum information over a subspace and requires
that this information remains decoherence-free, then Eqs.~(\ref{eq:dfsS}),(\ref{eq:dfsM}) provide necessary and sufficient conditions for the existence
of such DFSs. The notion of a subspace which remains decoherence-free
throughout the evolution of a system is not, however, the most general
method for providing decoherence-free encoding of information in a quantum
system. Recently, Knill, Laflamme, and Viola \cite{Knill:99a} have
discovered a method for decoherence-free coding into {\em subsystems}
instead of into subspaces.

Decoherence-free subsystems \cite{Knill:99a,Zanardi:99d,DeFilippo:99a} are
most easily presented in the Hamiltonian formulation of decoherence. Let $
{\cal A}$ denote the associative algebra formed by the system Hamiltonian $
{\bf H}_{S}$ and the system components of the interaction Hamiltonian, the $
{\bf S}_{\alpha }$'s. To simplify our discussion we will assume that the
system Hamiltonian vanishes. (It is easy to incorporate the system
Hamiltonian into the ${\bf S}_{\alpha }$'s when one desires that the system
evolution preserves the decoherence-free subsystem.) We also assume that the
identity operator is included as ${\bf S}_{0}={\bf I}_{S}$ and ${\bf B}_{0}=
{\bf I}_{B}$. This will have no observable consequence but allows for the
use of an important representation theorem. ${\cal A}$ consists of linear
combinations of products of the ${\bf S}_{\alpha }$'s. Because the
Hamiltonian is Hermitian the ${\bf S}_{\alpha }$'s must be closed
under Hermitian conjugation: ${\cal A}$ is a $\dagger $-closed operator
algebra. A basic theorem of such operator algebras which include the
identity operator states that, in general, ${\cal A}$ will be a reducible
subalgebra of the full algebra of operators on ${\cal H}_{S}$ \cite
{Landsman:98a}. This means that the algebra is isomorphic to a direct sum of
$d_{J}\times d_{J}$ complex matrix algebras, each with multiplicity $n_{J}$:
\begin{equation}
{\cal A}\cong \bigoplus_{J\in {\cal J}}{\bf I}_{n_{J}}\otimes {\cal M}
(d_{J}, \CC ).  \label{eq:repthm}
\end{equation}
Here ${\cal J}$ is a finite set labeling the irreducible components of $
{\cal A}$, and ${\cal M}(d_{J},\CC)$ denotes a $d_{J}\times
d_{J} $ complex matrix algebra. It is also useful at this point to introduce
the {\em commutant} ${\cal A}^{\prime }$ of ${\cal A}$. This is the set of
operators which commutes with the algebra ${\cal A}$, ${\cal A}^{\prime
}=\left\{ {\bf X}:[{\bf X},{\bf A}]=0,\forall {\bf A}\in {\cal A}\right\} $.
They also form a $\dagger $-closed algebra, which is reducible to
\begin{equation}
{\cal A}^{\prime }=\bigoplus_{J\in {\cal J}}{\cal M}(n_{J},\CC )\otimes {\bf I}_{d_{J}}
\label{eq:commutant}
\end{equation}
over the same basis as ${\cal A}$ in Eq.~(\ref{eq:repthm}).

The structure implied by Eq.~(\ref{eq:repthm}) is illustrated schematically as
follows, for some system operator ${\bf S}_{\alpha }$:
\begin{equation}
{\bf S}_{\alpha }=\left[
\begin{tabular}{cccc}
\cline{1-1}
\multicolumn{1}{|c}{$J=1$} & \multicolumn{1}{|c}{} &  &  \\ \cline{1-2}
& \multicolumn{1}{|c}{$J=2$} & \multicolumn{1}{|c}{} &  \\ \cline{2-2}
&  & $\ddots $ &  \\ \cline{4-4}
&  &  & \multicolumn{1}{|c|}{$J=|{\cal J}|$} \\ \cline{4-4}
\end{tabular}
\right]
\end{equation}
where a typical block with given $J$ has the structure:
\begin{equation}
J\text{ given}\text{:\quad }\left[
\begin{tabular}{ccccccccccc}
\cline{1-3}
\multicolumn{1}{|c}{} &  &  & \multicolumn{1}{|c}{} &  &  &  &  &  &  &  \\
\multicolumn{1}{|c}{} & $M_{\alpha }$ &  & \multicolumn{1}{|c}{} &  &  &  &
&  &  & $\lambda =0$ \\
\multicolumn{1}{|c}{} &  &  & \multicolumn{1}{|c}{} &  &  & $\mu $ &  &  &
&  \\ \cline{1-6}
&  &  & \multicolumn{1}{|c}{} &  &  & \multicolumn{1}{|c}{$0$} &  &  &  &
\\
&  &  & \multicolumn{1}{|c}{} & $M_{\alpha }$ &  & \multicolumn{1}{|c}{$
\vdots $} &  &  &  & $\lambda =1$ \\
&  &  & \multicolumn{1}{|c}{} &  &  & \multicolumn{1}{|c}{$d_{J}-1$} &  &  &
&  \\ \cline{4-6}
&  & $\mu ^{\prime }:$ & $0$ & $\cdots $ & $d_{J}-1$ & $\ddots $ &  &  &  &
\\ \cline{8-10}
&  &  &  &  &  &  & \multicolumn{1}{|c}{} &  &  & \multicolumn{1}{|c}{} \\
&  &  &  &  &  &  & \multicolumn{1}{|c}{} & $M_{\alpha }$ &  &
\multicolumn{1}{|c}{$\lambda =n_{J}-1$} \\
&  &  &  &  &  &  & \multicolumn{1}{|c}{} &  &  & \multicolumn{1}{|c}{} \\
\cline{8-10}
\end{tabular}
\right]
\end{equation}
Associated with this decomposition of the algebra ${\cal A}$ is the
decomposition over the system Hilbert space:
\begin{equation}
{\cal H}_{S}=\sum_{J\in {\cal J}}\CC^{n_{J}}\otimes \CC^{d_{J}}.
\label{eq:repspc}
\end{equation}
Decoherence-free subsystems are defined as the situation in which
information is encoded in a single subsystem space $\CC^{n_{J}}$ of Eq.~(\ref
{eq:repspc}) (thus the dimension of the decoherence-free subsystem is $n_{J}$). The decomposition in Eq.~(\ref{eq:repthm}) reveals that information
encoded in such a subsystem will always be affected as identity on the
subsystem space $\CC^{n_{J}}$, and thus this information will not decohere. It should be
noted that the tensor product nature which gives rise to the name subsystem
in Eq.~(\ref{eq:repthm}) is a tensor product over a direct sum, and
therefore will not in general correspond to the natural tensor product of
qubits. Further, it should be noted that the subsystem nature of the
decoherence implies that the information should be encoded in a separable
way. Over the tensor structure of Eq.~(\ref{eq:repspc}) the density matrix
should split into two valid density matrices: ${\bf \rho }_{S}(0)={\bf \rho }
\otimes {\bf \gamma }$ where ${\bf \rho }$ is the decoherence-free subsystem
and ${\bf \gamma }$ is the corresponding component of the density matrix
which {\em does} decohere. Finally it should be pointed out that not all of
the subsystems in the different irreducible representations can be
simultaneously used: (phase) decoherence will occur between the different
irreducible components of the Hilbert space labeled by $J\in {\cal J}$. For
this reason, from now on we restrict our attention to the subspace defined
by a {\em given} $J$.

Decoherence-free subspaces are now easily connected to decoherence-free
subsystems. Decoherence-free subspaces correspond to decoherence-free
subsystems possessing {\em one-dimensional} irreducible matrix algebras: $
{\cal M}(1,\CC)$. The multiplicity of these one-dimensional irreducible algebras
is the dimension of the decoherence-free subspaces. In fact it is easy to
see how the decoherence-free subsystems arise out of a non-commuting
generalization of the decoherence-free subspace conditions. Let $\{|\lambda
_{\mu }\rangle \}$, $1\leq \lambda \leq n_{J}$ and $1\leq \mu \leq d_{J}$,
denote a subspace of ${\cal H}_{S}$ with given $J$. Then the condition for
the existence of an irreducible decomposition as in Eq.~(\ref{eq:repthm}) is
\begin{equation}
{\bf S}_{\alpha }|\lambda _{\mu }\rangle =\sum_{\mu ^{\prime
}=1}^{d_{J}}M_{\mu \mu ^{\prime },\alpha }|\lambda _{\mu ^{\prime }}\rangle ,
\label{eq:subcon}
\end{equation}
for all ${\bf S}_{\alpha }$, $\lambda $ and $\mu $. Notice that $M_{\mu \mu
^{\prime },\alpha }$ is {\em not} dependent on $\lambda $, in the same way
that $c_{\alpha }$ in Eq.~(\ref{eq:dfsS}) is not the same for all $|\tilde{
\imath}\rangle $ (there $\mu =1$ and fixed). Thus for a fixed $\lambda $,
the subspace spanned by $|\lambda _{\mu }\rangle $ is acted upon in some
nontrivial way. However, because $M_{\mu \mu ^{\prime },\alpha }$ is not
dependent on $\lambda $, each subspace defined by a fixed $\mu $ and running
over $\lambda $ is acted upon in an {\em identical manner} by the
decoherence process.

At this point it should be noted that the generalization of the Lindblad
master equation Eq.~(\ref{eq:mastereq}) with a decoherence-free subspace to
the corresponding master equation for a decoherence-free system is not
trivial. This is because, as above, the ${\bf F}_{\alpha }$ operators in
Eq.~(\ref{eq:mastereq}) are (for all practical purposes) not required to be
closed under conjugation. The representation theorem Eq.~(\ref{eq:repthm})
is hence not directly applicable. We will show, however, that the master
equation analog of Eq.~(\ref{eq:subcon})
\begin{equation}
{\bf F}_{\alpha }|\lambda _{\mu }\rangle =\sum_{\mu ^{\prime
}=1}^{d_{J}}M_{\mu \mu ^{\prime },\alpha }|\lambda _{\mu ^{\prime }}\rangle
\label{eq:subcond2}
\end{equation}
provides a necessary and sufficient condition for the preservation of
decoherence-free subsystems.

As above, we consider a subspace of the system Hilbert space spanned by $
|\lambda _{\mu }\rangle $, with $1\leq \lambda \leq n_{J}$ and $1\leq \mu
\leq d_{J}$. Our notation will be significantly simpler if we explicitly
write out the formal tensor product over this subspace: $|\lambda _{\mu
}\rangle =|\lambda \rangle \otimes |\mu \rangle $. In the subsystem
notation, we claim that the decoherence-free subsystem condition is
\begin{equation}
{\bf F}_{\alpha }|\lambda \rangle \otimes |\mu \rangle =|\lambda \rangle
\otimes {\bf M}_{\alpha }|\mu \rangle .
\end{equation}
A proper decomposition of the system Hilbert space requires, as noted above,
that the system density matrix is a tensor product of two valid (Hermitian,
positive) density matrices:
\begin{equation}
{\bf \rho }_{S}(0)=\sum_{\lambda \lambda ^{\prime },\mu \mu ^{\prime }}\rho
_{\lambda \lambda ^{\prime }}(0)\gamma _{\mu \mu ^{\prime }}(0)|\lambda
_{\mu }\rangle \langle \lambda _{\mu ^{\prime }}^{\prime }|={\bf \rho }
(0)\otimes {\bf \gamma }(0),
\end{equation}
where $\rho (0)$ contains the information which will remain
decoherence-free, and $\gamma (0)$ is an arbitrary but valid density matrix.

In general the operators ${\bf F}_{\alpha }$ will not be decomposable as a
single tensor product corresponding to ${\bf \rho }(0)\otimes {\bf \gamma }
(0)$. Rather, they will be a sum over such tensor products, corresponding to
an expansion over an operator basis: ${\bf F}_{\alpha }=\sum_{p}{\bf N}
_{\alpha }^{p}\otimes {\bf M}_{\alpha }^{p}$. The decohering generator of
evolution (\ref{eq:mastereq}) thus becomes
\begin{eqnarray}
{\tt L}_{D}[{\bf \rho }_{S}(0)] &=&{\frac{1}{2}}\sum_{\alpha \beta
}a_{\alpha \beta }\sum_{pq}\left( 2{\bf N}_{\alpha }^{p}{\bf \rho }(0){\bf N}
_{\beta }^{q\dagger }\otimes {\bf M}_{\alpha }^{p}{\bf \gamma }(0){\bf M}
_{\beta }^{q\dagger }-{\bf N}_{\beta }^{q\dagger }{\bf N}_{\alpha }^{p}{\bf
\rho }(0)\otimes {\bf M}_{\beta }^{q\dagger }{\bf M}_{\alpha }^{p}{\bf
\gamma }(0)\right.  \nonumber \\
&-&\left. {\bf \rho }(0){\bf N}_{\beta }^{q\dagger }{\bf N}_{\alpha
}^{p}\otimes {\bf \gamma }(0){\bf M}_{\beta }^{q\dagger }{\bf M}_{\alpha
}^{p}\right) .
\end{eqnarray}
Tracing over the $\gamma $ component, and using the cyclic nature of the
trace allows one to factor out a common $m_{\alpha \beta }^{pq}\equiv {{\rm
Tr}_{\gamma }}({\bf M}_{\alpha }^{p}{\bf \gamma }(0){\bf M}_{\beta
}^{q\dagger })$, yielding:
\[
{\rm Tr}_{\gamma }\{{\tt L}_{D}[{\bf \rho }_{S}(0)\}={\frac{1}{2}}
\sum_{\alpha \beta ,pq}a_{\alpha \beta }m_{\alpha \beta }^{pq}\left( 2{\bf N}
_{\alpha }^{p}{\bf \rho }(0){\bf N}_{\beta }^{q\dagger }-{\bf N}_{\beta
}^{q\dagger }{\bf N}_{\alpha }^{p}{\bf \rho }(0)-{\bf \rho }(0){\bf N}
_{\beta }^{q\dagger }{\bf N}_{\alpha }^{p}\right)] .
\]
The evolution of the $\rho $ component of the density matrix thus satisfies
the standard master equation (\ref{eq:mastereq}), for which it is known that
the evolution is decoherence-free \cite{Lidar:98a} if and only if
\begin{equation}
{\bf N}_{\alpha }^{q}|\lambda \rangle =c_{\alpha ,q}|\lambda \rangle \quad
\forall \alpha .
\end{equation}
This implies that the necessary and sufficient condition for a
decoherence-free subsystem is
\begin{equation}
{\bf F}_{\alpha }=\sum_{q}c_{\alpha ,q}{\bf I}\otimes {\bf M}_{\alpha }^{q}=
{\bf I}\otimes \sum_{q}c_{\alpha ,q}{\bf M}_{\alpha }^{q}={\bf I}\otimes
{\bf M}_{\alpha },
\end{equation}
which is the claimed generalization of the Hamiltonian condition of
decoherence-free subsystems, Eq.~(\ref{eq:subcond2}).

We will use the acronym DFS to denote both decoherence-free subsystems and
their restriction, decoherence-free subspaces, whenever no confusion can
arise. When we refer to DF subspaces we will be specifically referring to
the one-dimensional version of the DF subsystems.

\section{The Stabilizer Formalism and Error-Correction}

\label{stabichapter}

In the theory of quantum error correcting codes (QECCs) it proved fruitful
to study properties of a code by considering its stabilizer ${\cal S}$. This
is the group formed by those system operators which leave the codewords
unchanged, i.e., they ''stabilize'' the code. Properties of stabilizer codes
and the theory of quantum computation on these stabilizer codes have been
developed in \cite{Gottesman:97}. In the framework of QECCs, the stabilizer
allows on the one hand to identify the errors the code can detect and
correct. On the other hand it also permits one to find a set of universal,
fault-tolerant gates by analyzing the centralizer of ${\cal S}$, defined as
the set of operations that commute with all elements in ${\cal S}$ (equal to
the normalizer - the set of operations that preserve ${\cal S}$ under conjugation -- in the case of the Pauli group). In the context of
QECCs, the stabilizer ${\cal S}$ is restricted to elements in the
Pauli-group, i.e., the group of tensor products of ${\bf I,X,Y,Z}$, and is a
{\em finite abelian} group.

The extension of stabilizer theory yields much insight into DFSs. We do this
here by defining a {\em non-abelian}, and in certain cases {\em infinite}
stabilizer group. The observation that DFSs are highly degenerate QECCs \cite
{Lidar:99a} will appear naturally from this formalism. Such a generalized
stabilizer has already been defined in previous work dealing with
decoherence-free subspaces \cite{Bacon:99b}, and its normalizer shown there
to lead to identification of local gates for universal computation. A key
consequence of this approach was the observation that the resulting gates do
not take the system out of the DFS {\em during the entire switching time of
the gate}.

We now review and extend the results in \cite{Bacon:99b} to analyze the
error detection and correction properties of DFSs and QECCs. We shall
incorporate DFSs and QECCs into a unified framework, similarly to the
representation-theoretic approach of \cite{Knill:99a,Zanardi:99d}. The
question of performing quantum computation on a specific DFS will be
addressed in the next section.

\subsection{The Stabilizer - General Theory}

An operator ${\bf S}$ is said to stabilize a code ${\cal C}$ if
\begin{equation}
|\Psi \rangle \in {\cal C}\quad {\rm iff}\quad {\bf S}|\Psi \rangle =|\Psi
\rangle \quad \forall {\bf S}\in {\cal S}.  \label{Stabicond}
\end{equation}
The set of operators $\{{\bf S}\}$ form a group ${\cal S}$, known as the
stabilizer of the code \cite{Gottesman:97}. Clearly, ${\cal S}$ is closed
under multiplication. In the theory of QECC the stabilizers that have been
studied are subgroups of the Pauli-group (tensor products of ${\bf I,X,Y,Z}$
). Since any two elements of the Pauli group either commute or anticommute,
a subgroup (and in particular the stabilizer), in this case is always {\em
abelian }\cite{Lidar:99b}. The code is thus the common eigenspace of the
stabilizer elements with eigenvalue $1$.

In general an error-process can be described by the Kraus operator-sum
formalism \cite{Kraus:83a,Knill:97b}: $\rho \rightarrow \sum_{\mu }{\bf A}
_{\mu }\rho {\bf A}_{\mu }^{\dagger }$. The Kraus-operators ${\bf A}_{\mu }$
can be expanded in a basis $\{{\bf E}_{\alpha }\}$ of ``errors''. The
standard QECC error-model assumes that errors affect single qubits {\em
independently}. Therefore the theory of QECCs has focused on searching for
codes that make quantum information robust against 1, 2,... or more
erroneous qubits. {\em Detection and correction procedures must then be
implemented at a rate higher than the intrinsic error rate}. In the QECC
error-model, the independent errors are spanned by single-qubit elements ($
{\bf X,Y,Z}$). An analysis of the error-correction properties can then be
restricted to correction of combinations of these basic errors (which are
also members of the Pauli group) acting on a certain number of qubits
simultaneously.

The {\em distance }$d${\em \ }of a QECC is the number of single-qubit errors
that have to occur in order to transform one codeword in ${\cal C}$ to
another codeword in ${\cal C}$. An error $E$ is {\em detectable} if it takes
a codeword to a subspace of the Hilbert space that is orthogonal to the
space spanned by ${\cal C}$ (this can be observed by a non-perturbing
orthogonal von Neuman measurement). A distance $d$ code can detect up to $
d-1 $ errors. In order to be able to {\em correct} an error on a certain codeword the error (up to a degenerate action of different errors) also needs to be identified, so that it can be
undone. Hence errors on different codewords have to take the
codewords to different orthogonal subspaces. The above translates to the
QECC-condition \cite{Knill:97b}:

A QECC ${\cal C}$ can correct errors ${\cal E}=\{{\bf E}_{\alpha }\}$ if and
only if
\begin{equation}
\langle \Psi _{j}|{\bf E}_{\beta }^{\dagger }{\bf E}_{\alpha }|\Psi
_{i}\rangle =c_{\alpha \beta }\delta _{ij}\quad \forall {\bf E_{\alpha
},E_{\beta }}\in {\cal E}.  \label{QECCcond}
\end{equation}

The stabilizer of a QECC allows identification of the errors which the code
can detect and correct \cite{Gottesman:97a}. Two types of errors can be
dealt with by stabilizer codes: (i) errors ${\bf E}_{\alpha }^{\dagger }{\bf
E}_{\beta }\neq {\bf I}$ that anticommute with an ${\bf S}\in {\cal S}$, and
(ii) errors that are part of the stabilizer (${\bf E}_{\alpha }\in {\cal S}$
). It is straightforward to see that both (i) and (ii) imply the QECC
condition Eq.~(\ref{QECCcond}). For case (i), if ${\bf E}_{i}^{\dagger }{\bf
E}_{j}{\bf S}=-{\bf SE}_{i}^{\dagger }{\bf E}_{j}$, then $\langle \Psi _{j}|
{\bf E}_{\beta }^{\dagger }{\bf E}_{\alpha }|\Psi _{i}\rangle =\langle \Psi
_{j}|{\bf E}_{\beta }^{\dagger }{\bf E}_{\alpha }{\bf S}|\Psi _{i}\rangle
=-\langle \Psi _{j}|{\bf SE}_{\beta }^{\dagger }{\bf E}_{\alpha }|\Psi
_{i}\rangle =-\langle \Psi _{j}|{\bf E}_{\beta }^{\dagger }{\bf E}_{\alpha
}|\Psi _{i}\rangle $. Hence $\langle \Psi _{j}|{\bf E}_{\beta }^{\dagger }
{\bf E}_{\alpha }|\Psi _{i}\rangle =0$ and $c_{\alpha \beta }=\delta
_{\alpha \beta }$. Errors of type (ii), ${\bf E_{\alpha }}\in {\cal S}$,
leave the codewords unchanged and therefore trivially lead to Eq.~(\ref
{QECCcond}). The first class, (i), are errors that require active
correction. The second class, (ii), are ``degenerate'' errors that do not
affect the code at all. QECCs can be regarded as (passive) DFSs for the
errors in their stabilizer \cite{Lidar:99b}. Conversely, being passive,
highly degenerate codes \cite{Lidar:99a}, DFSs can be viewed as a class of
stabilizer codes that protect against type (ii) errors [i.e., where the $
{\bf A}_{\mu }$ are linear combinations of elements generated (under
multiplication) by the stabilizer], and against the (usually small)\ set of
errors that anticommute with the DFS stabilizer. The stabilizer thus provides a unified tool to identify the errors that a
given code can deal with, as a DFS {\em and} as a QECC. An analysis of the
properties of DFSs with a stabilizer in the Pauli-group has been carried out
in \cite{Lidar:99b}.

\subsection{DFS-Stabilizer}

Most of the DFSs stemming from physical error-models will not have a
stabilizer in the Pauli-group, i.e., they are {\em non-additive codes}. The stabilizer may even be infinite. In
particular, the codes obtained from a noise model where errors arise from a
symmetric coupling of the system to the bath and that form the focus of this
paper, are of this type.

As discussed in the previous section, a DFS is completely specified by the
condition:
\begin{equation}
{\bf S}_{\alpha }|\mu \rangle \otimes |\lambda \rangle =|\mu \rangle \otimes
{\bf M}_{\alpha }|\lambda \rangle ,  \label{eq:DFScond}
\end{equation}
arising from the splitting of the algebra generated by the ${\bf S}_{\alpha
} $'s: ${\cal A}=\bigoplus_{J\in {\cal J}}{\bf I}_{n_{J}}\otimes {\cal M}
(d_{J},\CC)$. This splitting of the algebra has allowed both DFSs and QECCs to be put
into a similar framework \cite{Knill:99a,Zanardi:99d}. We will now show that
the DFS condition on the algebra ${\cal A}$ generated by the ${\bf S}
_{\alpha }$ can be converted into a stabilizer condition on the complex Lie
algebra generated by the ${\bf S}_{\alpha }$'s. We define the continuous
DFS-stabilizer ${\bf D}(\vec{v})$ as
\begin{equation}
{\bf D}(v_{1},v_{2},\dots v_{N})=\exp \left[ \sum_{\alpha }v_{\alpha }\left(
{\bf S}_{\alpha }-{\bf I}\otimes {\bf M}_{\alpha }\right) \right] ,\quad
v_{\alpha }\in \CC.  \label{eq:DFSstabi}
\end{equation}
Clearly, if the DFS condition Eq.~(\ref{eq:DFScond}) is fulfilled for a set
of states $|\mu \rangle \otimes |\lambda \rangle $, then
\begin{equation}
{\bf D}(\vec{v})|\mu \rangle \otimes |\lambda \rangle =|\mu \rangle \otimes
|\lambda \rangle \quad \forall v_{\alpha }\in \vec{v}.
\label{eq:stable}
\end{equation}
Thus the DFS condition implies that the ${\bf D}(\vec{v})$ {\em stabilize}
the DFS. Further if Eq.~(\ref{eq:stable}) holds then in particular it must
hold for a $\vec{v}$ which has only one non-vanishing component $v_{\beta }$. Thus Eq.~(\ref{eq:stable}) implies that ${\bf D}(0,\dots ,0,v_{\beta
},0\dots 0)|\mu \rangle \otimes |\lambda \rangle =|\mu \rangle \otimes
|\lambda \rangle $. Recalling that $\exp [\cdot ]$ is a one-to-one mapping
from a neighborhood of the zero matrix to a neighborhood of the identity
matrix, it follows that there must exist a small enough $
v_{\alpha }$ such that Eq.~(\ref{eq:stable}) implies the DFS condition Eq.~(\ref{eq:DFScond}). Thus we see that we can convert the DFS condition into a
condition on the stabilizer of the complex Lie algebra generated by the $
{\bf S}_{\alpha }-{\bf I}\otimes {\bf M}_{\alpha }$'s:
\begin{equation}
|\Psi \rangle \in {\rm DFS}\quad {\rm iff}\quad {\bf D}(\vec{v})|\Psi
\rangle =|\Psi \rangle \quad \forall \vec{v}\in \CC^{N}
\end{equation}
In some cases we will be able to pick a {\em finite} subgroup from elements
of ${\bf D}(\vec{v})$ which constitutes a stabilizer. We will mention these
instances in the following sections. However, apart from the conceptual
framework, our main motivation to introduce the stabilizer for a DFS is to
be able to analyze the errors which a DFS (i) {\em detects/corrects} (as a
QECC), and (ii) those which it {\em avoids} (passive error correction). The
continuous stabilizer provided in Eq.~(\ref{eq:DFSstabi}) will be sufficient
to study these errors.

As in the previous paragraph, errors ${\bf E}_{\alpha }$ (i) that
anticommute with an element in the stabilizer will take codewords to
subspaces that are orthogonal to the code. These errors will be detectable
(and correctable if ${\bf E}_{\beta }^{\dagger }{\bf E}_{\alpha }$
anticommutes with a stabilizer element) \cite{Gottesman:97a}. In order
to identify the
QECC-properties of a DFS, it will be convenient to look for elements of the
Pauli group among the ${\bf D}(\vec{v})$.

A code ${\cal C}$ with stabilizer ${\bf D}(\vec{v})$ will avoid errors of
type (ii) in its stabilizer in the sense that, if all of the Kraus operators
of a given decoherence process can be expanded over stabilizer elements $
{\bf A}_{i}(t)=\int_{\CC^{N}}e_{i}(\vec{v},t){\bf D}(\vec{v})d\vec{v}$, then
\begin{equation}
\rho (t)=\sum_{i}{\bf A}_{i}(t)\rho (0){\bf A}_{i}^{\dagger
}(t)=\sum_{i}\left| \int_{\CC^{N}}e_{i}(\vec{v},t)d\vec{v}
\right| ^{2}\rho (0).
\end{equation}
The normalization condition $\sum_{i}{\bf A}_{i}(t)^{\dagger }{\bf A}
_{i}(t)= {\bf I}$ then implies that $\sum_{i}\left| \int\limits_{\CC^{N}}e_{i}(\vec{v},t)d\vec{v}
\right| ^{2}=1$. Consequently, as expected, the DFS does not evolve. Hence
we see that the stabilizer provides an efficient method for identifying the
errors which a code {\em avoids}. In later sections we analyze the concrete form of the stabilizer Eq.~(\ref
{eq:DFSstabi}) for the error models studied in this paper.

\section{The Commutant and Universal Quantum Computation on a DFS}

\label{universalchapter}

A DFS is a promising way to {\em store} quantum information in a robust
fashion \cite{Bacon:99a}. From the perspective of quantum computation
however, it is even more important to be able to {\em controllably
transform} states in a DFS, if it is to be truly useful for quantum
information processing. More specifically, to perform quantum algorithms on a DFS one has to be able to
perform {\em universal quantum computation} using decoherence-free states. The notion of
universal computation is the following: with a restricted set of operations
or interactions at hand, one wishes to implement any unitary transformation
on the given Hilbert space, to an arbitrary degree of accuracy. From a
physical implementation perspective it seems clear that the operations used
(gates) should be limited to at most two-body interactions. In particular we wish to identify a finite set of
such gates that is universal on a DFS.

Since we do not wish to implement active QECC, we impose a very
stringent requirement on the operations we shall allow for computation
using DFSs: we do not allow gates that ever take the decoherence free
states outside the DFS, where the states would decohere under the noise- process
considered.\footnote{We shall lift this requirement in
Sec.~\ref{concatchapter}.} As a first step towards this goal we thus
need to be able to identify the physical operations which perform
transformations entirely within the DFS.

\subsection{Operations that Preserve the DFS}

\label{DFSpreserve}

\noindent There are essentially two equivalent approaches to identify
the ``encoded''
operations that preserve a DFS. One is via the normalizer of the stabilizer of a code \cite
{Bacon:99b}; the second is via the commutant of the ${\bf \dagger }$-closed
algebra generated by the error operators \cite{Knill:99a}. Both will be
briefly reviewed here.

{\em Computation on a stabilizer DFS:} The stabilizer formalism is very
useful for identifying allowed gates that take codewords to codewords \cite
{Gottesman:97}. An operation ${\bf U}$ keeps code-words $|\Psi
\rangle $ inside the code-space, if and only if the transformed state ${\bf
U }|\Psi \rangle $ is an element of the code ${\cal C}$. Thus, using the
stabilizer condition (\ref{Stabicond}) for codes with stabilizer ${\cal S}$
and ${\cal C}=\{|\Psi \rangle :{\bf S}|\Psi \rangle =|\Psi \rangle \,\forall
{\bf S}\in {\cal S}\}$, we have
\begin{equation}
{\bf U}|\Psi \rangle \in {\cal C}\quad {\rm iff}\quad {\bf SU}|\Psi \rangle
= {\bf U}|\Psi \rangle \quad \forall {\bf S}\in {\cal S}.
\end{equation}
This implies ${\bf U^{-1}SU}|\Psi \rangle =|\Psi \rangle $ and so ${\bf
U^{-1}SU}\in {\cal S}$: Allowed operations ${\bf U}$ transform stabilizer
elements ${\bf S}$ by conjugation into stabilizer elements; ${\bf U}$ is in
the {\em normalizer} of ${\cal S}$ (if ${\cal S}$ is a group). If we
restrict the allowed operations to gates in the Pauli-group (as is done in
\cite{Gottesman:97}), then the allowed gates ${\bf U}$ will fix the
stabilizer {\em pointwise }(element by element). In the case of DFS with a
continuous stabilizer ${\bf D}(\vec{v})$, the above translates to the
following condition \cite{Bacon:99b}
\begin{equation}
{\bf U}{\bf D}(\vec{v}){\bf U}^{\dagger }={\bf D}(\vec{v}^{\prime }(\vec{v}
)),  \label{eq:DFSnor1}
\end{equation}
together with the requirement that ${\bf D}(\vec{v}^{\prime }(\vec{v}))$
must cover ${\cal S}$. To satisfy the covering condition, it is sufficient
to have $\vec{v}^{\prime }(\vec{v})$ be a one-to-one mapping.

Eq.~(\ref{eq:DFSnor1}), derived by generalizing concepts from the theory of
stabilizers in the Pauli group, is a condition that allows one to identify
gates ${\bf U}$ that transform codewords to codewords. In a physical
implementation these gates will be realized by turning on Hamiltonians ${\bf
H}$ between physical qubits for a certain time $t$: ${\bf U}(t)=e^{-it{\bf H}
} $. So far we only required that the action of the gate preserve the
subspace at the {\em conclusion} of the gate operation, but not that the
subspace be preserved {\em throughout the entire duration} of the gate
operation. The stabilizer approach allows us to further identify the more
restrictive set of Hamiltonians that keep the states within the DFS {\em
throughout the entire switching time of the gate}. As a result, in the limit
of ideal gates, the entire system is free from noise at all times. This is
different from QECC, since there errors continuously take the codewords
outside of the code-space \cite{Paz:98}, and hence error correction needs to
be applied frequently {\em even in the limit of perfect gate operations}.
Imperfections in gate operations can be dealt with in the DFS approach by
concatenation with a QECC \cite{Lidar:99a}, as shown explicitly for the
exchange interaction in \cite{Lidar:99c}.

By rewriting condition (\ref{eq:DFSnor1}) as ${\bf U}(t){\bf D}(\vec{v})=
{\bf D}(\vec{v}^{\prime }(\vec{v},t)){\bf U}(t)$, taking the derivative with
respect to $t$ and evaluating at $t=0$ we obtain ${\bf H}{\bf D}(\vec{v})=
{\bf D}(\vec{v}^{\prime }(\vec{v},0)){\bf H}+i\left. \frac{\partial {\bf D}}{
\partial \vec{v}^{\prime }}\frac{d\vec{v}^{\prime }}{dt}\right| _{t=0}$, so
that:

{\it Theorem 1}{\bf ---} A sufficient condition for the generating
Hamiltonian to keep a state at all times entirely within the DFS is ${\bf H}
{\bf D}(\vec{v})={\bf D}(\vec{v}^{\prime }(\vec{v})){\bf H}$ where $\vec{v}
^{\prime }(\vec{v})$ is one-to-one and time-independent.

For most applications we will only need gates that {\em commute} with all
stabilizer elements. The condition for the generating Hamiltonian then
simplifies to ${\bf H}{\bf D}(\vec{v})={\bf D }(\vec{v}){\bf H}$.

{\em Computation on irreducible subspaces:} We can derive conditions to
identify allowed gates on a DFS by using the representation theoretic
approach developed in \cite{Knill:99a}, \cite{Zanardi:99d} and section \ref
{Chapter1}. Recall that the decomposition of the algebra ${\cal A}\cong \bigoplus_{J\in {\cal J}}{\bf I}_{n_{J}}\otimes
{\cal M}(d_{J},\CC)$ generated by
the errors $\{ {\bf S}_\alpha \}$ induces a splitting of the Hilbert space ${\cal H}_{S}=\sum_{J\in
{\cal J}}\CC
^{n_{J}}\otimes \CC^{d_{J}}$ into subspaces possessing a tensor product structure
suitable to isolate decoherence-free subsystems. The set of operators in the
commutant of ${\cal A}$, ${\cal A}^{\prime }=\left\{ {\bf X}:[{\bf X},{\bf A}
]=0,\forall {\bf A}\in {\cal A}\right\} ={\cal A}^{\prime }=\bigoplus_{J\in
{\cal J}}{\cal M}(n_{J},\CC)\otimes {\bf I}_{d_{J}}$,
obviously generate transformations that affect the codespace only. In
particular, they take states in a DFS to other states in that same DFS. $
{\cal A}^{\prime }$ is generated by operators which commute with the ${\bf S}
_{\alpha }$. Again, our goal is to find gates that act within a DFS {\em
during the entire switching time}, and to this end we need to identify
Hermitian operators {\bf H} in ${\cal A}^{\prime }$ to generate an evolution
${\bf U}(t)=\exp [-it{\bf H}]$ on the DFS.

{\it Theorem 2--- }A sufficient condition for a Hamiltonian ${\bf H}$ to
generate dynamics ${\bf U}(t)=\exp [-it{\bf H}]$ which preserves a DFS is
that ${\bf H}$ be in the commutant of the algebra ${\cal A}$.

However, because we can only use one particular DFS (corresponding to a
specific $K\in {\cal J}$) to store quantum information (the coherences
between superpositions of different DFSs are not protected), the operators
which commute with the ${\bf S}_{\alpha }$'s are not the only operators
which perform non-trivial operations on a specific DFS. The operations in $
{\cal A}^{\prime }$ preserve all DFSs in parallel. However, if we restrict
our system to only one such DFS, we do not need any constraints on the
evolution of the other subspaces. It is then possible to construct a
necessary and sufficient condition for a Hamiltonian by modifying the
commutant to:

\begin{equation}
{\cal T}\cong \left( {\cal M}(n_{K},\CC)\otimes {\bf I}
_{d_{K}}\right) \oplus {\cal M}(d-d_{K}n_{K},\CC)  \label{eq:T}
\end{equation}
where $d_{K}={\rm dim}({\cal H}_{S})$ and just leaves the specific DFS
(${K}$) invariant.

{\it Theorem 3---} A necessary and sufficient condition for a Hamiltonian $
{\bf H}$ to generate dynamics which preserves a DFS corresponding to the
irreducible representation $K\in {\cal J}$, is ${\bf H}\in {\cal T}$.

We will use both the stabilizer and the commutant approaches, to find
a set of universal gates for decoherence processes of physical
relevance. In the cases discussed in this paper, any one of the two
approaches is clearly sufficient and we do not need all theorems in full
generality. However we provide here a general framework and the tools
required to analyze DFS and QECC stemming from {\em any} error model.

Finally we should point out again that from a practical perspective, it is
crucial to look for the Hermitian operations which perform nontrivial
operations on the DFS and which correspond to {\em only one or two-body
physical interactions}. Without this requirement, it is clear that one can
{\em always} \cite{Zanardi:99c} construct a set of Hamiltonians
(satisfying the conditions of Theorem 2) which span
the allowed operations on a DFS. A
primary goal of this paper is therefore to construct such one and two-body
Hamiltonians for specific decoherence mechanisms, in order to achieve
true universal computation on the corresponding DFSs.

\subsection{Universality and Composition of Allowed Operations}

\label{allowed}

Using the tools developed in the previous subsection, we can now find local
one-and-two qubit gates that represent encoded operations on DFSs. However,
in general, a discrete set of gates applied in alternation is not sufficient
to generate a universal set of gates. Nor is it sufficient to obtain every
encoded unitary operation exactly. Furthermore, for analysis of the
complexity of computations performed with a given universal set of gates, it
is essential to keep under control the number of operations needed to achieve a
certain gate within a desired accuracy. In the theory of universality (e.g.,
\cite{Aharonov:99a}) the composition laws of operations have been analyzed
extensively. We will review the essential results relevant for our purposes
here.

Let us assume that we have a set of (up to two-body) Hamiltonians ${\sf H}
=\{ {\bf H}_{i}:i=1,\ldots ,M\}$ that take DFS states to DFS states. We will
construct gates using the following composition laws:

\begin{enumerate}
\item  {{\bf Arbitrary phases:} Any interaction can be switched on for an
arbitrary time. Thus any gate of the form ${\bf U}(t)=\exp \left(-it{\bf H}
_{i}\right) $ can be implemented.}

\item  {{\bf Trotter formula:} Gates performing sums of Hamiltonians are
implemented by using the short-time approximation to the Trotter formula }$
\exp \left[ i(t_{1}{\bf H}_{i}+t_{2}{\bf H}_{j})\right] ${$
=\lim_{n\rightarrow \infty }\left[ \exp \left( i\frac{t_{1}}{n}{\bf H}
_{i}\right) \exp \left( i\frac{t_{2}}{n}{\bf H}_{j}\right) \right] ^{n}$:}
\begin{equation}
e^{i(t_{1}{\bf H}_{i}+t_{2}{\bf H}_{j})/n}=e^{it_{1}{\bf H}_{i}/n}e^{it_{2}
{\bf H}_{j}/n}+O(\frac{1}{n^{2}}).
\end{equation}
{This is achieved by quickly turning on and off the two interactions $
{\bf H}_{i},{\bf H}_{j}$ with appropriate ratios of duration times. An
alternative, direct, way of implementing this gate is to switch on the two
interactions {\em simultaneously} for the appropriate time-intervals. }

\item  {{\bf Commutator:} It is possible to implement the commutator of
operations that are already achievable. This is a consequence of the Lie
product formula
\[
\exp [{\bf H}_{i},{\bf H}_{j}]=\lim_{n\rightarrow \infty }
\left[ \exp \left( i{\bf H}_{i}/\sqrt{n}\right) \exp \left( i{\bf H}_{j}/
\sqrt{n}\right) \exp \left( -i{\bf H}_{i}/\sqrt{n}\right) \exp \left( -i{\bf
H}_{j}/\sqrt{n}\right) \right] ^{n} ,
\]
which has the short-time approximation}
\begin{equation}
e^{t{[{\bf H}_{i},{\bf H}_{j}]/n}}{=}e^{it{\bf H}_{i}/\sqrt{n}}e^{it{\bf H}
_{j}/\sqrt{n}}e^{-it{\bf H}_{i}/\sqrt{n}}e^{-it{\bf H}_{j}/\sqrt{n}}+O(\frac{
1}{n\sqrt{n}}).
\end{equation}
Again, the gate $e^{it(-i{[{\bf H}_{i},{\bf H}_{j}]})}$ can be implemented
to high precision by alternately switching on and off the appropriate two
interactions with a specific duration ratio.\footnote{Note that in order
to implement $e^{-it{\bf A}}$ we would use $e^{i\vartheta {\bf A}}=I$ and
implement $e^{i(\vartheta -t){\bf A}}$ instead. This depends on ${\bf A}$
having rationally related eigenvalues, which will always be the case for the
Hamiltonians of interest to us.}

\item  {\bf Conjugation by unitary evolution:} Another useful action in
constructing universal sets of gates comes from the observation that if a
specific gate ${\bf U}$ and its inverse ${\bf U}^{\dagger }$ can be
implemented, then any Hamiltonian ${\bf H}$ which can be implemented can be
modified by performing ${\bf U}$ before, and ${\bf U}^{\dagger }$ after the
gate $\exp (-it{\bf H})$. This gives rise to the transformed
Hamiltonian
\begin{equation}
{\bf U}\exp (-it{\bf H}){\bf U}^{\dagger }=\exp (-it{\bf U}{\bf H}{\bf U}
^{\dagger })=\exp (-it{\bf H}_{{\rm eff}}).
\label{eq:Heff}
\end{equation}
\end{enumerate}

Note that the laws (1-3) correspond to closing the set of allowed
Hamiltonians as a Lie-algebra (scalar multiplication, addition and
Lie-commutators can be obtained out of the given Hamiltonians).

If (a subset of) the composition laws (1-4) acting on the set ${\sf H}$ give rise to
a set of gates that is dense in the group $SU(d_{K})$ (via successive
application of these gates), where $d_{K}$ is the dimension of the DFS, then
we shall refer to ${\sf H}$ as a {\em universal set of generators}.
Equivalently, this means that ${\sf H}$ generates the Lie-algebra $su(d_{K})$
(traceless matrices) via {\em scalar multiplication, addition,
Lie-commutators, and conjugation by unitaries}. The generators of this
algebra can be obtained from ${\sf H}$ by these operations.

For all
practical applications and implementations of algorithms, we will only be
interested to approximate a certain gate sequence with a given accuracy.
Note that the composition laws (2) and (3) use only repeated applications of
(1) in order to {\em approximate} a certain gate. We can replace the
requirement to perform an arbitrary phase, (1), by noting that $e^{i\tilde{t}
{\bf H}_{i}}$ is dense in the group $\{ \exp(it{\bf H}_{i}) : t\in
\lbrack 0,2\pi )\}$ if $\tilde{t}$ is an irrational multiple of $\pi $. Repeated
application of that gate can then approximate an arbitrary phase to any
desired accuracy. Thus we can in principle restrict our available gates to $
\{ \exp(i\tilde{t}_{i}{\bf H}_{i}) \}$, together with fixed irrational
switching times $\tilde{t}_{i}$. Repeated application of these gates can
then be used to approximate any operation in $SU(d_{K})$ to arbitrary accuracy.

In order to prove that a set ${\sf H}$ generates a {\em universal set} of
Hamiltonians, we use the fact that a large group of universal sets have
already been identified \cite{DiVincenzo:95a,Lloyd:95a,Barenco:95a}. It
suffices to show that ${\sf H}$ generates one of these sets, in order to
prove that ${\sf H}$ is a universal set of generators. We will use the fact
that the set of one qubit operations $SU(2)$ is generated by any two
arbitrary rotations with irrational phase, around two non-parallel axes.
Alternatively, if we are given these two rotations with {\em any} phase,
then an Euler-angle construction can be used to yield any gate in $SU(2)$ by application
of a small number of rotations (three if the axes are orthogonal). In addition we will use (and prove) a lemma ({\em Enlarging Lemma}, Appendix~\ref{appB}) that allows extension to $
su(n+1) $ of a given $su(n)$ acting on an $n$-dimensional subspace of a
Hilbert space of dimension $n+1$, with the help of an additional $su(2)$.

In order to use this approach to universality, it is crucial to have bounds
on the length of the gate sequences approximating a certain gate in terms of
the desired accuracy. This is all the more important if one universal set is
to be replaced by any other with only polynomial overhead in the number of
gates applied, for otherwise the complexity classes would not be robust
under the exchange of one set for another. The whole notion of universality
would then by questionable. The following key theorem proved independently
by Solovay and Kitaev (see \cite{Aharonov:99a}) establishes the equivalence
of universal sets, and provides bounds on the length of gate sequences for a
desired accuracy of approximation. In order to quantify the accuracy of an
approximation, we need to define a distance on matrices. Since our matrices
act in a space of given (finite) dimension $d_{K}$, any metric is as good as
any other. For example, we can use the trace-norm $d({\bf U},{\bf V})=\sqrt{
1-\frac{1}{d_{K}}{\rm Re}\left[ {\rm Tr}({\bf U}^{\dagger }{\bf V})\right] }$
. A matrix ${\bf V}$ is then said to approximate a transformation ${\bf U}$
to accuracy $\epsilon $ if $d({\bf U},{\bf V})\leq \epsilon $.

{\it Theorem (Solovay-Kitaev) ---} Given a set of gates that is dense
in $SU(2^{k})$ and closed under Hermitian conjugation, any gate ${\bf U}$ in $SU(2^{k})$ can be approximated to an
accuracy $\epsilon $ with a sequence of ${\rm poly}\left[ \log (1/\epsilon )\right]
$ gates from the set.

{\it DFS-Corollary to the Solovay-Kitaev Theorem---} Assume that the DFS
encodes a $d_{K}$-dimensional system into $n$ physical qubits. Given that
one can exactly implement the gate set $\{e^{i\tilde{t}_{i}{\bf H_{i}}}:{\bf
H}_{i}\in {\sf H}\}$, [$\tilde{t}_{i}$ are (fixed) irrational multiples of $
\pi $, and ${\sf H}$ is a universal generating set] it is possible to
approximate any gate in $SU(d_{K})$ (any encoded operation) using $m={\rm poly}
\left[ \log (1/\epsilon )\right] $ gates.

Furthermore, if we can only implement the given gates approximately, say to
an accuracy $\delta $, we will still be able to approximate the target gate:
It is known that a sequence of $m$ $\delta $-imprecise unitary matrices is
(in some norm) at most distance $m\delta $ far from the desired gate. If a
sequence of exactly implemented gates ${\bf U}_{1},\ldots, {\bf U}
_{m}$ approximates a target gate ${\bf U}$ up to $\epsilon $, and instead of
${\bf U}_{1},\ldots ,{\bf U}_{m}$, we use gates that are at most some
distance $\delta $ apart, then the total sequence will be at most $\epsilon
+m\delta =\epsilon +{\rm poly}\left[ \log (1/\epsilon )\right] \delta $ apart from
${\bf U}$. If we make sure that $\delta <\epsilon {\rm poly}\left[ \log
(1/\epsilon )\right] $ then the $\delta $-faulty sequence will still
approximate ${\bf U}$ to a precision $2\epsilon $.

If we further assume that the physical interaction that we switch on and off
is given by the device and is unlikely to change its form, then the
imprecision of the gate comes entirely through the coupling strength and the
interaction time, i.e. a faulty gate is of the form ${\bf U}_{f}=e^{i(\phi
+\Delta \phi ){\bf H}}$, where ${\bf U}=e^{i\phi {\bf H}}$ is the
unperturbed gate. The distance
\begin{eqnarray}
d({\bf U},{\bf U}_{f})&=&\sqrt{1-\frac{1}{d_{K}} \mathop{\rm Re} \left[ {\rm
Tr}(e^{i\Delta \phi {\bf H}})\right] }=\sqrt{1-\frac{1}{d_{K}} \mathop{\rm
Re} \left[ {\rm Tr}(\cos \Delta \phi 1+i\sin \Delta \phi {\bf H})\right] }
\nonumber \\
&=&\sqrt{1-\cos \Delta \phi }=\sqrt{2}\sin (\Delta \phi /2)\approx \frac{
\Delta \phi }{\sqrt{2}}\equiv \delta ,
\end{eqnarray}
is proportional to the error $\Delta \phi $ of the product of
coupling-strength and interaction time. This translates to (nearly) linear
behavior in the desired final accuracy $\epsilon $.

\section{Collective Decoherence}

\label{collectivechapter}

We now focus on a particularly interesting and useful model of a DFS. This
is the case of collective decoherence on $n$ qubits. We distinguish
between two forms of collective decoherence. The first, and simpler, type of
collective decoherence is {\em weak collective decoherence} (WCD). We define
the collective operators as

\begin{equation}
{\bf S}_{\alpha }\equiv \sum_{j=1}^{n}\sigma
_{\alpha }^{j} ,
\end{equation}
where $\sigma_{\alpha }^{j}$ denotes a tensor product of
the $\alpha ^{{\rm th}}$ Pauli matrix, $\alpha =x,y,z$,
\begin{equation}
\sigma _{x}=\left(
\begin{array}{cc}
0 & 1 \\
1 & 0
\end{array}
\right) \quad \sigma _{y}=\left(
\begin{array}{cc}
0 & -i \\
i & 0
\end{array}
\right) \quad \sigma _{z}=\left(
\begin{array}{cc}
1 & 0 \\
0 & -1
\end{array}
\right)
\end{equation}
(in the basis spanned by $\sigma _{z}$ eigenstates $|0\rangle $ and $
|1\rangle $) operating on the $j^{{\rm th}}$ qubit, and the identity on all
of the other qubits. WCD is the situation in which only one collective
operator ${\bf S}_{\alpha }$ is involved in the coupling to the bath, i.e., $
{\bf H}_{I}={\bf S}_{\alpha }\otimes {\bf B}$.

The second, more general type of collective decoherence is {\em strong
collective decoherence} (SCD). We define SCD as the general situation in
which the interaction Hamiltonian is given by ${\bf H}_{I}=\sum_{\alpha
=1}^{3}{\bf S}_{\alpha }\otimes {\bf B}_{\alpha }$. The ${\bf S}_{\alpha }$
provide a representation of the Lie algebra $su(2)$:
\begin{equation}
\lbrack {\bf S}_{\alpha },{\bf S}_{\beta } \rbrack=-2i\epsilon _{\alpha \beta
\gamma }{\bf S}_{\gamma }
\end{equation}
The ${\bf B}_{\alpha }$'s are not required to be linearly independent.

Both of these collective decoherence mechanisms are expected to arise from
the physical condition that {\em the bath cannot distinguish the system
qubits} \cite{Palma:96,Duan:97b,Zanardi:97c,Lidar:98a}.
If there are $n$ qubits interacting with a bath, the most general
interaction Hamiltonian linear in the $\sigma _{\alpha }^{i}$ is given by
\begin{equation}
{\bf H}_{I}=\sum_{i=1}^{n}\sum_{\alpha =x,y,z}\sigma _{\alpha }^{i}\otimes
{\bf B}_{i,\alpha }
\end{equation}
where the ${\bf B}_{i,\alpha }$ are bath operators. If the bath cannot
distinguish between the system qubits, then ${\bf B}_{i,\alpha }$ should not
depend on $i$ and the Hamiltonian becomes ${\bf H}_{I}=\sum_{\alpha =x,y,z}
{\bf S}_{\alpha }\otimes {\bf B}_{\alpha }$, i.e., strong collective
decoherence.

As a concrete example of such collective decoherence, consider the situation
in which the bath is the electromagnetic field, and the wavelength of the
transition between the states of the qubits is larger than the spacing
between the qubits. The electromagnetic field will interact with each of
these qubits in an identical manner, because the field strength over a
single wavelength will not vary substantially. This gives rise to the well-
known phenomena of Dicke super- and sub-radiance \cite{Dicke:54}. Whenever
the bath is a field whose energy is dependent on its wavelength and this
wavelength is much greater than the spacing between the qubits, one should
expect collective decoherence to be the dominant decoherence
mechanism. It is natural to expect this to be the case for
condensed-phase high-purity materials at low temperatures. However, to
the best of our knowledge at present a rigorous study quantifying the
relevant parameter ranges for this interesting condition to hold in
specific materials is still lacking (see Refs. \cite{Zanardi:98c,Zanardi:99a} for an
application to quantum dots, though).

\section{The Abelian Case: Weak Collective Decoherence}

\label{weakchapter}

For a decoherence mechanism with only one operator ${\bf S}_{\alpha }$
coupling to the bath, the implementation and discussion of universal
computation with local interactions is simpler than in the general
case, because we can work in the basis that diagonalizes ${\bf \ S}_{\alpha
} $ (${\bf S}_{\alpha }$ is necessarily Hermitian in the Hamiltonian model
we consider here). The algebra generated by ${\bf S}_{\alpha }$ is{\em \
abelian} and reduces to one-dimensional (irreducible) subalgebras
corresponding to the eigenvalues of ${\bf S}_{\alpha }$. More specifically, $
{\cal A}_{1}=\bigoplus_{\lambda _{J}}{\bf I}_{n_{J}}\otimes {\cal M}(\lambda
_{J})$, where $\lambda _{J}$ is the $J^{{\rm th}}$ eigenvalue with
degeneracy $n_{J}$, and ${\cal M}(\lambda _{J})$ is the algebra generated by
$\lambda _{J}$. ${\cal M}(\lambda _{J})$ acts by multiplying the
corresponding vector by $\lambda _{J}$. In this situation the DF subsystems
are only of the DF subspace type. This simpler case of weak collective
decoherence allows us to present a treatment with examples, that will make
the general case of strong collective decoherence (SCD) more intuitive.

In the following we will, without loss of generality, focus on the case $
{\bf S}_{\alpha }\equiv {\bf S}_{z}=\sum_{k=1}^{n}\sigma _{z}^{k}$.\footnote{
The cases $\alpha =x$ ($y$) follow by applying a bitwise Hadamard
(Hadamard+phase) transform to the code.} This operator is already diagonal
in the computational basis (the eigenstates are bitstrings of qubits in
either $|0\rangle $ or $|1\rangle $). Since $\sigma _{z}^{k}$ acting on the $
k^{\rm th}$ qubit contributes $1$ if the qubit is $|0\rangle $, and $-1$ if the
qubit is $|1\rangle $, the eigenvalue of a bitstring is $\#0-\#1$ (the
number of zeroes minus the number of ones), and the eigenvalues of ${\bf S}
_{z}$ are $\{n,n-2,\ldots ,-n+2,-n\}$.

The degeneracy $n_{J}$ of the eigenspace corresponding to an eigenvalue
\begin{equation}
\lambda _{J}=n-2J
\end{equation}
is
\begin{equation}
n_{J}={\ {
{n  \choose J}
} }
\end{equation}
(the number of different bitstrings with $n-J$ zeroes and $J$ ones). The
abelian algebra generated by ${\bf S}_{z}$ thus splits into one-dimensional
subalgebras with degeneracy $n_{J}$. The largest decoherence-free subspaces
in this situation correspond to the space spanned by bitstring-vectors where
the number of zeroes and the number of ones are either the same ($n$ even),
or differ by one ($n$ odd).

\subsection{The Stabilizer and Error Correction Properties}

Following the formalism developed in Section (\ref{stabichapter}) we find,
using Eq.~(\ref{eq:DFSstabi}) with $v=i\theta $ ($\theta $ can be complex),
the stabilizer for the weak case corresponding to a DFS with eigenvalue $
\lambda _{J}$ to be
\begin{equation}
{\bf Z}_{J}^{\otimes n}(\theta )=\exp [i\theta ({\bf S}_{z}-\lambda _{J}{\bf
I})]=\bigotimes_{k=1}^{n}e^{-i\lambda _{J}\theta }({\bf I}\cos \theta
+\sigma _{z}^{k}i\sin \theta )=e^{-i\lambda _{J}\theta }{\bf P}(\theta
)^{\otimes n}  \label{eq:WCDstabi}
\end{equation}
where
\begin{equation}
{\bf P}(\theta )=\left(
\begin{array}{cc}
e^{i\theta } & 0 \\
0 & e^{-i\theta }
\end{array}
\right) .
\end{equation}
For strictly real $\theta $ some of the errors which are protected against
are simply collective rotations about the $\sigma _{z}$ axis (and an
irrelevant global phase). For strictly imaginary $\theta $ we find that the
errors which are protected against are contracting collective errors of the
form ${\rm diag}(e^{\theta },e^{-\theta })$, i.e., they result in loss of
norm of the wave function. Any physical process with Kraus operators that
are linear combinations of these errors will therefore not affect the DFS.

This is the right framework in which to present another form of the
stabilizer. We note that in the case of weak collective decoherence, we can
find a stabilizer group with a {\em finite} number of elements. Define
\begin{equation}
Z_{\frac{1}{n}}=\exp \left( {\frac{2\pi i}{n}\sigma _{z}}\right) =\left(
\begin{array}{cc}
\exp ({\frac{2\pi i}{n}}) & 0 \\
0 & \exp ({-\frac{2\pi i}{n}})
\end{array}
\right) .
\end{equation}
Then the $n$-element group ${\cal Z}_{n}$ generated by $\exp \left( {-i}2\pi
\lambda _{J}/n\right) Z_{1/n}^{\otimes n}$ is a stabilizer for the DFS
corresponding to the eigenvalue $\lambda _{J}$. To see that
\begin{equation}
\exp \left( {-\frac{2\pi i\lambda _{J}}{n}}\right) Z_{\frac{1}{n}}^{\otimes
n}|\Psi \rangle =|\Psi \rangle \quad {\rm iff}\quad |\Psi \rangle \in {\rm
DFS}(\lambda _{J}),  \label{eq:easy}
\end{equation}
note that a $Z_{1/n}$ acting on a $|0\rangle $ contributes $\exp (2\pi i/n)$ to the total phase, whereas $Z_{1/n}$ acting on a $
|1\rangle $ contributes $\exp(-2\pi i/n)$. So $
Z_{1/n}^{\otimes n}$ gives a total phase of $\exp \left( {2\pi i(\#|0\rangle
-\#|1\rangle )/n}\right) =\exp(2\pi i\lambda _{J}/n) $
when acting on a bitstring. This stabilizer and Eq.~(\ref{eq:easy}) provide
a simple criterion to check whether a state is in a DFS or not.

Let us now briefly comment on the error-correction and detection properties of the code
in the WCD case. The stabilizer elements are all diagonal, and equal to a
tensor product of identical 1-qubit operators. The element ${\bf Z}^{\otimes
n}$ is in the stabilizer and anticommutes with odd-number ${\bf X}$ and ${\bf
Y}$ errors. So odd-number qubit bit-flips are detectable errors. However the
code is not able to detect any form of error involving ${\bf Z}$'s and
even-number ${\bf X}$'s and ${\bf Y}$'s, since any such error commutes with
all elements in the stabilizer.

\subsection{Nontrivial Operations}

Observe that the algebra ${\cal A}$ in the WCD case is generated entirely by
${\bf S}_{z}$. Hence, by Theorem 2, the DFS-preserving operations are those
that are in the commutant of ${\bf S}_{z}$. For single-body Hamiltonians it
is easy to see that the only non-trivial such set is formed by interactions
proportional to $\sigma _{z}^{i}$ operators. As for two-qubit Hamiltonians,
it is simpler to use Theorem 1 and the expression (\ref{eq:WCDstabi}) for
the stabilizer. We are then looking for $4\times 4$ Hermitian matrices that
commute with ${\bf P}(\theta )^{\otimes 2}$; these are of the form
\begin{equation}
{\bf T}_{ij}(z_{1},z_{2},z_{3},z_{4},h)=\left(
\begin{array}{cccc}
z_{1} & 0 & 0 & 0 \\
0 & z_{2} & h & 0 \\
0 & h^{\ast } & z_{3} & 0 \\
0 & 0 & 0 & z_{4}
\end{array}
\right)
\end{equation}
where ${\bf T}_{ij}$ acts on qubits $i$ and $j$ only. Here $z_{i}$ is real, $
h$ is complex, and the row space is spanned by the $i^{\rm th}$ and $j^{\rm th}$ qubit
basis $\{|00\rangle,|01\rangle,|10\rangle,|11\rangle\} $. We
note that systems with an internal Hamiltonian of the Heisenberg type,
\begin{equation}
{\bf H}_{{\rm Heis}}=\sum_{j=1}^{n}\epsilon _{j}\sigma _{z}^{j}+\frac{1}{2}
\sum_{i,j=1}^{n}J_{ij}\vec{{\sigma }}_{i}\cdot \vec{{\sigma }}_{j},
\end{equation}
have exactly the correct form for any pair of spins $i$,$j$. Indeed, it is
not hard to see that $[{\bf H}_{{\rm Heis}},{\bf S}_{z}]=0$ \cite
{Lidar:99c}. The Heisenberg Hamiltonian is ubiquitous, and appears, e.g., in
NMR. This means that the natural evolution of NMR systems under WCD
preserves the DFS, and implements a non-trivial computation.

The specific case
\begin{equation}
{\bf E}_{ij}\equiv {\bf T}_{ij}(1,0,0,1,1)=\left(
\begin{array}{cccc}
1 & 0 & 0 & 0 \\
0 & 0 & 1 & 0 \\
0 & 1 & 0 & 0 \\
0 & 0 & 0 & 1
\end{array}
\right) ,
\label{eq:exchange}
\end{equation}
which flips the two states $|01\rangle $ and $|10\rangle $ of qubits $i$ and
$j$ and leaves the other two states invariant, is especially important: it is
the {\em exchange interaction}. The other interactions we employ are
\begin{eqnarray}
{\bf T}_{ij}^{P}& \equiv {\bf T}_{ij}(1,0,0,0,0)={\rm diag}(1,0,0,0)
\nonumber \\
{\bf T}_{ij}^{Q}& \equiv {\bf T}_{ij}(0,0,0,1,0)={\rm diag}(0,0,0,1),
\end{eqnarray}
which introduce a {\em phase} on the state $|00\rangle $ ($P$) and $
|11\rangle $ ($Q$) of qubits $i$ and $j$; and
\begin{equation}
\bar{{\bf Z}}_{12}\equiv {\bf T}_{12}(0,0,1,0,0)\}={\rm diag}(0,0,1,0) .
\end{equation}
In the following we show that these special interactions are sufficient to obtain a universal generating
set operating entirely {\em within} a weak-collective DFS.

\subsection{Universal Quantum Computation inside the Weak-Collective DFS}

\begin{figure}[!htb]
\hspace{0.2\textwidth}
\psfig{figure=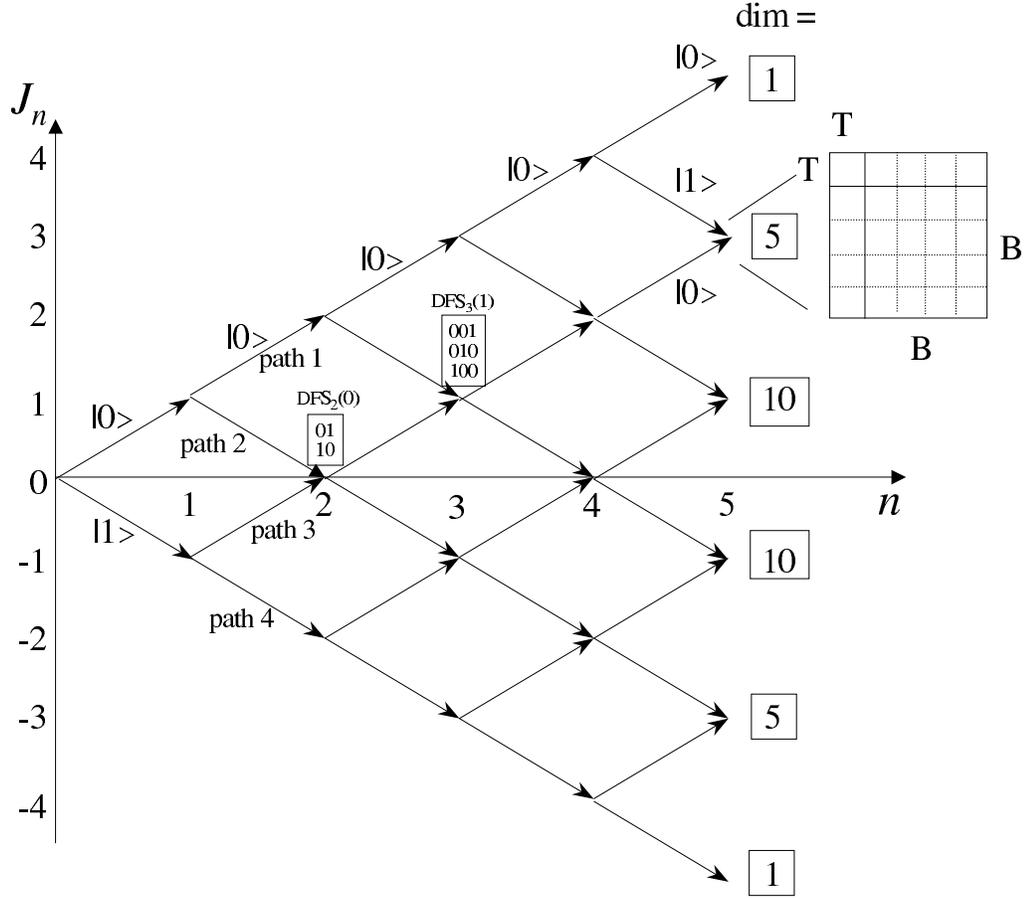,width=1.\textwidth,angle=270} 
\vspace{0.5cm}
\caption{Weak collective decoherence. The horizontal axis is the number of
qubits; the vertical axis is the number of $0$'s minus the number of $1$'s in
each state ($J_n$). Each state in the standard basis thus corresponds to a path
from the origin following the indicated arrows. The dimension of a DFS
corresponds to the multiple pathways through which one can arrive at the same
$J_n$. The insert shows the matrix structure of operators acting on DFS$_5(3)$,
in terms of Top (T) and Bottom (B) states. Note that there is only one T-state
entering DFS$_5(3)$, whence the action of exchange is represented by a $1\times
1$ block.} \label{figure1}
\end{figure}

Let DFS$_{n}$($K$) denote the DFS on $n$ physical qubits with
eigenvalue $K$. We show here that
\begin{equation}
{\sf H}=\{{\bf E}_{i,i+1},{\bf T}_{i,i+1}^{P},{\bf T}_{i,i+1}^{Q}:i=1,\ldots
,n-1,\bar{{\bf Z}}_{12}\}
\label{eq:weakugs}
\end{equation}
is a universal generating set for any of the DFSs occurring in a system of $
n $ physical qubits. It is convenient to work directly with the
Hamiltonians, and to show that ${\sf H}$
gives rise to the Lie-algebra $su(d_{K})$ on each DFS$_{n}$($K$) [via scalar
multiplication, addition, and Lie-commutator; see the
allowed compositions of operations (1-3) in Section \ref{allowed}].
Exponentiation then gives the group $SU(d_{K})$ on the DFS. We will proceed
by induction on $n$, the number of physical qubits, building the DFS-states
of $n$ qubits out of DFS-states for $n-1$ qubits. A graphical representation
of this construction is useful (and will also generalize to the strong case
presented in the following section \ref{strongchapter}): see
Fig.~(\ref{figure1}).

We have seen that in the WCD case the DFS states are simply bitstrings of $n$
qubits in either $|0\rangle $ or $|1\rangle $. The different $n$-qubit DFSs
are labeled by their eigenvalue
\begin{equation}
\lambda _{J}=\#0-\#1\equiv J_{n}.
\end{equation}
To obtain a DFS-state of $n$ qubits out of a DFS-state of $n-1$ qubits
corresponding to $J_{n-1}$ we can either add the $n^{{\rm th}}$ qubit as $
|0\rangle $ ($J_{n}=J_{n-1}+1$) or as $|1\rangle $ ($J_{n}=J_{n-1}-1$). Each
DFS-state can be built sequentially from the first qubit onward by adding
successively $|0\rangle $ or $|1\rangle $, and is uniquely defined by a
sequence $J_{1},\ldots ,J_{n}$ of eigenvalues. In the graphical
representation of Fig.~(\ref{figure1}) the horizontal axis marks $n$, the
number of qubits up to which the state is already built, and the vertical
axis shows $J_{n}$, the difference $\#0-\#1$ up to the $n^{{\rm th}}$ qubit.
Adding a $|0\rangle $ at the $n+1^{{\rm th}}$ step will correspond to a line
pointing upwards, adding a $|1\rangle $ to a line pointing down. {\em Each
DFS-state of }$n${\em qubits with eigenvalue }$\lambda _{J}=J_{n}${\em \
is thus in one-to-one correspondence with a path on the lattice from the origin to} $(n,J_{n})$.

Consider the first non-trivial case, $n=2$, which gives rise to one
DFS-qubit: DFS$_{2}$($0$). This corresponds to the two states $|0_{L}\rangle
=|01\rangle $ [path 2 in Fig.~(\ref{figure1})] and $|1_{L}\rangle =|10\rangle $ (path 3) with $
J_{2}=0$. The remaining Hilbert space is spanned by the one-dimensional DFS$
_{2}$($2$) $|00\rangle $ (path 1) corresponding to $J_{2}=2$, and DFS$_{2}$($
-2$) $|11\rangle $ (path 4) corresponding to $J_{2}=-2$. The exchange ${\bf
E }_{12}$ flips $|0_{L}\rangle $ and $|1_{L}\rangle $ (path 2 and 3), and
leaves the other two paths unchanged. The interaction ${\bf A}_{12}={\rm diag}
(0,0,1,0) $ induces a phase on $|1_{L}\rangle =|10\rangle $ (path 3). Their commutator forms an encoded $\sigma_y$ {\em acting entirely within the} DFS$_{2}$($0$) {\em subspace}. Its commutator with ${\bf
E }_{12}$ in turn forms an encoded $\sigma_z$ with the same property. Together they form the
(encoded) Lie algebra $su(2)$ acting entirely within this DFS. The Lie algebra is completed by forming the commutator between these $\bar{{\bf Y}}$ and $\bar{{\bf Z}}$ operations. To summarize:
\begin{eqnarray}
\bar{{\bf Y}}_{12} &=& i[\bar{{\bf A}},{\bf E}_{12}]=\left(
\begin{array}{cccc}
0 & 0 & 0 & 0 \\
0 & 0 & -i & 0 \\
0 & i & 0 & 0 \\
0 & 0 & 0 & 0
\end{array}
\right) \\  \nonumber
\bar{{\bf Z}}_{12} &\equiv& i[{\bf E}_{12},\bar{{\bf Y}}_{12}]  \nonumber \\
\bar{{\bf X}}_{12} &\equiv& i[\bar{{\bf Y}}_{12},\bar{{\bf Z}}_{12}]
\end{eqnarray}
We call the property of acting entirely within the specified DFS {\em independence},
meaning that the corresponding Hamiltonian has zero entries in the rows and
columns corresponding to the other DFSs [DFS$_{2}$($2$)=$|00\rangle $ and
DFS$_{2}$($-2$)=$|11\rangle $ in this case]. When the Hamiltonian is
exponentiated, the corresponding gate will act as identity on all DFSs
except DFS$_{2}$(0).

To summarize these considerations, the Lie-algebra formed by ${\sf H}
_{0}^{2}=\{\bar{{\bf X}},\bar{{\bf Z}}\}$ is $su(2)$, and generates $SU(2)$
on DFS$_{2}$(0) by exponentiation. In addition, this is an independent $
SU(2) $, namely, these operations act as identity on the other DFSs: when
written as matrices over the basis of DFS-states, their generators in ${\sf
H }_{0}^{2}$ have zeroes in the rows and columns corresponding to all other
DFSs.

In the following we show how this construction generalizes to $n>2$ qubits,
by proving the following theorem:

{\it Theorem 4---} For any $n\geq 2$ qubits undergoing weak collective
decoherence, there exist sets of Hamiltonians ${\sf H}_{J_{n}}^{n}$
[obtained from ${\sf H}$ of Eq.~(\ref{eq:weakugs}) via scalar
multiplication, addition, and Lie-commutator] acting as $
su(d_{J_{n}})$ on the DFS corresponding to the eigenvalue $J_{n}$.
Furthermore each set acts {\em independently} on this DFS only (i.e., with
zeroes in the matrix representation corresponding to their action on the
other DFSs).

Before proving this theorem, we first explain in detail the steps taken in
order to go from the $n=2$ to the $n=3$ case, so as to make the general
induction procedure more transparent.

The structure of the DFSs for $n=2$ and $3$ qubits is:
\begin{eqnarray}
{\rm DFS}_{2}(2)&=&\{|00\rangle \},\quad {\rm DFS}_{2}(0)=\left\{
\begin{array}{c}
|01\rangle \\
|10\rangle
\end{array}
\right. ,\quad {\rm DFS}_{2}(-2)=\{|11\rangle \}  \nonumber \\
{\rm DFS}_{3}(3)&=&\{|000\rangle \},\quad {\rm DFS}_{3}(1)=\left\{
\begin{array}{c}
|001\rangle \\
|010\rangle \\
|100\rangle
\end{array}
\right. ,\quad {\rm DFS}_{3}(-1)=\left\{
\begin{array}{c}
|011\rangle \\
|101\rangle \\
|110\rangle
\end{array}
\right. ,\quad {\rm DFS}_{3}(-3)=\{|111\rangle \}.
\end{eqnarray}
DFS$_{3}$($3$) is obtained by appending a $|0\rangle $ to DFS$_{2}$($2$).
Similarly DFS$_{3}$($-3$) is obtained by appending a $|1\rangle $ to DFS$
_{2} $($-2$). Graphically, this corresponds to moving along the only allowed
pathway from DFS$_{2}$($2$) [DFS$_{2}$($-2$)] to DFS$_{3}$($3$) [DFS$_{3}$($
-3$)], as shown in Fig.~(\ref{figure1}). The lowest and highest $\lambda _{J}$
for $n$ qubits will always be made up of the single pathway connecting the
lowest and highest $\lambda _{J}$ for $n-1$ qubits. The structure of DFS$
_{3} $($\pm 1$) is only slightly more complicated. DFS$_{3}$($1$) is made up
of one state, $|001\rangle $, which comes from appending a $|1\rangle $
(moving down) to DFS$_{2}$($2$). We call $|001\rangle $ a ``Top-state''\ in
DFS$_{3}$($1$). The two other states, $|010\rangle $ and $|100\rangle $,
come from appending $|0\rangle $ (moving up) to DFS$_{2}$($0$). Similarly,
we call $|010\rangle $ and $|100\rangle $ ``Bottom-states'' in\ DFS$_{3}$($1$). DFS$_{3}$($-1$) is constructed in an analogous manner (Fig.~\ref{figure1}).

We showed above that it is possible to perform independent $su(2)$
operations on DFS$_{2}$($0$). DFS$_{2}$($\pm 2$) are also both acted upon
independently, but because they are one-dimensional subspaces, independence
implies that $su(2)$ operations annihilate them. Since the states $
\{|010\rangle ,|100\rangle \}\in $ DFS$_{3}$($1$) and the states $
\{|011\rangle ,|101\rangle \}\in $ DFS$_{3}$($-1$) both have $\{|01\rangle
,|10\rangle \}\in $ DFS$_{2}$($0$) as their first two qubits, one immediate
consequence of the independent action on DFS$_{2}$($0$) is that one can {\em
simultaneously} perform $su(2)$ operations on the corresponding daughter
subspaces created by expanding DFS$_{2}$($0$) into DFS$_{3}$($\pm 1$). The
first step in the general inductive proof is to eliminate this simultaneous
action, and to act independently on each of these subspaces (the ``{\em
independence step}''). To see how this is achieved, it is convenient to
represent the operators acting on the $8$-dimensional Hilbert space of $3$
qubits in the basis of the $4$ DFSs:

\begin{center}
\begin{tabular}{cccccccc}
$000$ & \multicolumn{1}{|c}{$001$} & $010$ & $100$ & \multicolumn{1}{|c}{$
011 $} & $101$ & $110$ & \multicolumn{1}{|c}{$111$} \\ \hline\hline
\multicolumn{1}{|c}{$M_{3}$} & \multicolumn{1}{|c}{} &  &  &  &  &  &  \\
\cline{1-4}
& \multicolumn{1}{|c}{} &  &  & \multicolumn{1}{|c}{} &  &  &  \\
& \multicolumn{1}{|c}{} & $M_{1}$ &  & \multicolumn{1}{|c}{} &  &  &  \\
& \multicolumn{1}{|c}{} &  &  & \multicolumn{1}{|c}{} &  &  &  \\ \cline{2-7}
&  &  &  & \multicolumn{1}{|c}{} &  &  & \multicolumn{1}{|c}{} \\
&  &  &  & \multicolumn{1}{|c}{} & $M_{-1}$ &  & \multicolumn{1}{|c}{} \\
&  &  &  & \multicolumn{1}{|c}{} &  &  & \multicolumn{1}{|c}{} \\ \cline{5-8}
&  &  &  &  &  &  & \multicolumn{1}{|c|}{$M_{-3}$} \\ \cline{8-8}
\end{tabular}
\end{center}

The simultaneous action on DFS$_{3}$($\pm 1$) can now be visualized in terms
of both $M_{\pm 1}$ being non-zero. Let us show how to obtain an action
where, say, just $M_{1}$ is non-zero. This can be achieved by applying the
commutator of two operators with the property that their intersection has
non-vanishing action just on $M_{1}$. This is true for the ${\bf T}_{23}^{P}$
and $\bar{{\bf X}}_{12}$ Hamiltonians: ${\bf T}_{23}^{P}$ annihilates every
state except those that are $|00\rangle $ over qubits $2$ and $3$, namely $
|100\rangle \in $ DFS$_{3}$($1$) and $|000\rangle \in $DFS$_{3}$($3$). This
implies that the only non-zero blocks in its matrix are
\begin{equation}
M_{3}({\bf T}_{23}^{P})=1,\quad M_{1}({\bf T}_{23}^{P})=\left(
\begin{array}{ccc}
0 & 0 &  \\
0 & 0 &  \\
&  & 1
\end{array}
\right) .
\end{equation}
On the other hand, $\bar{{\bf X}}_{12}$ is non-zero only on those states
that are $|01\rangle $ or $|10\rangle $ on qubits $1$ and $2$. Therefore it
will be non-zero on all $3$-qubit states that have $|01\rangle $ or $
|10\rangle $ as ``parents''. This means that in its matrix representation $
M_{\pm 3}=0$ and
\begin{equation}
M_{1}(\bar{{\bf X}}_{12})=\left(
\begin{array}{ccc}
0 &  &  \\
& 0 & 1 \\
& 1 & 0
\end{array}
\right) ,\quad M_{-1}(\bar{{\bf X}}_{12})=\left(
\begin{array}{ccc}
0 & 1 &  \\
1 & 0 &  \\
&  & 0
\end{array}
\right) .
\end{equation}
Clearly, taking the product of ${\bf T}_{23}^{P}$ and $\bar{{\bf X}}_{12}$
leaves non-zero just the lower $2\times 2$ block of $M_{1}$, and this is the
crucial point:\ it shows that an independent action on DFS$_{3}$($1$) can be
obtained by forming their commutator. Specifically, since the lower $2\times
2$ block of $M_{1}({\bf T}_{23}^{P})$ is just $\frac{1}{2}\left( {\bf I}
-\sigma _{z}\right) $:
\begin{equation}
i[{\bf T}_{23}^{P},\bar{{\bf X}}_{12}]=\bar{{\bf Y}}_{\{|100\rangle
,|010\rangle \}},
\end{equation}
i.e., this commutator acts as an encoded $\sigma _{y}$ inside the $
\{|100\rangle ,|010\rangle \}$ subspace of DFS$_{3}$($1$). Similarly, ${\bf
\bar{Z}}_{\{|100\rangle ,|010\rangle \}}=\frac{i}{2}[\bar{{\bf Y}}
_{\{|100\rangle ,|010\rangle \}},\bar{{\bf X}}_{12}]$. Together $\{\bar{{\bf
Y}}_{\{|100\rangle ,|010\rangle \}},\bar{{\bf Z}}_{\{|100\rangle
,|010\rangle \}}\}$ generate $su(2)$ acting {\em independently} on the $
\{|100\rangle ,|010\rangle \}$ subspace of DFS$_{3}$($1$), which we achieved
by subtracting out the action on DFS$_{3}$($-1$).

In an analogous manner, an independent $su(2)$ can be produced on the $
\{|011\rangle ,|101\rangle \}$ subspace of DFS$_{3}$($-1$) by using the
Hamiltonians acting on DFS$_{2}$($0$) in conjunction with ${\bf T}_{23}^{Q}$
to subtract out the $su(2)$ action on DFS$_{3}$($1$).\footnote{
Since ${\bf T}_{23}^{Q}$ annihilates every state except those that are $
|11\rangle $ over qubits $2$ and $3$, namely $|011\rangle \in $ DFS$_{3}$($
-1 $) and $|111\rangle \in $DFS$_{3}$($-3$), the only non-zero blocks
in its matrix are
\[
M_{-3}({\bf T}_{23}^{Q})=1,\quad M_{-1}({\bf T}_{23}^{Q})=\left(
\begin{array}{ccc}
1 &  &  \\
& 0 & 0 \\
& 0 & 0
\end{array}
\right) .
\]
} Thus we can obtain independent action for each of the daughters of DFS$
_{2} $($0$), i.e., separate actions on the subspace spanned by $
\{|010\rangle ,|100\rangle \}$ and $\{|011\rangle ,|101\rangle \}$.

\begin{figure}[!htb]
\hspace{0.2\textwidth}
\psfig{figure=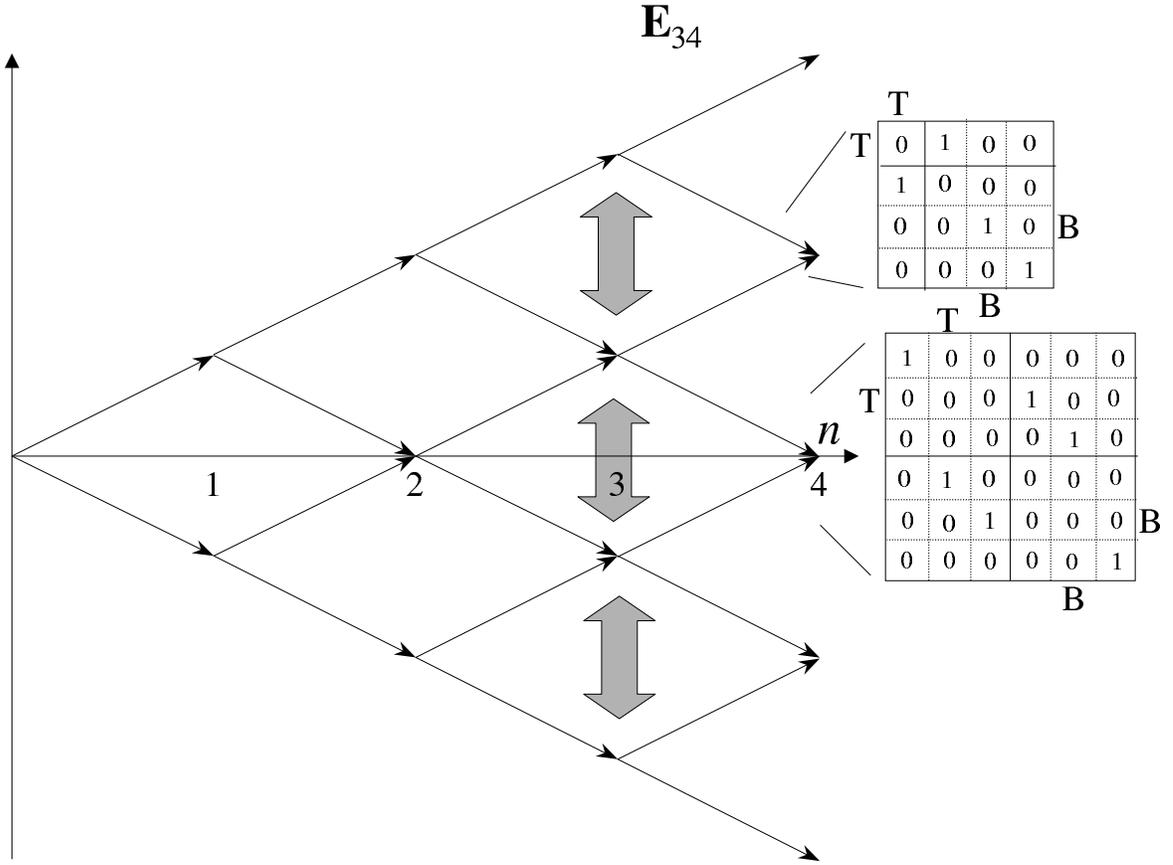,width=1.\textwidth,angle=270}
\vspace{0.5cm} \caption{Action of exchange is to simultaneously flip different
paths. Matrices are the representations of ${\bf E}_{34}$ on DFS$_4(0)$ (lower)
and DFS$_4(2)$ (upper).} \label{figure2}
\end{figure}

Having established independent action on the two {\em subspaces} of DFS$_{3}$($1$) and DFS$_{3}$($-1$) arising from DFS$_{2}$($0$), we need only show
that we can obtain the full action on DFS$_{3}$($1$) and DFS$_{3}$($-1$).
For DFS$_{3}$($1$) we need to {\em mix} the subspace $\{|010\rangle
,|100\rangle \}$ over which we can already perform independent $su(2)$, with
the $|001\rangle $ state. To do so, note that the effect of the exchange
operation ${\bf E}_{23}$ is to flip $|001\rangle $ and $|010\rangle $, and
leave $|100\rangle $ invariant. I.e., the matrix representation of ${\bf E}
_{23}$ is
\begin{equation}
M_{1}({\bf E}_{23})=\left(
\begin{array}{ccc}
0 & 1 &  \\
1 & 0 &  \\
&  & 1
\end{array}
\right) .
\end{equation}
Unfortunately, ${\bf E}_{23}$ has a simultaneous action on DFS$_{3}$($-1$).
This, however, is not a problem, since we have already constructed an
independent $su(2)$ on DFS$_{3}$($1$) elements. Thus we can eliminate the
simultaneous action by simply forming commutators with these $su(2)$
elements. The Lie algebra generated by these commutators will act
independently on {\em all} of DFS$_{3}$($1$). In fact we claim this Lie
algebra to be all of $su(3)$ (see Appendix~\ref{appA} for a general proof).
In other words, the Lie algebra spanned by the $su(2)$ elements $\{\sigma
_{x},\sigma _{y},\sigma _{z}\}$ acting on the subspace $\{|100\rangle
,|010\rangle \}$, together with the exchange operation ${\bf E}
_{23}$, generate all of $su(3)$ independently on DFS($1$). A similar argument
holds for DFS$_{3}$($-1$). This construction illustrates the induction
step: we have shown that it is possible to perform independent $su(d_{K})$
actions on all four of the DFS$_{3}$($K$) ($K=\pm 3,\pm 1$), given that we
can perform independent action on the three DFS$_{2}$($K$) ($K=\pm
1,0$). In Fig.~(\ref{figure2}) we have further illustrated these
considerations by depicting the action of exchange on two the
$4$-qubit DFSs. Let us now proceed to the general proof.

{\it Proof---} By induction.

The case $n=2$ already treated above will serve to initialize the induction.
Assume now that the theorem is true for $n-1$ qubits and let us show that it
is then true for $n$ qubits as well.

\begin{figure}[!htb]
\hspace{0.2\textwidth}
\psfig{figure=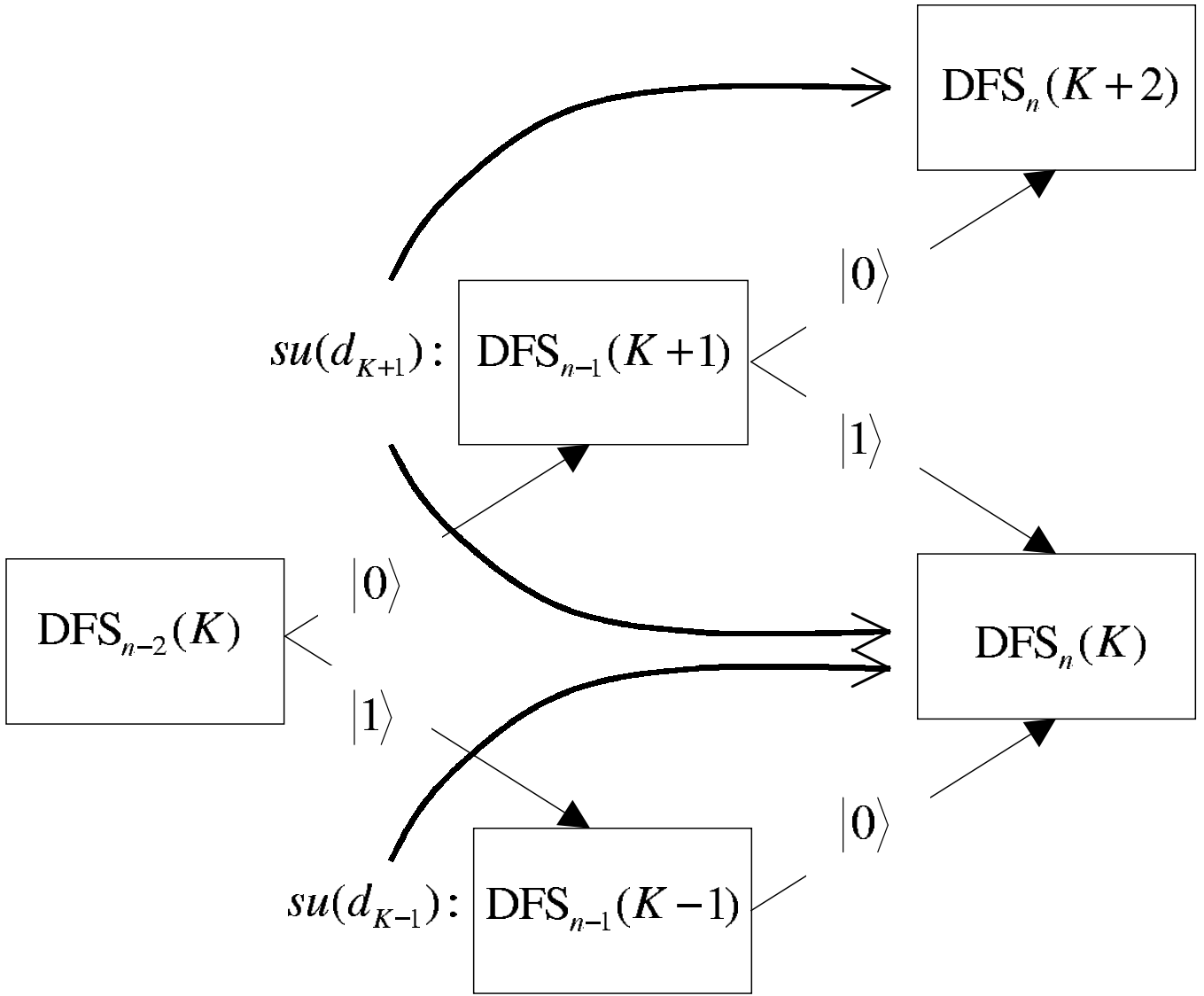,width=1.\textwidth}
\vspace{0.5cm}
\caption{The structure of the pathways connecting adjacent DFSs in the
weak collective decoherence case. The action of the different $su$ Lie
algebras is indicated by heavy arrows.}
\label{figure3}
\end{figure}

First note that each DFS$_{n}$($K$) is constructed either from the DFS$
_{n-1}$($K-1$) (to its lower left) by adding a $|0\rangle $ for the $n^{{\rm
th}}$ qubit, or from DFS$_{n-1}$($K+1$) (to its upper left) by adding a $
|1\rangle $: the states in DFS$_{n}$($K$) correspond to all paths ending in $
(n,K)$ that either come from below (B) or from the top (T). See
Fig.~(\ref{figure3}).

If we apply a certain gate ${\bf U}=\exp (i{\bf H}t)$ to DFS$_{n-1}$($K+1$),
then this operation will induce the same ${\bf U}$ on DFS$_{n}$($K$), by
acting on all paths (states) entering DFS$_{n}$($K$) from above. At the same
time ${\bf U}$ is induced on DFS$_{n}$($K+2$) by acting on all paths
entering this DFS from below. So, ${\bf U}$ affects two DFSs {\em
simultaneously}. In other words, the set of valid Hamiltonians ${\sf H}
_{K+1}^{n-1}$ [acting on $n-1$ qubits and generating $su(d_{K+1})$] on DFS$
_{n-1}$($K+1$), that we are given by the induction hypothesis, induces a
{\em simultaneous} action of $su(d_{K+1})$ on DFS$_{n}$($K$) (on the paths
coming from above only) and DFS$_{n}$($K+2$) (on the paths coming from below
only). Additionally, it does not affect any other $n$-qubit DFS, since we
assumed that the action on DFS$_{n-1}$($K+1$) was {\em independent}, and the
only $n$-qubit DFSs built from DFS$_{n-1}$($K+1$) are DFS$_{n}$($K$) and DFS$
_{n}$($K+2$). These considerations are depicted schematically in
Fig.~(\ref{figure3}).

We now show how to annihilate, for a given non-trivial (i.e., dimension $>1$) DFS$_{n}$($K$), the unwanted simultaneous action on other DFSs (the ``{\em
independence step''}). Then we proceed to obtain the full $su(d_{K})$, by
using the $su(d_{K\pm 1})$ on DFS$_{n-1}$($K\pm 1$) that are given by the
induction hypothesis (the ``{\em mixing step''}).

\subsubsection{\noindent Independence}

Let us call all the $t_{K}$ paths converging on DFS$_{n}$($K$) from above
``Top-states'', or T-states for short, and the $b_{K}$ paths converging from
below ``Bottom- (or B) states'' (recall that there is a 1-to-1
correspondence between paths and states). The total number of paths
converging on a given DFS is exactly its dimension, so $d_{K}=t_{K}+b_{K}$.
By using the induction hypothesis on DFS$_{n-1}$($K+1$) we can obtain $
su(t_{K})$ (generated by ${\sf H}_{K+1}^{n-1}$) on the T-states of DFS$_{n}$($K$), which will simultaneously affect the B-states in the higher lying DFS$_{n}(K+2)$ as $su(b_{K+2}$) (note that $t_{K}=b_{K+2}$). The set ${\sf H}
_{K+1}^{n-1}$ is non-empty only if $n-3\geq K+1\geq -(n-3)$ [because the
``highest'' and ``lowest'' DFS are always one-dimensional and $su(1)=0$]. If
this holds then DFS$_{n}$($K+2$) ``above'' DFS$_{n}$($K$) is non-trivial
(dimension $>1$), and there are paths in DFS$_{n}$($K$) ending in $
|11\rangle $ (``down, down''). This is exactly the situation in which we can
use ${\bf T}_{n-1,n}^{Q}$ to wipe out the unwanted action on DFS$_{n}$($K+2$): recall that ${\bf T}_{n-1,n}^{Q}$ annihilates all states except those
{\em ending} in $|11\rangle $, and therefore affects non-trivially only
these special T-states in each DFS. Since the operations in ${\sf H}
_{K+1}^{n-1}$ affect only B-states on DFS$_{n}$($K+2$), ${\bf T}_{n-1,n}^{Q}$
commutes with ${\sf H}_{K+1}^{n-1}$ on DFS$_{n}$($K+2$). Therefore the
commutator of ${\bf T}_{n-1,n}^{Q}$ with elements in ${\sf H}_{K+1}^{n-1}$
annihilates all states not in DFS$_n$($K$).\footnote{
The argument thus far closely parallels the discussion above showing how to
generate an independent $su(2)$ on the $\{|011\rangle ,|101\rangle \}$
subspace of DFS$_{3}$($-1$), starting from the $su(2)$ on DFS$_{2}$($0$) and
${\bf T}_{23}^{Q}$.} To show that commuting ${\bf T}_{n-1,n}^{Q}$ with ${\sf
H}_{K+1}^{n-1}$ generates $su(t_{K})$ on the T-states of DFS$_n$($K$) we need the following lemma, which shows how to form $su(d)$ from an overlapping $su(d-1)$ and $su(2)$:

{\it Enlarging Lemma---} Let ${\cal H}$ be a Hilbert space of dimension $d$
and let $|i\rangle \in {\cal H}$. Assume we are given a set of Hamiltonians $
{\sf H}_{1}$ that generates $su(d-1)$ on the subspace of ${\cal H}$ that
does not contain $|i\rangle $ and another set ${\sf H}_{2}$ that generates $
su(2)$ on the subspace of ${\cal H}$ spanned by $\{|i\rangle ,|j\rangle \}$,
where $|j\rangle $ is another state in ${\cal \ H}$. Then $[{\sf H}_{1},
{\sf H}_{2}]$ (all commutators) generates $su(d)$ on ${\cal H}$ under closure as a Lie-algebra
(i.e., via scalar multiplication, addition and Lie-commutator).

{\it Proof---} See Appendix~\ref{appB}

Now consider two states $|i\rangle ,|j\rangle \in $DFS$_{n}$($K$) such that $
|i\rangle $ ends in $|11\rangle $ and $|j\rangle $ is a\ T-state, but does
not end in $|11\rangle .$ Then we can generate $su(2)$ on the subspace
spanned by $\{|i\rangle ,|j\rangle \}$ as follows: (i) We use the exchange
interaction $\bar{{\bf X}}_{ij}=|i^{\prime }\rangle \langle j^{\prime
}|+|j^{\prime }\rangle \langle i^{\prime }|$ [a prime indicates the
bitstring with the last bit (a $1$ in this case)\ dropped] in $su(t_{K})\in
{\sf H}_{K+1}^{n-1}$ to generate a simultaneous action on DFS$_{n}$($K$) and
DFS$_{n}$($K+2$). This interaction is represented by a $2\times 2$ $\sigma
_{x}$-matrix in the subspace spanned by $\{|i\rangle ,|j\rangle \}$. (ii) $
{\bf T}_{n-1,n}^{Q}$ is represented by the $2\times 2$ matrix ${\rm diag}
(1,0)=\frac{1}{2}\left( {\bf I}+\sigma _{z}\right) $ in the same subspace,
and commutes with $\bar{{\bf X}}_{ij}$ on DFS$_{n}$($K+2$) (since $\bar{{\bf
X}}_{ij}$ affects only B-states in DFS$_{n}$($K+2$)$,$ and ${\bf T}
_{n-1,n}^{Q}$ is non-zero only on states ending in $|11\rangle $). Thus we
can use it to create an independent action on DFS$_{n}$($K$) alone: ${\bf
\bar{Y}}_{ij}=i[{\bf T}_{n-1,n}^{Q},\bar{{\bf X}}_{ij}]$, $\bar{{\bf Z}}_{ij}=\frac{i}{
2}[\bar{{\bf Y}}_{ij},\bar{{\bf X}}_{ij}]$.

Together $\{\bar{{\bf Y}}_{ij},\bar{{\bf Z}}_{ij}\}$ generate $su(2)$
independently on $\{|i\rangle ,|j\rangle \}\in$ DFS$_{n}$($K$). Since these operators vanish
everywhere except on DFS$_{n}$($K$), their commutators with elements in $
{\sf H}_{K+1}^{n-1}$ [acting as $su(t_{K})$] will annihilate all other DFSs.
Therefore, using the Enlarging Lemma, in this way all operations in
$su(t_{K})$ acting on DFS$_n$($K$) {\em only} can be generated.

So far we have shown how to obtain an independent $su(t_{K})$ on the
T-states of DFS$_{n}$($K$) using ${\sf H}_{K+1}^{n-1}$ (for $K\leq n-4$). To
obtain an independent $su(b_{K})$ on the {\em B-states} of DFS$_{n}$($K)$ we
use Hamiltonians in ${\sf H}_{K-1}^{n-1}$ (acting on DFS$_{n-1}$($K-1)$ --
the DFS from below). This will generate a simultaneous $su(b_{K})$ in DFS$
_{n}$($K$) and $su(t_{K-2})$ in DFS$_{n}$($K-2$). To eliminate the unwanted
action on DFS$_{n}$($K-2$) we apply the previous arguments almost
identically, except that now we use ${\bf T}_{n-1,n}^{P}$ to wipe out the
action on all states except those ending in $|00\rangle $. We thus get an
independent $su(b_{K})$ on DFS$_{n}$($K$). Together, the ``above'' and
``below'' constructions respectively provide independent $su(t_{K})$ and $
su(b_{K})$ on DFS$_{n}$($K$). Finally, note that we did not really need both ${\bf T}_{ij}^{P}$ and ${\bf T}_{ij}^{Q}$, since once we established independent action on the T-states, we could have just subtracted out this action when considering the B-states. Also, the specific choice of ${\bf T}_{ij}^{P,Q}$ was rather arbitrary (though convenient): in fact almost any other diagonal interaction would do just as well.

\subsubsection{Mixing} In order to induce operations between the two sets
of paths (from ``above'' and from ``below'') that make up DFS$_{n}$($K$)
consider the effect of ${\bf E}_{n-1,n}$. This gate does not affect any
paths that ``ascend'' two steps to $(n,K)$ (corresponding to bitstrings
ending in $|00\rangle $) and paths that ``descend'' two steps (ending in $
|11\rangle $), but it flips the paths that pass from $(n-2,K)$ via $
(n-1,K+1) $ with the paths from $(n-2,K)$ via $(n-1,K-1)$ [see
Fig.~(\ref{figure3})]. It does this for all DFSs simultaneously.

In order to get a full $su(d_{K})$ on DFS$_{n}$($K$) we need to ``mix'' $
su(t_{K})$ (on the T-states) and $su(b_{K})$ (on the B-states) which we already
have. We show how to obtain an {\em independent} $su(2)$ between a
T-state and a B-state. By the Enlarging Lemma this generates $su(d_{K})$.

Since $n\geq 3$ DFS$_{n}$($K$) contains states terminating in $|00\rangle $
and/or $|11\rangle $. Let us assume, w.l.o.g., that states terminating in $
|00\rangle $ are present, and let $|i\rangle $ be such a state (B-state).
Let $|j\rangle $ be a B-state not terminating in $|00\rangle $, and let $
|k\rangle ={\bf E}_{n-1,n}|j\rangle $ ($|k\rangle $ is a T-state). Let ${\bf
\bar{Z}}_{ij}=|i\rangle \langle i|-|j\rangle \langle j|\in su(b_{K})$, and
recall that we have {\em independent} $su(b_{K})$. Then as is easily
checked, $i[{\bf E}_{n-1,n},\bar{{\bf Z}}_{ij}]\equiv \bar{{\bf Y}}_{jk}$ yields $
\sigma _{y}$ between $|j\rangle $ and $|k\rangle $ {\em only}.\footnote{
Since ${\bf E}_{n-1,n}=|i\rangle \langle i| + |k\rangle \langle j| + |j\rangle \langle k| + O$, where $O$ is some action on an orthogonal subspace.} In addition, $ \bar{{\bf Z}}_{jk} \equiv \frac{i}{2}[{\bf E}_{n-1,n},
\bar{{\bf Y}}_{jk}] $ gives $\sigma _{z}$ between $|j\rangle $
and $|k\rangle $, thus completing a generating set for $su(2)$ on the
B-state $|j\rangle $ and the T-state $|k\rangle $, that affects these two
states only and annihilates all other states. This completes the proof.

To summarize, we have shown {\em constructively} that it is possible to
generate the entire Lie algebra $su(d_{K})$ on a given weak
collective-decoherence DFS$_{n}$($K$) of dimension $d_{K}$, from the
elementary composition of the operations of scalar multiplication, addition,
Lie-commutators (conjugation by unitaries was not necessary in the WCD case). Moreover, this $su(d_{K})$
can be generated independently on each DFS, implying that universal quantum
computation can be performed inside {\em each} DFS$_{n}$($K$). Naturally,
one would like to do this on the largest DFS. Since given the number of
qubits $n$ the dimensions of the DFSs are $d_{K}= {
{n  \choose K}
} $, the largest DFS is the decoherence-free sub{\em space} $K=0$. In
principle it is possible, by virtue of the independence result, to
universally quantum compute {\em in parallel} on all DFSs.

\subsection{State Preparation and Measurement on the Weak Collective
Decoherence DFS}
\label{WCDprep}

To make use of a DFS for encoding information in a quantum computer,
in addition to the universal 
quantum computation described above, it must also be possible to
initially prepare encoded states and to decode the quantum 
information on the DFS at the end of a computation. Encoding requires
that the density matrix of 
the prepared states should have a large overlap with the DFS. Note
that it is {\em not} necessary to prepare states that have support
exclusively within the DFS. 
This follows from the fact that in our construction, while a
computation is performed there is no mixing of states inside and
outside of the DFS. If an initially prepared state is ``contaminated''
(has some support outside the DFS we want to compute on), then the
result of the computation will have the same amount of contamination,
i.e., the initial error does not spread.

For example, suppose we can prepare the state $
\rho=(1-p)|\psi\rangle \langle \psi| + p|\psi_\perp \rangle \langle
\psi_\perp|$ where $|\psi\rangle$ is a state of a particular DFS and
$|\psi_\perp\rangle$ is a state outside of this DFS.  Then the computation will
proceed independently on the DFS and the states outside of the DFS.  Readout
will then obtain the result of the computation with probability $1-p$.
Repeated application of the quantum computation will give the desired
result to arbitrary confidence level. 

There are many choices for the initial states of a computation and the decision as to which states to prepare should be guided by the available gates and measurements and the accuracy that is achievable. For efficient computation one should try to maximize the overlap of the prepared state with the desired initial DFS state. 

For the WCD case preparation of initial pure states is very simple.
Suppose we are concerned with the ${\bf S}_z$ error WCD-DFS. Pure state
preparation into such a DFS then corresponds to the ability to prepare a state
which has support over states with a specific number of $|0\rangle$ and
$|1\rangle$ (eigenstates of the $\sigma_z$ operator). This is
particularly simple if measurements in the $\sigma_z$ basis
($|0\rangle$ , $|1\rangle$) 
as well as $\sigma_x$ gates (to ``flip'' the bits) are available.

The second crucial ingredient for computation on a DFS (in addition to
preparation) is the decoding or readout of quantum information
resulting from a computation. Once again, there are many options 
for how this can be performed. For example, in the WCD case one can make a
measurement which distinguishes all of the DFSs and all of the states within
this DFS by simply making a measurement in the $\sigma_z$ basis on every qubit.
Further, all measurements with a given number of distinct eigenvalues can be
performed by first rotating the observable into one corresponding to a
measurement in the computational basis (which, in turn, corresponds to
a unitary operation on the DFS) and then performing the 
given measurement in the $\sigma_z$ basis, and finally rotating back. There are other situations where one would like to, say,
make a measurement of an observable over the DFS which has only two different
eigenvalues. This type of measurement can be most easily performed by a {\em
concatenated measurement} \cite{Bacon:99b}. In this scheme, one attaches 
another DFS to the original DFS, forming a single larger DFS. Then, assuming
universal quantum computation over this larger DFS one can always perform
operations which allow a measurement of the first DFS by entangling it
with the second DFS, and reading out (destructively as described for the WCD
above) the second DFS. For example, suppose the first DFS encodes two
bits of quantum 
information, $|k,l\rangle_L$, $k,l=\{0,1\}$, 
and the second DFS encodes a single bit of quantum information
$\{|0\rangle_L$, $|1\rangle_L\}$. Then one can make a measurement of
the observable $\sigma_z \otimes {\bf I}$ on the first DFS by
performing an encoded controlled-{\sc NOT}
operation between the first and the second DFS, and reading out the second
DFS in the encoded $\sigma_z$ basis. For the WCD case the ability to make this
destructive measurement on the ancilla (not on the code) simply
corresponds to the ability to measure single $\sigma_z$ operations.

Finally, we note that for a WCD-DFS there is a destructive measurement
which distinguishes between different DFSs (corresponding to a
measurement of the number of $|1\rangle$'s). One can fault-tolerantly
prepare a WCD-DFS state by repeatedly performing such a measurement to
guarantee that the 
state is in the proper DFS. The concatenated measurement procedures described
above for any DFS are naturally fault-tolerant in the sense that they can be
repeated and are non-destructive \cite{Bacon:99b,Gottesman:97}. Thus
fault-tolerant preparation and decoding is available for the WCD-DFS.

\section{Strong Collective Decoherence}

\label{strongchapter}

\begin{figure}[!htb]
\hspace{0.2\textwidth}
\psfig{figure=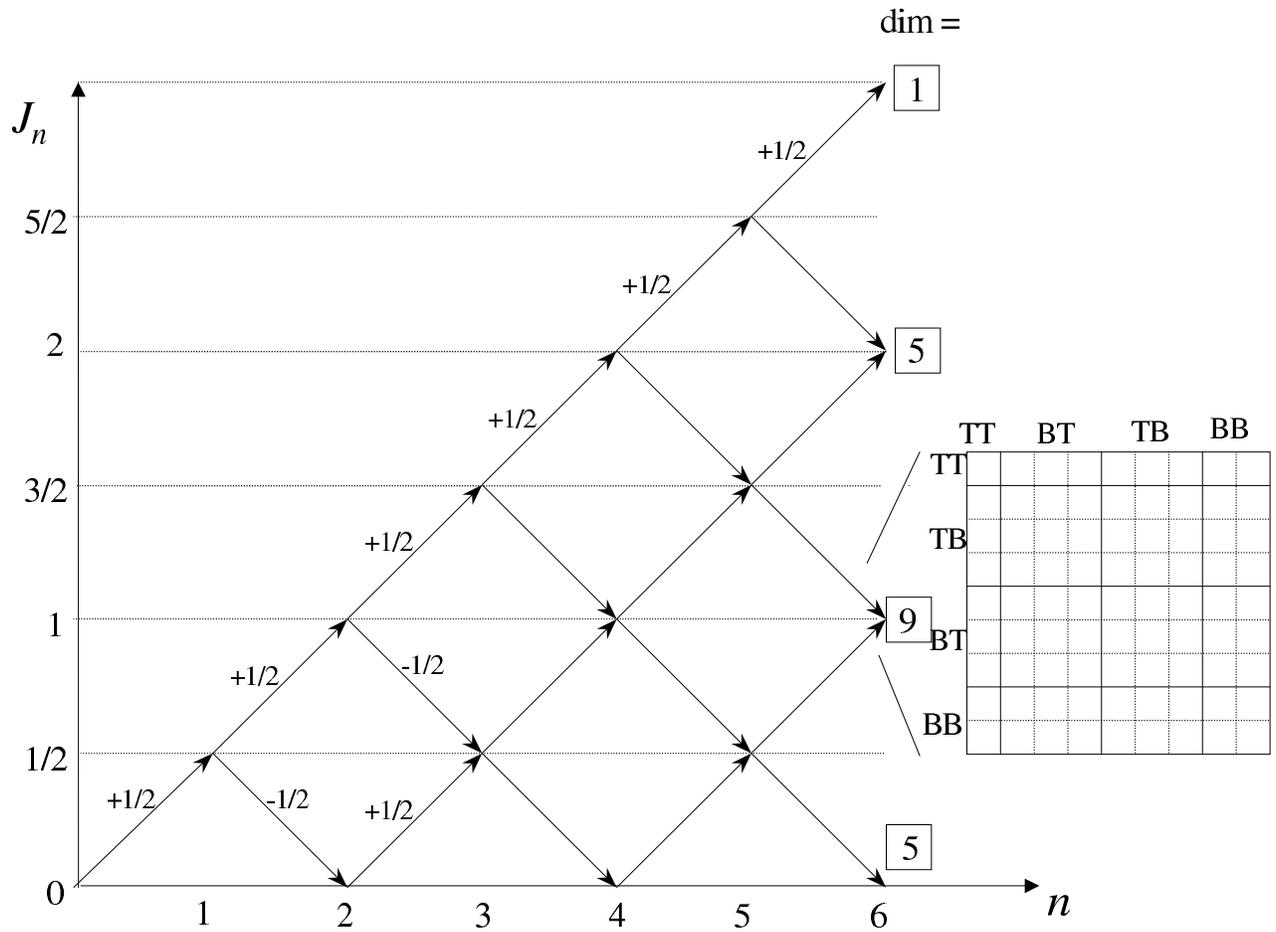,width=1.\textwidth,angle=270}
\vspace{0.5cm}
\caption{Strong collective decoherence. The horizontal axis is the number of
qubits and the vertical axis is the total angular momentum. Each state in
the DFS is represented by a pathway from the origin along the arrows as
indicated. The insert shows the matrix structure  of operators acting
on DFS$_6(1)$, in terms of TT, TB, BT, and BB-states.}
\label{figure4}
\end{figure}

Strong collective decoherence on $n$ qubits is characterized by the three
system operators ${\bf S}_{x}$, ${\bf S}_{y}$ and ${\bf S}_{z}$. These
operators form a representation of the semisimple Lie algebra $su(2)$. The
algebra ${\cal A}$ generated by these operators can be decomposed as\footnote{
Note that as a complex algebra $\{{\bf S}_{x},{\bf S}_{y},{\bf S}_{z}\}$ span all
of $gl(2)$, not just $su(2)$.}
\begin{equation}
{\cal A}\cong \bigoplus_{J=0(1/2)}^{n/2}{\bf I}_{n_{J}}\otimes gl(2J+1,\CC)
\label{eq:Agl2}
\end{equation}
where $J$ labels the total angular momentum of the corresponding Hilbert
space decomposition (and hence the $0$ or $1/2$ depending on whether $n$ is
even or odd respectively) and $gl(2J+1,\CC)$ is the general linear
algebra acting on a space of size $2J+1$. The resulting decomposition of the
system Hilbert space
\begin{equation}
{\cal H}_{S}\cong \bigoplus_{J=0(1/2)}^{n/2}\CC_{n_{J}}\otimes \CC_{2J+1}
\label{eq:Hsdecomp}
\end{equation}
is exactly the reduction of the Hilbert states into different Dicke states
\cite{Dicke:54,Mandel:95a}. The degeneracy for each $J$ is given by \cite
{Mandel:95a}:
\begin{equation}
n_{J}={\frac{(2J+1)n!}{(n/2+J+1)!(n/2-J)!}}.
\label{eq:nJ}
\end{equation}
Eq.~(\ref{eq:Agl2}) shows that given $J$, a state $|J,\lambda ,\mu \rangle $
is acted upon as identity on its $\lambda $ component. Thus a DFS is defined
by fixing $J$ and $\mu $. As we will show later, $\lambda$ corresponds
to the paths leading to a given point $(n,J)$ on the diagram of
Fig.~(\ref{figure4}).

The DFSs corresponding to the different $J$ values for a given $n$ can be
computed using standard methods for the addition of angular momentum. We use
the convention that $|1\rangle $ represents a $|j=1/2,m_{j}=1/2\rangle $
particle and $|0\rangle $ represents a $|j=1/2,m_{j}=-1/2\rangle $ particle
in this decomposition although, of course, one should be careful to treat
this labeling as strictly symbolic and not related to the physical angular
momentum of the particles.

The smallest $n$ which supports a DFS and encodes at least a qubit of
information is $n=3$ \cite{Knill:99a}. For $n=3$ there are two possible
values of the total angular momentum: $J=3/2$ or $J=1/2$. The four $J=3/2$ states $|J,\lambda,\mu \rangle = |3/2,0,\mu \rangle$ ($\mu=m_J = \pm 3/2, \pm 1/2$) are singly degenerate; the $J=1/2$ states have degeneracy
$2$. They can be constructed by either adding a $J_{12}=1$ (triplet) or a $
J_{12}=0$ (singlet) state to a $J_{3}=1/2$ state. These two possible methods
of adding the angular momentum to obtain a $J=1/2$ state are exactly the
degeneracy of the algebra. The four $J=1/2$ states are:
\begin{eqnarray}
|0_{L}\rangle &=&\left\{
\begin{array}{l}
|{\frac{1}{2}},0,0\rangle =|0,0\rangle \otimes |{\frac{1}{2}},-{\frac{1}{2}}
\rangle ={\frac{1}{\sqrt{2}}}\left( |010\rangle -|100\rangle \right) \\
|{\frac{1}{2}},0,1\rangle =|0,0\rangle \otimes |{\frac{1}{2}},{\frac{1}{2}}
\rangle ={\frac{1}{\sqrt{2}}}\left( |011\rangle -|101\rangle \right)
\end{array}
\right.  \nonumber \\
|1_{L}\rangle &=&\left\{
\begin{array}{l}
|{\frac{1}{2}},1,0\rangle ={\frac{1}{\sqrt{3}}}\left( -\sqrt{2}|1,-1\rangle
\otimes |{\frac{1}{2}},{\frac{1}{2}}\rangle +|0,0\rangle \otimes |{\frac{1}{
2 }},-{\frac{1}{2}}\rangle \right) ={\frac{1}{\sqrt{6}}}\left( -2|001\rangle
+|010\rangle +|100\rangle \right) \\
|{\frac{1}{2}},1,1\rangle ={\frac{1}{\sqrt{3}}}\left( \sqrt{2}|1,1\rangle
\otimes |{\frac{1}{2}},-{\frac{1}{2}}\rangle -|1,0\rangle \otimes |{\frac{1}{
2}},{\frac{1}{2}}\rangle \right) ={\frac{1}{\sqrt{6}}}\left( 2|110\rangle
-|101\rangle -|011\rangle \right)
\end{array}
\right.
\label{eq:3dfs}
\end{eqnarray}
where in the first column we indicated the grouping forming a logical qubit;
in the second we used the $|J,\lambda ,\mu \rangle $ notation; in the third
we used tensor products of the form $|J_{12},m_{J_{12}}\rangle \otimes
|J_{3},m_{J_{3}}\rangle $; and in the fourth the states are expanded in
terms the single-particle $|j=1/2,m_{j}=\pm 1/2\rangle $ basis using
Clebsch-Gordan coefficients. These states form a decoherence-free subsystem:
the decomposition of Eqs.~(\ref{eq:Agl2}),(\ref{eq:Hsdecomp}) ensures that
the states $\{|{\frac{1}{2}},0,0\rangle ,|{\frac{1}{2}},0,1\rangle \}$ are
acted upon identically, and so are the states $\{|{\frac{1}{2}},1,0\rangle
,| {\frac{1}{2}},1,1\rangle \}$. Thus information of a qubit $\alpha
|0_{L}\rangle +\beta |1\rangle $ should be encoded into these states as
\begin{equation}
\rho = \mathrel{\mathop{\underbrace{\left[ |{\frac{1}{2}}\rangle \langle
{\frac{1}{2}}|\right] }}\limits_{J}} \mathrel{\mathop{\underbrace{\left[
(\alpha ^{\ast }|0_{L}\rangle +\beta ^{\ast }|1_{L}\rangle )(\alpha \langle
0_{L}|+\beta \langle 1_{L}|)\right] }}\limits_{\lambda }}
\mathrel{\mathop{\underbrace{\left[ \gamma _{00}|0\rangle \langle 0|+\gamma
_{01}|0\rangle \langle 1|\gamma _{01}+\gamma _{10}|1\rangle \langle
0|+\gamma _{11}|1\rangle \langle 1|\right] }}\limits_{\mu }} .
\label{eq:rhocoll}
\end{equation}
where $\gamma _{ij}$ form the components of a valid density matrix (unity
trace and positive). Using Eq.~(\ref{eq:Agl2}) It follows that each of the $
{\bf S}_{\alpha }$'s act on $\rho $ in such a manner that only the $\lambda $
component is changed. Indeed, the ${\bf S}_{\alpha }$'s act like a
corresponding $\sigma _{\alpha }$ in the $\mu $-basis because this basis is
two-dimensional, and $\sigma _{\alpha }$ are the two dimensional irreducible
representations of $su(2)$. These considerations are illustrated in detail
for the exchange interaction in Sec.~\ref{n=3:exchange}.

The smallest decoherence-free subspace (as opposed to subsystem) supporting
a full encoded qubit comes about for $n=4$. Subspaces for the SCD mechanism
correspond to the degeneracy of the zero total angular momentum eigenstates
(there are also two decoherence-free subsystems with degeneracy $1$ and $3$). This subspace is spanned by the states:
\begin{eqnarray}
|0_{L}\rangle &=&|0,0,0\rangle =|0,0\rangle \otimes |0,0\rangle ={\frac{1}{2}
}|(|01\rangle -|10\rangle )(|01\rangle -|10\rangle )  \nonumber \\
|1_{L}\rangle &=&|0,1,0\rangle ={\frac{1}{\sqrt{3}}}(|1,1\rangle \otimes
|1,-1\rangle -|1,0\rangle \otimes |1,0\rangle +|1,-1\rangle \otimes
|1,1\rangle )  \nonumber \\
&=&{\frac{1}{\sqrt{12}}}(2|0011\rangle +2|1100\rangle -|0101\rangle
-|1010\rangle -|0110\rangle -|1001\rangle ).
\label{eq:4dfs}
\end{eqnarray}
The notation is the same as in Eq.~(\ref{eq:3dfs}), except that in the
second column we used the notation $|J_{12},m_{J_{12}}\rangle \otimes
|J_{34},m_{J_{34}}\rangle $ which makes it easy to see how the angular
momentum is added.

As seen from Eqs.~(\ref{eq:3dfs}) and (\ref{eq:4dfs}), there is a variety of useful bases which one can choose for the
SCD-DFSs. We now show how the generic basis $|J,\lambda ,\mu\rangle $
can be given both a graphical and an angular momentum
interpretation. Consider the addition of angular momentum as more
particles are included, similar to
the construction we used in the WCD case. To construct the $n$ qubit SCD-DFS
for a specific $J$, denoted in this section as DFS$_{n}$($J$), one takes DFS$
_{n-1}$($J-1/2$) and DFS$_{n-1}$($J+1/2$) and uses the angular momentum
addition rules to add another qubit ($j=1/2$). Table~(\ref{tab1}) presents
the degeneracy of the $J^{{\rm th}}$ irreducible representation for $n$
qubits. The entries are obtained just as in Pascal's triangle, except that
half of the triangle [the bottom according to the scheme of
Table~(\ref{tab1})] is missing.

\begin{table}[tbp]
\caption{Strong collective decoherence DFS dimensions.} \label{tab1}
\begin{tabular}{ccccccc}
$J={3}$ &  &  &  &  &  & $1$ \\ $J={\frac{5}{2}}$ &  &  &  &  & $1$ &  \\ $J=2$
&  &  &  & $1$ &  & $5$ \\ $J={\frac{3}{2}}$ &  &  & $1$ &  & $4$ &  \\ $J=1$ &
& $1$ &  & $3$ &  & $9$ \\ $J={\frac{1}{2}}$ & $1$ &  & $2$ &  & $5$ &  \\
$J=0$ &  & $1$ &  & $2$ &  & $5$ \\ & $n=1$ & $n=2$ & $n=3$ & $n=4$ & $n=5$ &
$n=6$
\end{tabular}
\end{table}

Table~(\ref{tab1}) demonstrates how the degeneracies of the $(n-1)$-qubit $
J\pm 1/2$ irreducible representations (irreps), i.e., the dimensions of DFS$
_{n-1}(J\pm 1/2)$, add to determine the dimension of DFS$_{n}(J)$. This method
of addition of the angular momentum leads to a natural interpretation of the
$|J,\lambda,\mu \rangle$ basis for the SCD-DFSs which we now present.

Define the partial collective operators
\begin{equation}
{\bf S}_{\alpha }^{k}\equiv {\bf S}_{\alpha }^{(1,2,\dots
,k)}=\sum_{i=1}^{k}\sigma _{\alpha }^{i}.
\label{eq:Spart}
\end{equation}
This can be used to find a set of mutually commuting
operators for the SCD-DFSs: the partial total angular momentum operators
\begin{equation}
({\bf S}^{k})^{2}=\sum_{\alpha =x,y,z}\left( {\bf S}_{\alpha }^{k}\right)
^{2}.
\end{equation}
As shown in Appendix~\ref{appD}:
\begin{equation}
\lbrack ({\bf S}^{k})^{2},({\bf S}^{l})^{2} \rbrack=0\quad \forall k,l.
\end{equation}
Thus the $\{({\bf S}^{k})^{2}\}$ can be used to label the SCD-DFSs by their
eigenvalues $J_{k}$.

In order to make the connection between the addition of angular momentum and
the Dicke states one should, however, use
\begin{equation}
{\bf s}_{\alpha }^{k}\equiv \sum_{i=1}^{k}{\frac{1}{2}}\sigma _{\alpha }^{i}=
{\frac{1}{2}}{\bf S}_{\alpha }^{k}.
\end{equation}
With this definition $({\bf s}^{k})^{2}=\sum_{\alpha }({\bf s}_{\alpha
}^{k})^{2}$ is just the operator whose eigenvalue for the $J^{{\rm th}}$
irrep of the $k$ qubit case is $J_{k}(J_{k}+1)$.
We label the basis determined by the eigenvalues of $({\bf s}^{k})^{2}$ by
\begin{equation}
|J_{1},J_{2},J_{3},\dots ,J_{n-1},J;m_{J}\rangle ,
\end{equation}
where
\begin{equation}
({\bf s}^{k})^{2}|J_{1},J_{2},J_{3},\dots ,J_{n-1},J;m_{J}\rangle
=J_{k}(J_{k}+1)|J_{1},J_{2},J_{3},\dots ,J_{n-1},J;m_{J}\rangle ,
\end{equation}
and where for consistency with the $|J,\lambda ,\mu \rangle $ notation we
use $J$ for $J_{n}$. As in the WCD case, the degeneracy which leads to the
SCD-DFS can be put into a one-to-one correspondence with a graphical
representation of the addition of angular momentum. Here, however, each step
does not simply correspond to adding a $|0\rangle $ or $|1\rangle $ state
but instead corresponds to combining the previous spin $J$ particle with a
spin $1/2$ particle to create a $J+1/2$ or $|J-1/2|$ particle (note the
absolute value so that the total spin is positive). In the graphical
representation of Fig.~(\ref{figure4}) the horizontal axis counts qubits,
and the vertical axis corresponds to the total angular momentum $J_{i}$ up
to the $i^{{\rm th}}$ qubit [note the similarity to Table~(\ref{tab1})].
Each SCD-DFS state then corresponds to a path constructed by successively
moving up or down $1/2$ unit of angular momentum, starting from a single
qubit with $J_{1}=1/2$ . For example, the two DFS$_{3}$($1/2$) states are $
\{|1/2,0,1/2;\pm 1/2\rangle ,|1/2,1,1/2;\pm 1/2\rangle \}$ (corresponding,
respectively, to the paths ``up,down,up'' and ``up,up,down'' and $
m_{J_{3}=1/2}=\pm 1/2$), and the two DFS$_{4}$($0$) states are $
\{|1/2,0,1/2,0;0\rangle ,|1/2,1,1/2,0;0\rangle \}$. Clearly, the set of
paths ${\bf J}_{n}\equiv \{J_{1},J_{2},J_{3},\dots ,J_{n-1},J_{n}\}$ with
fixed $J_{n}$ counts the degeneracy of DFS$_{n}(J_{n})$. Therefore we can
identify the general degeneracy index $\lambda $ (of $|J,\lambda ,\mu
\rangle $) with ${\bf J}_{n}$. Similarly, the dimensionality index $\mu $
can now be identified with $m_{J_{n}}$. Finally, as claimed above $J$ is
just the final $J_{n}$.

\subsection{The Stabilizer and Error Correction Properties}

Note from Eq.~(\ref{eq:Spart}) that the system operators ${\bf S}_{\alpha }={\bf S}_{\alpha }^{n}$.
Therefore they can only affect the last component $|J_{n};m_{J_{n}}\rangle $
of the DFS states. By the identification of the degeneracy index $\lambda $
with the paths $\{J_{1},\dots ,J_{n-1},J\}$, and from the
general expression (\ref{eq:Agl2}) for the action of the ${\bf S}_{\alpha }$
, we know that ${\bf S}_{\alpha }$ acts only on the dimensionality
component:
\begin{equation}
{\bf S}_{\alpha }|J_{1},\dots ,J_{n-1},J;m_{J}\rangle =|J_{1},\dots
,J_{n-1},J\rangle \otimes ({\bf P}_{\alpha }|m_{J}\rangle ),
\label{eq:Sa-action}
\end{equation}
where the ${\bf P}_{\alpha }$ are a $2J+1$ dimensional representation of
$su(2)$ acting directly on the $|m_{J}\rangle $ components of the DFS. The
corresponding DFS stabilizer is
\begin{equation}
{\bf D}(\vec{v})={\bf D}(v_{x},v_{y},v_{z})=\exp \left[ \sum_{\alpha
=x,y,z}v_{\alpha }({\bf S}_{\alpha }-{\bf I}\otimes {\bf P}_{\alpha })\right]
.
\end{equation}
For the $J=0$ DFSs this reduces to all collective rotations+contractions
\cite{Bacon:99a}:
\[
{\bf D}(\vec{v})=\exp \left[ \sum_{\alpha =x,y,z}v_{\alpha }{\bf S}_{\alpha }
\right] =\bigotimes_{i=1}^{n}\exp \left[ \vec{v}\cdot \vec{{\sigma }}_{i}
\right] =\left[ {\bf I}\cos ||\vec{v}||+{\frac{\vec{{\sigma }}\cdot \vec{
v}}{||\vec{v}||}}\sin ||\vec{v}||\right] ^{\otimes n},
\]
where $||v||\equiv (\sum_{\alpha }v_{\alpha }^{2})^{1/2}$ may be complex.
Thus\ DFS$_{n}(0)$ protects against all processes described by Kraus
operators that are linear combinations of collective rotations+contractions $
\exp \left[ \vec{v}\cdot \vec{\sigma} \right] $. The situation for $J\neq 0$ is
more complicated to calculate analytically.

Let us now comment briefly on the error-correction and detection properties
of DFS$_{n}(0)$: The stabilizer elements are tensor products of identical
1-qubit operators, including the following elements of the Pauli group: $
{\bf X}^{\otimes n}$, ${\bf Y}^{\otimes n}$ and ${\bf Z}^{\otimes n}$. Thus,
for any odd--multiple $2k-1<n$ of single qubit errors ${\bf X}$, ${\bf Y}$
and ${\bf Z}$ there is an element in the stabilizer that anticommutes with
it: The code can{\em \ detect} any such error. The $J=0$ SCD-DFS is {\em an
error correcting code of distance} $2$.

\subsection{Nontrivial Operations}

Are there any single-qubit operators which preserve a SCD-DFS (and thus
allow for nontrivial operations on the DFS)? There are no nontrivial
single-qubit operators that commute with {\em all} ${\bf S}_{\alpha }$
operators, since
\begin{equation}
\left[ {\bf S}_{\alpha },\sigma _{\beta }^{j}\right] =\sum_{i}\left[ \sigma
_{\alpha }^{i},\sigma _{\beta }^{j}\right] =i\sum_{i}\delta _{ij}\varepsilon
_{\alpha \beta \gamma }\sigma _{\gamma }^{i}
\end{equation}
which vanishes iff $\alpha =\beta $. Therefore there are no single-qubit operators which preserve all
SCD-DFSs simultaneously.

As for two-qubit operators, the {\em only} such Hermitian operators which
commute with the ${\bf S}_{\alpha }$ are those that are proportional to the
exchange interaction [Eq.~(\ref{eq:exchange})]: ${\bf E}_{ij}|k\rangle
_{i}|l\rangle _{j}=|l\rangle _{i}|k\rangle _{j}$, where $i,j$ label the qubits
acted upon \cite{Zanardi:99c}. In both the single- and two-qubit cases, there could
be additional operators in the generalized commutant ${\cal T}$ (e.g., for
$n=4$ qubits there is an operator which mixes the different $J$'s and preserves
DFS$_{4}$($0$): ${\bf T}=|J=1,\lambda _{1},\mu _{1}\rangle \langle J=2,\lambda
_{1},\mu _{1}|+H.c$.). We will not be concerned with such operations as they
are not needed in order to demonstrate universality, and since we will
show that the exchange operator is sufficient for any SCD-DFS. Our task is
thus to show that exchange interactions alone suffice to generate the entire
$SU(N)$ group on each $N$-dimensional DFS, in the SCD case.

\subsection{Quantum Computation on the $n=3$ and $n=4$ qubit SCD-DFS}

\label{n=3:exchange}

We begin our discussion of universal quantum computation on SCD-DFSs by
examining the simplest SCD-DFS which supports encoding of quantum
information: the $n=3$ decoherence-free subsystem. We label these states as
in Eq.~(\ref{eq:3dfs}) by $|J,\lambda ,\mu \rangle $. Recall that the $J=3/2$
irrep is not degenerate and the $J=1/2$ irrep has degeneracy $2$. The
$J=3/2$ states can be written as $|\frac{3}{2},0,\mu \rangle $, with
$\mu = m_J =\pm 3/2, \pm 1/2$. Since the action of exchange does not
depend on $\mu$ (recall that it affects paths, i.e., the $\lambda$
component only) it suffices to consider the action on the
representative $\mu = 3/2$ only: $|111\rangle$. Let us then
explicitly calculate the action of exchanging the first two physical qubits
on this state and the four $J=1/2$ states. Using Eq.~(\ref{eq:3dfs}):
\begin{eqnarray}
{\bf E}_{12}|\frac{3}{2},0,\frac{3}{2} \rangle &=&{\bf E}_{12}|111\rangle =|\frac{3}{
2},0,\frac{3}{2} \rangle  \nonumber \\
{\bf E}_{12}|{\frac{1}{2}},0,0\rangle &=&{\bf E}_{12}{\frac{1}{\sqrt{2}}}
\left( |010\rangle -|100\rangle \right) ={\frac{1}{\sqrt{2}}}\left(
|100\rangle -|010\rangle \right) =-|{\frac{1}{2}},0,0\rangle  \nonumber \\
{\bf E}_{12}|{\frac{1}{2}},0,1\rangle &=&{\bf E}_{12}{\frac{1}{\sqrt{2}}}
\left( |011\rangle -|101\rangle \right) ={\frac{1}{\sqrt{2}}}\left(
|101\rangle -|011\rangle \right) =-|{\frac{1}{2}},0,1\rangle  \nonumber \\
{\bf E}_{12}|{\frac{1}{2}},1,0\rangle &=&{\bf E}_{12}{\frac{1}{\sqrt{6}}}
\left( -2|001\rangle +|010\rangle +|100\rangle \right) =|{\frac{1}{2}}
,1,0\rangle  \nonumber \\
{\bf E}_{12}|{\frac{1}{2}},1,1\rangle &=&{\bf E}_{12}{\frac{1}{\sqrt{6}}}
\left( 2|110\rangle -|101\rangle -|011\rangle \right) =|{\frac{1}{2}}
,1,1\rangle .
\end{eqnarray}
Focusing just on the $J=1/2$ states, the exchange action on $|\lambda
\rangle \otimes |\mu \rangle $ can thus be written as:
\begin{equation}
{\bf E}_{12}=-\sigma _{z}\otimes {\bf I}.
\end{equation}
Since the action of the ${\bf S}_{\alpha }$ operators on the $J=1/2$ states
is ${\bf I}_{n_{1/2}}\otimes gl(2)$ according to Eq.~(\ref{eq:Agl2}), this
explicit form for ${\bf E}_{12}$ confirms that is has the expected structure
of operators in the commutant of the algebra spanned by the ${\bf S}_{\alpha
}$. It can also be seen that\ quantum information should be encoded in
the $|\lambda \rangle $ component, as discussed before
Eq.~(\ref{eq:rhocoll}).

Using similar algebra it is straightforward to verify that the effect of the
three possible exchanges on the $n=3$ DFS states is given by:
\begin{equation}
{\bf E}_{12}=\left(
\begin{array}{ccc}
1 & 0 & 0 \\
0 & -1 & 0 \\
0 & 0 & 1
\end{array}
\right) \quad {\bf E}_{23}=\left(
\begin{array}{ccc}
1 & 0 & 0 \\
0 & {\frac{1}{2}} & -{\frac{\sqrt{3}}{2}} \\
0 & -{\frac{\sqrt{3}}{2}} & -{\frac{1}{2}}
\end{array}
\right) \quad {\bf E}_{13}=\left(
\begin{array}{ccc}
1 & 0 & 0 \\
0 & {\frac{1}{2}} & {\frac{\sqrt{3}}{2}} \\
0 & {\frac{\sqrt{3}}{2}} & -{\frac{1}{2}}
\end{array}
\right) ,
\end{equation}
where the rows and columns of these matrices are labelled by the basis
elements $\{|J=3/2,\lambda =0\rangle ,|J=1/2,\lambda =0\rangle
,|J=1/2,\lambda =1\rangle \}$. As expected from general properties of the
commutant, the exchange operators do not mix the different $J$ irreps. Now,
\begin{eqnarray}
{\frac{1}{3}}({\bf E}_{12}+{\bf E}_{13}+{\bf E}_{23}) &=&\left(
\begin{array}{ccc}
1 & 0 & 0 \\
0 & 0 & 0 \\
0 & 0 & 0
\end{array}
\right)  \nonumber \\
{\frac{1}{2}}(-{\bf E}_{12}+{\bf E}_{13}+{\bf E}_{23}) &=&\left(
\begin{array}{ccc}
0 & 0 & 0 \\
0 & 1 & 0 \\
0 & 0 & -1
\end{array}
\right)  \nonumber \\
{\frac{1}{\sqrt{3}}}({\bf E}_{13}-{\bf E}_{23}) &=&\left(
\begin{array}{ccc}
0 & 0 & 0 \\
0 & 0 & 1 \\
0 & 1 & 0
\end{array}
\right) ,
\end{eqnarray}
showing that the last two linear combinations of exchanges look like the
Pauli $\sigma _{z}$ and $\sigma _{x}$ on DFS$_{3}(1/2)$. Using a standard
Euler angle construction it is thus possible to perform any $SU(2)$ gate on
this DFS. Moreover, it is possible to act independently on DFS$_{3}(3/2)$
and DFS$_{3}(1/2)$. In other words, we can perform $U(1)$ on DFS$_{3}(3/2)$
alone, and $SU(2)$ on DFS$_{3}(1/2)$ alone. Note, however, that at this
point we cannot yet claim universal quantum computation on a register
composed of clusters of DFS$_{3}(J)$'s ($J$ constant) because we have not
shown how to couple such clusters.

For $n=4$ the Hilbert space splits up into one $J=2$-irrep [DFS$_{4}(2)$],
three $J=1$-irreps [DFS$_{4}(1)$], and two $J=0$-irreps [DFS$_{4}(0)$] --
see Table (\ref{tab1}). Direct calculation of the effect of exchange on
these DFSs shows that we can independently perform $su(1)$ (i.e. zero), $
su(3)$, and $su(2)$. In particular, we find that \cite{Lidar:99c,Bacon:99b}:
\begin{equation}
{\bf X}={\frac{1}{\sqrt{3}}}({\bf E}_{23}-{\bf E}_{13})\quad {\bf Y}={\frac{
i }{2\sqrt{3}}}[{\bf E}_{23}-{\bf E}_{13},{\bf E}_{34}]\quad {\bf Z}={\frac{i
}{ 2}}[{\bf Y},{\bf X}]=-{\bf E}_{12}
\end{equation}
act as the corresponding $su(2)$ Pauli operators on DFS$_{4}(0)$ only.
Further, the following operators act independently on the $J=1$-irreps (rows
and columns are labelled by $\lambda =0,1,2$. The action occurs
simultaneously on all three $\mu $ components corresponding to a given $
\lambda $):
\begin{eqnarray}
{\bf Y}_{13} &=&{\frac{3i}{2\sqrt{2}}}[{\bf E}_{12},{\bf E}_{34}]=\left(
\begin{array}{ccc}
0 & 0 & -i \\
0 & 0 & 0 \\
i & 0 & 0
\end{array}
\right) ,\quad {\bf X}_{13}={\frac{i}{2}}[{\bf E}_{12},{\bf Y}_{13}]=\left(
\begin{array}{ccc}
0 & 0 & 1 \\
0 & 0 & 0 \\
1 & 0 & 0
\end{array}
\right) ,  \nonumber \\
\quad {\bf Z}_{13} &=&{\frac{i}{2}}[{\bf Y}_{13},{\bf X}_{13}]=\left(
\begin{array}{ccc}
1 & 0 & 0 \\
0 & 0 & 0 \\
0 & 0 & -1
\end{array}
\right) ,\quad {\bf Y}_{23}={\frac{2i}{\sqrt{3}}}[{\bf E}_{23},{\bf Z}
_{13}]=\left(
\begin{array}{ccc}
0 & 0 & 0 \\
0 & 0 & -i \\
0 & i & 0
\end{array}
\right) .
\end{eqnarray}
These operators clearly generate $su(3)$, and hence we have an independent $
SU(3)$ action on DFS$_{4}(1)$.

\subsection{Universal Quantum Computation on the $n\geq 5$ qubit SCD-DFSs}

We are now ready to prove our central result: that {\em using only the
two-body exchange Hamiltonians} every unitary operation can be performed on
a SCD-DFS. More specifically:

{\it Theorem 5---} For any $n\geq 2$ qubits undergoing strong collective
decoherence, there exist sets of Hamiltonians ${\sf H}_{J}^{n}$ obtained
from exchange interactions only via scalar multiplication, addition,
Lie-commutator and unitary conjugation, acting as $su(d_{J})$ on the DFS
corresponding to the eigenvalue $J$. Furthermore each set acts {\em
independently} on this DFS only (i.e., with zeroes in the matrix
representation corresponding to their action on the other DFSs).

In preparation for the proof of this result let us note several useful facts:

(i) The exchange operators do not change the value of $m_{J}$, because they
are in the commutant of ${\cal A}=\{S_{\alpha }\}$ [and recall Eq.~(\ref
{eq:Sa-action})]. Therefore in order to evaluate the action of the exchange
operators on the different DFS$_{n}(J)$ ($n$ given) it is convenient to fix $
m_{J}$, and in particular to work in the basis given by the maximal $m_{J}$
value ($m_{J}=J$). Expressions for these ``maximal'' states in terms of $
|J_{1},J_{2},\dots ,J_{n-2};m_{J}\rangle $ and the single qubit states of
the last two qubits are given in Appendix~\ref{appA}.

(ii) Every $({\bf s}^{k})^{2}$ can be written as a sum of exchange operators
and the identity operation \cite{Lidar:99c}. This follows from Eq.~(\ref
{eq:jpauli}) and noting that the exchange operator can be expanded as
\begin{equation}
{\bf E}_{ij}={\frac{1}{2}}\left( {\bf I}+\sigma _{x}^{i}\sigma
_{x}^{j}+\sigma _{y}^{i}\sigma _{y}^{j}+\sigma _{z}^{i}\sigma
_{z}^{j}\right) ,
\end{equation}
so that:
\begin{equation}
({\bf s}^{k})^{2}=k\left(1- {\frac{k}{4}}\right) {\bf I}+{\frac{1}{2}}
\sum_{i\neq j=1}^{k}{\bf E}_{ij}.
\end{equation}
Thus $({\bf s}^{k})^{2}$ is a Hamiltonian which is at our disposal.

We are now ready to present our proof by induction. Recall the
DFS-dimensionality formula for $n_{J}$, Eq.~(\ref{eq:nJ}). We assume that it is
possible to perform $su(n_{J})$ on each of the different DFS$_{n-1}(J)$ {\em
independently} using only exchange operators and the identity Hamiltonian.
Our construction above proves that this is true for $3$ and $4$
qubits. The assumption that the actions we can perform can be
performed independently translates
into the ability to construct Hamiltonians which annihilate all of the DFSs
except a desired one on which they act as $su(n_{J})$.

As in the WCD case a specific DFS$_{n}(J)$ of dimension $n_{J}$ splits into
states which are constructed by the subtraction of angular momentum from DFS$
_{n-1}(J+1/2)$ (T-states), or by the addition of angular momentum to DFS$
_{n-1}(J-1/2)$ (B-states) [see Fig.~(\ref{figure5})]. Performing $su(n_{J+1/2})$ on DFS$_{n-1}(J+1/2)$
will simultaneously act on DFS$_{n}(J)$ and DFS$_{n}(J+1)$. In other words, $
su(n_{J+1/2})$ on DFS$_{n-1}(J+1/2)$ acts on both the B-states of DFS$
_{n}(J+1)$ and on the T-states of DFS$_{n}(J)$. We split the proof into
three steps. In the first step we obtain an $su(2)$ set of operators which
acts only on DFS$_{n}(J)$ and mixes particular B- and T-states. In the second
step we expand the set of operators which mix B- and T-states to cover all
possible $su(2)$ algebras between any two B- and T-states. Finally, in the
third step we apply a Mixing Lemma which shows that we can obtain the full $
su(n_{J})$ (i.e., also mix B-states and mix T-states).

\subsubsection{T- and B-Mixing}
\begin{figure}[!htb]
\hspace{0.2\textwidth}
\psfig{figure=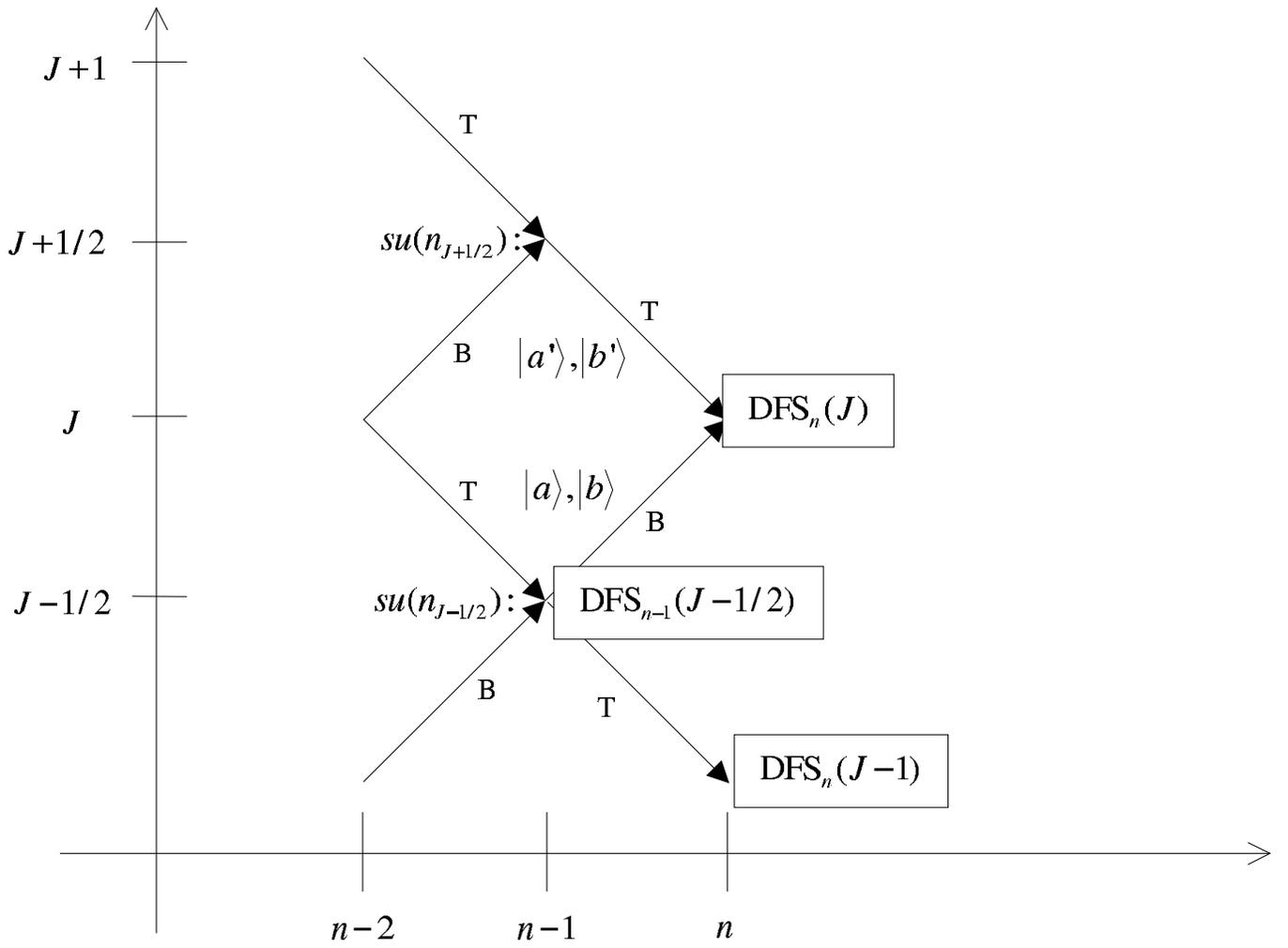,width=1.\textwidth}
\vspace{0.5cm}
\caption{Scheme to visualize the inductive proof for the strong
collective decoherence case. TB- and BT-states of DFS$_n(J)$ are
indicated. $su(n_{J-1/2})$ acts on DFS$_n(J-1)$ and DFS$_n(J)$ via
DFS$_{n-1}(J-1/2)$. See text for details.}
\label{figure5}
\end{figure}

There are two simple instances where there
is no need to show independent action in our proof: (i) The (upper) $J=n/2$
-irrep is always $1$-dimensional, so the action on it is always trivial
(i.e., the Hamiltonian vanishes and hence the action is independent by
definition); (ii) For odd $n$ the ``lowest'' DFS$_{n}(1/2)$ is acted upon
independently by the $su(n_{0})$ from DFS$_{n-1}(0)$ [i.e., $su(n_{0})$
cannot act ``downward'']. In order to facilitate our construction we extend
the notion of T and B-states one step further in the construction of the
DFS. TB-states are those states which are constructed from T-states on
($n-1$)-qubits and from the B-states on $n$-qubit states [see
Fig.~(\ref{figure5})]. Similarly we can define the BT, TT, and
BB-states:
\begin{eqnarray}
|{\rm TT}\rangle  &\equiv &|J_{1},\dots ,J_{n-3},J_{n}+1,J_{n}+{\frac{1}{2}}
,J_{n};m_{J}=J_{n}\rangle =
\begin{array}{rl}
\searrow  &  \\
& \searrow
\end{array}
\\
|{\rm BT}\rangle  &\equiv &|J_{1},\dots ,J_{n-3},J_{n},J_{n}+{\frac{1}{2}}
,J_{n};m_{J}=J_{n}\rangle =\nearrow \searrow   \nonumber \\
|{\rm TB}\rangle  &\equiv &|J_{1},\dots ,J_{n-3},J_{n},J_{n}-{\frac{1}{2}}
,J_{n};m_{J}=J_{n}\rangle =\searrow \nearrow   \nonumber \\
|{\rm BB}\rangle  &\equiv &|J_{1},\dots ,J_{n-3},J_{n}-1,J_{n}-{\frac{1}{2}}
,J_{n};m_{J}=J_{n}\rangle =
\begin{array}{rl}
& \nearrow  \\
\nearrow  &
\end{array}
.
\end{eqnarray}
Every DFS$_{n}(J)$ can be broken down into a direct sum of TT, BT, TB, and
BB-states; e.g., as seen in Fig.~(\ref{figure4}), in DFS$_{6}(1)$ there
are 1 TT, 3 TB, 3 BT and 2 BB states. Note that for $J=n/2-1$ there are no
TT-states, for $J=0$ there are no BB and BT-states, for $J=1/2$ there are no
BB-states, and otherwise there are as many TB as there are BT states,

At this point it is useful to explicitly give the action of exchange on
the last two qubits of a SCD-DFS. Using Eq.~(\ref{eq:twodeep}) we find
(assuming the existence of the given states, i.e., $n$ large enough and $J$
not too large) the representation
\begin{equation}
{\bf E}_{n,n-1}=\left(
\begin{array}{cccc}
1 & 0 & 0 & 0 \\
0 & -\cos (\theta _{J+1}) & \sin (\theta _{J+1}) & 0 \\
0 & \sin (\theta _{J+1}) & \cos (\theta _{J+1}) & 0 \\
0 & 0 & 0 & 1
\end{array}
\right)
\begin{array}{c}
{\rm TT} \\
{\rm BT} \\
{\rm TB} \\
{\rm BB}
\end{array}
\label{eq:lastex}
\end{equation}
where $\tan (\theta _{J})=2\sqrt{J(J+1)}$. Thus exchange acts to
transform the BT and TB states entering a given DFS into linear combinations
of one another, while leaving invariant the BB and TT states.

Let us now consider the action of $su(n_{J-1/2})$ from DFS$_{n-1}(J-1/2)$
[see Fig.~(\ref{figure5})]. It acts on DFS$_{n}(J-1)$ and DFS$_{n}(J)$
simultaneously. However, since the T-states of DFS$_{n}(J-1)$ and the B-states
of DFS$_{n}(J)$ share the same set of quantum numbers
$\{J_{1},...,J_{n-1}\}$, the action of the $su(n_{J-1/2})$ operators
is identical on these two sets of states.

We first deal with the case where the number of BT-states of DFS$_{n}(J)$
is greater than 1. As can be inferred from Fig.~(\ref{figure4}), this condition
corresponds to $J<n/2-1$ and $n>4$. We will separately deal with the $
J=n/2-1$ case at the end of the proof. Let $|a\rangle $ and $|b\rangle $
be any two orthogonal BT-states of DFS$_{n}(J)$ (i.e., states differing
only by the paths on the first $n-2$ qubits). Corresponding to these are $
\{|a^{\prime }\rangle ,|b^{\prime }\rangle \}$: a pair of orthogonal
BT-states of DFS$_{n}(J)$. One of the elements in $su(n_{J-1/2})$ is the
traceless operator ${\bf C}=|a\rangle \langle a|-|b\rangle \langle b|$,
which we have at our disposal by the induction hypothesis. Consider $i[{\bf E
}_{n,n-1},{\bf C}]$: since ${\bf E}_{n,n-1}$ acts as identity on BB states,
even though ${\bf C}$ has an action on DFS$_{n}(J-1)$ the commutator acting
on the BB states of DFS$_{n}(J-1)$ vanishes. The action of $i[{\bf E}
_{n,n-1},{\bf C}]$ on the BT and TB states can be calculated by observing,
using Eq.~(\ref{eq:lastex}), that the matrix representations of ${\bf C}$
and ${\bf E}_{n,n-1}$ are, in the ordered $\{|a^{\prime }\rangle
,|b^{\prime }\rangle ,|a\rangle ,|b\rangle \}$ basis:
\begin{eqnarray}
{\bf C}&=&{\rm diag}(0,0,1,-1)=\frac{1}{2}\left( {\bf I}\otimes
{\sigma }_{z} - {\sigma }_{z}\otimes {\sigma }_{z}\right)   \nonumber \\
{\bf E}_{n,n-1}&=&\left(
\begin{array}{cccc}
-\cos (\theta _{J}) & 0 & \sin (\theta _{J}) & 0 \\
0 & -\cos (\theta _{J}) & 0 & \sin (\theta _{J}) \\
\sin (\theta _{J}) & 0 & \cos (\theta _{J}) & 0 \\
0 & \sin (\theta _{J}) & 0 & \cos (\theta _{J})
\end{array}
\right) =-\cos (\theta _{J}){\sigma }_{z}\otimes {\bf I}+\sin (\theta
_{J}){\sigma }_{x}\otimes {\bf I}.
\end{eqnarray}
This yields:
\begin{equation}
i[{\bf E}_{n,n-1},{\bf C}]=-\sin (\theta _{J}){\sigma }_{y}\otimes {\bf
\sigma }_{z}=i\sin (\theta _{J})\left( -|a\rangle \langle a^{\prime
}|+|a^{\prime }\rangle \langle a|+|b\rangle \langle b^{\prime }|-|b^{\prime
}\rangle \langle b|\right) .
\end{equation}
Now let $|c\rangle $ be a TT-state of DFS$_{n}(J)$. Such a state always
exists unless $J=n/2-1$, which is covered at the end of the proof. Then
there is an operator ${\bf D}=|a^{\prime }\rangle \langle a^{\prime }|-|c\rangle \langle c|$ in $
su(n_{J+1/2})$.\footnote{
We need to subtract $|c\rangle \langle c|$ in order to obtain a traceless
operator.} It follows that:
\begin{equation}
{\bf X}_{aa^{\prime }}\equiv {\frac{1}{\sin (\theta _{J})}}i[i[{\bf E}
_{n,n-1},{\bf C}],{\bf D}]=|a\rangle \langle a^{\prime }|+|a^{\prime
}\rangle \langle a|,
\end{equation}
acts like an encoded ${\sigma }_{x}$ on $|a\rangle $ and $|a^{\prime
}\rangle $ and annihilates all other states. Further, one can implement the
commutator
\begin{equation}
{\bf Y}_{aa^{\prime }}=i[{\bf X}_{aa^{\prime }},{\bf D}]=i\left( |a\rangle
\langle a^{\prime }|-|a^{\prime }\rangle \langle a|\right) ,
\end{equation}
which acts like an encoded ${\sigma }_{y}$ on $|a\rangle $ and $|a^{\prime
}\rangle $. Finally, one can construct ${\bf Z}_{aa^{\prime }}=i[{\bf X}
_{aa^{\prime }},{\bf Y}_{aa^{\prime }}]=|a\rangle \langle a|-|a^{\prime
}\rangle \langle a^{\prime }|$. Thus we have shown that for $J<n/2-1$ we
can validly (using only exchange Hamiltonians) perform $su(2)$ operations
between $|a\rangle $, a specific B-state and $|a^{\prime }\rangle $, its
corresponding T-state, on DFS$_n(J)$ only.

\subsubsection{Extending the $su(2)$'s}
We now show that by using the
operation of conjugation by a unitary we can construct $su(2)$ between {\em
any} two B and T-states. To see this recall Eq.~(\ref{eq:Heff}), which
allows one
to take a Hamiltonian ${\bf H}$ and turn it via conjugation by a
unitary gate into the new Hamiltonian ${\bf H}
_{{\rm eff}}={\bf U}{\bf H}{\bf U}^{\dagger }$. By the induction hypothesis
we have at our disposal every $SU$ gate which acts on the T-states of DFS$_{n}(J)$ [and simultaneously acts on the B-states of DFS$_{n}(J+1)$] and
also every $SU$ gate which acts on the B-states of DFS$_{n}(J)$ [and
simultaneously acts on the T-states of DFS$_{n}(J-1)$]. Above we have shown
how to construct ${\bf X}$, ${\bf Y}$, and ${\bf Z}$ operators between
specific T- and B-states: $|a^{\prime }\rangle $ and $|a\rangle $. Let $
|i^{\prime }\rangle $ and $|i\rangle $ be some other T- and B-states of DFS$_{n}(J)$,
respectively. Then we have at our disposal the gate ${\bf P}_{i^{\prime }i}=|a^{\prime
}\rangle \langle i^{\prime }|+|i^{\prime }\rangle \langle a^{\prime }|+|a\rangle \langle
i|+|i\rangle \langle a|+{\bf O}$ where ${\bf O}$ is an operator which acts
on a DFS other than DFS$_{n}(J)$ (included to make ${\bf P}_{i^{\prime }i}$ an $SU$
operator). It is simple to verify that
\begin{equation}
{\bf X}_{i^{\prime }i}={\bf P}_{i^{\prime }i}{\bf X}_{aa^{\prime }}{\bf P}_{i^{\prime }i}^{\dagger
}=|i^{\prime }\rangle \langle i|+|i\rangle \langle i^{\prime }|,
\end{equation}
which acts as an encoded $\sigma _{x}$ between $|i^{\prime }\rangle $ and $|i\rangle $.
Note that because ${\bf X}_{aa^{\prime }}$ only acts on DFS$_{n}(J)$, $
{\bf X}_{i^{\prime }i}$ will also only act on the same DFS. Similarly one can
construct ${\bf Y}_{i^{\prime }i}={\bf P}_{i^{\prime }i}{\bf Y}_{aa^{\prime }}{\bf P}
_{i^{\prime }i}^{\dagger }$ and ${\bf Z}_{i^{\prime }i}={\bf P}_{i^{\prime }i}{\bf Z}_{aa^{\prime }}{\bf P}
_{i^{\prime }i}^{\dagger }$ which act, respectively, as encoded $\sigma _{y}$ and $
\sigma _{z}$ on $|i^{\prime }\rangle $ and $|i\rangle $. Thus we have shown that one
can implement every $su(2)$ between any two T- and B-states in DFS$_{n}(J)$.
Each of these $su(2)$ operations is performed {\em
independently} on DFS$_{n}(J)$.

\subsubsection{Mixing T- and B-States}
Next we use a Lemma proved in Appendix~\ref{appB}:

{\it Mixing Lemma}: Given is a Hilbert space ${\cal H}$ $={\cal H}_{1}\oplus
{\cal H}_{2}$ where $\dim {\cal H}_{j}=$ $n_{j}$. Let $\{|i_{1}\rangle \}$
and $\{|i_{2}\rangle \}$ be orthonormal bases for ${\cal H}_{1}$ and ${\cal H
}_{2}$ respectively. If one can implement the operators ${\bf X}
_{i_1 i_2}=|i_{1}\rangle \langle i_{2}|+|i_{2}\rangle \langle i_{1}|$, ${\bf Y}
_{i_1 i_2}=i|i_{1}\rangle \langle i_{2}|-i|i_{2}\rangle \langle i_{1}|$, and $
{\bf Z}_{i_1 i_2}=|i_{1}\rangle \langle i_{1}|-|i_{2}\rangle \langle i_{2}|$,
then one can implement $su(n_{1}+n_{2})$ on ${\cal H}$.

Above we have explicitly shown that we can obtain every ${\bf X}_{i_1 i_2}$, $
{\bf Y}_{i_1 i_2}$, and ${\bf Z}_{i_1 i_2}$ acting independently on DFS$_{n}(J)$. Thus
direct application of the Mixing Lemma tells us that we can perform $
su(n_{J})$ {\em independently} on this DFS.

{\it Special case of }$J=n/2-1$: We have neglected DFS$_{n}(n/2-1)$ because
it did not contain two different BT-states (nor a TT) state. The dimension
of this DFS is $n-1$. We now show how to perform $su(n-1)$ on this DFS using
the fact that we have already established $su(n_{J=n/2-2})$ on DFS$_{n}(n/2-2
$). First, note that by the induction hypothesis we can perform $
su(n_{J=n/2-3/2})$ independently on DFS$_{n-1}(n/2-3/2)$. As above, this
action simultaneously affects DFS$_{n}(n/2-1)$ and DFS$_{n}(n/2-2)$.
However, since we can perform $su(n_{J=n/2-2})$ on DFS$_{n}(n/2-2)$, we can
subtract out the action of $su(n_{J=n/2-3/2})$ on DFS$_{n}(n/2-2)$. Thus we
can obtain $su(n_{J=n/2-3/2})$ on all of the B-states of DFS$_{n}(n/2-1)$.
But the exchange operator ${\bf E}_{n,n-1}$ acts to mix the B-states with
the single T-state of DFS$_{n}(n/2-1)$. Thus we can construct an $su(2)$
algebra between that single-T state and a single B-state in a manner
directly analogous to the above proof for $J<n/2-1$. Finally, by the
Enlarging Lemma it follows that we can obtain $su(n-1)$ on DFS$_{n}(n/2-1)$.

This concludes the proof that the exchange interaction is universal
independently on each of the different strong-collective-decoherence DFSs.

\subsection{State Preparation and Measurement on the Strong Collective
Decoherence DFS}

At first glance it might seem difficult to prepare pure states of a
SCD-DFS, because these states are nontrivially entangled. However, it
is easy to see that every DF {\em subspace} contains a state which is
a tensor product of singlet states:
\begin{equation}
|0_D\rangle = \left({1 \over \sqrt{2}} \right)^{n/2} \otimes_{j=1}^{n/2} (|01
\rangle -|10\rangle),
\end{equation}
because these states have zero total angular momentum. Thus a supply
of singlet states is sufficient to prepare DF subspace
states. Further, DF {\em subsystems} always contain a state which is a
tensor product 
of a DF subspace and a pure state of the form $|1\rangle \otimes
\cdots \otimes |1\rangle$. This can be seen from Fig.~(\ref{figure4}),
where the lowest path leading to a specific DFS$_n(J)$ is composed of
a segment passing through a DF subspace (and is thus of the form
$|0_D\rangle$), and a segment going straight up from there to
DFS$_n(J)$. The corresponding state is equivalent to adding a spin-$0$
(DF subspace) 
and a spin-$J$ DF subsystem (the $|J,m_J=J\rangle$ state of the latter
is seen to be made up 
entirely of $|1\rangle \otimes \cdots \otimes |1\rangle$). In general,
addition of a spin-$0$ 
DFS and a spin-$J$ DFS simply corresponds to tensoring 
the two states. Note, however, that addition of two arbitrary DF
subsystems into a larger DFS is not nearly as simple: concatenation of
two $J\neq0$ DFSs does {\em not} correspond to tensoring.

Pure state preparation for a SCD-DFS can thus be as simple as the ability to
produce singlet states and $|1\rangle$ states (it is also possible to use the
$|J,m_J=-J\rangle=|0\rangle \otimes \cdots \otimes |0\rangle$ or any of the
other $|J,m_J\rangle$ states plus singlets). Other, more complicated pure state
preparation procedures are also conceivable, and the decision as to which
procedure to use is clearly determined by the available resources to manipulate
quantum states. The pure state preparation of singlets and computational basis
states has the distinct advantage that verification of these states should be
experimentally achievable. Such verification is necessary for fault-tolerant
preparation \cite{Gottesman:97}.

Measurements on the SCD-DFS can be performed by using the concatenated measurement
scheme detailed in the WCD-DFS discussion [Sec.~(\ref{WCDprep})]. In
particular, by attaching a SCD-DF {\em subspace} ancilla via
concatenation, one can construct 
any concatenated measurement scenario. All that remains to be shown is how to
perform a destructive measurement on such an ancilla. In
\cite{Bacon:99b,DiVincenzo:99a} such schemes are presented for the $n=4$ SCD-DF
subspace which encodes a single qubit of information. We will not
repeat the details of these schemes here, but note that they involve measurements of
single physical-qubit observables and thus are experimentally very reasonable.
Further, we note that the ability to perform a concatenated measurement scenario
by concatenating an ancilla DFS composed of a single encoded-qubit,
can be used to perform 
any possible concatenated DFS measurement scenario. As mentioned in
the WCD case, the concatenated measurement procedures are fault-tolerant. Thus we
have shown how to perform fault-tolerant preparation and decoding on 
the SCD-DFS.

\section{Universal Fault-Tolerant Computation on Concatenated Codes}

\label{concatchapter}

So far we have shown how to implement universal computation with local
Hamiltonians on a DFS corresponding to a single block of qubits. This
construction assumes that the only errors are collective. This is a very
stringent symmetry requirement, which obviously becomes less realistic the
larger the number of particles $n$ is. It is thus desirable to be able to
deal with perturbations that break the collective-decoherence (permutation)\
symmetry. To this end we have previously studied the effect of
symmetry-breaking perturbations on decoherence-free subspaces \cite
{Bacon:99a}, and have proposed a concatenation method to make DFSs robust in
the presence of such perturbations. This method embeds DFS blocks of four
particles (each block constituting a single encoded qubit) into a QECC \cite
{Lidar:99a}. The QECC in the outer layer then takes care of any single
encoded-qubit errors on each of its constituent DFS-blocks; in fact the code
can correct for any ``leakage'' error taking a state outside of the DFS, by
transforming this into a single encoded qubit error on the QECC. By choosing
an appropriate QECC it is thus possible to deal with any type of
non-collective error on the encoded DFS-qubits. In particular, by using the
``perfect'' 5-qubit code \cite{Laflamme:96} it is possible to correct all
independent errors between blocks of four particles.

The problem with this construction so far was, that in order to correct on
the outer QECC, it is necessary to perform encoded operations on the
constituent DFSs in a fault-tolerant way, using (realistic) local
interactions. Specifically, it is necessary to be able to implement all
single encoded-qubit operations on the DFS-qubits of the outer QECC, as well
as operations {\em between} two DFS-blocks (see \cite{Gottesman:97} for
details). Given that one can perform single qubit (or ``qupit'' for
higher-dimensional DFSs) operations on each DFS-block, the only additional
gate necessary to implement error-correction {\em and} universal quantum
computation on a concatenated QECC-DFS, is any non-separable
two-encoded-qubit gate ${\bf K}$ between any four states in the two
DFS-blocks. For instance, a controlled-phase operation, which gives a phase
of $-1$ to $|0_{L}0_{L}\rangle $ and leaves all other states unchanged. In
fact, it is sufficient to be able to perform this gate ${\bf K}$ between
neighboring blocks only \cite{Gottesman:99a}.

The above results give us the tools to perform single DFS-qubit (or qupit)
operations on a block. To construct an encoded ${\bf K}$ between two
neighboring blocks, we assume that the corresponding physical qubits are
spatially close together during the switching time of the gate. Since the
symmetry of collective decoherence arises from the spatial correlation of
the decoherence process, we can further assume that during this switching
time, both DFS-blocks couple to the same bath mode. This assumption is
physically motivated by the expectation that collective decoherence occurs
in the analog of the Dicke limit of quantum optics, where the qubits have
small spatial separations relative to the bath correlation length \cite
{Dicke:54}. Then the two DFS-clusters temporarily form a bigger DFS and we
can use the universal operations we have constructed previously on this big
DFS to implement the desired gate ${\bf K}$.

This makes the concatenated QECC-DFS fully workable as a code supporting
universal fault-tolerant quantum computation.

\section{Summary and Conclusions}

\label{concludechapter}

In this paper we have settled the issue of {\em quantum computation} with
realistic (few-body) means on both decoherence-free subspaces and
decoherence-free (noiseless) subsystems (DFSs) for two important forms of
decoherence: collective phase damping (``weak collective decoherence''), and
collective phase damping plus collective dissipation (``strong collective
decoherence''). This resolves an outstanding question as to whether
universal computation on these physically relevant DFSs by using just 1- and
2-body Hamiltonians is possible.

The implications of this result for the usefulness of DFSs are
drastic. They
put the theory of DFSs on an equal footing with the theory of quantum error
correction, in that the full repertoire of universal fault tolerant quantum
computation is now available on DFSs for collective decoherence: the
most important pertinent decoherence
process. Moreover, the strict assumption of
collective decoherence can be lifted by allowing for perturbing independent
qubit errors. As we proposed earlier it is possible to stabilize DFSs against
such errors by concatenation with a quantum error correcting code (QECC).
However, to be able to implement error-correction and fault-tolerant
universal computation on these concatenated codes a crucial (and so far
missing) ingredient was the ability to perform encoded operations on the
DFS-blocks fault-tolerantly. This paper settles that matter, showing
constructively that DFSs can be made robust.

Furthermore, this paper reports on a general framework incorporating both
DFSs and QECCs, and generalizes the theory of stabilizer codes to the
(non-abelian) DFS-case. This framework enabled us to identify the allowed
operations on a DFS and to show that these operations can be performed while
maintaining a very strong form of fault-tolerance: the states remain within
the DFS during the entire switching time of the gate. Our formalism should
be readily applicable for other non-additive codes.

There is an interesting duality between QECCs which are designed to correct
single (or greater) qubit errors and DFSs. In QECC the errors are all single
body interactions. The QECC condition therefore implies that any one or
two-body Hamiltonian must take codewords outside of the code space because
these interactions themselves look like errors. QECCs must leave their
codespace in order to perform quantum computation on the encoded operations.
This means that QECCs must have gates which act much faster than the
decoherence mechanism so that a perturbative treatment can be carried out.
QECC can correct small errors but the price paid for this is that gates
must be executed quickly (not to mention that fault-tolerant gates must also
be used). DFSs on the other hand, do not have the requirement of correcting
single qubit errors and we have found that a single two-body
interaction (exchange) is sufficient to generate universal quantum
computation fault-tolerantly. DFSs have larger errors but this allows
for an economy of Hamiltonians.

As corollaries to our results on weak and strong collective decoherence two
additional properties of the corresponding DFS encodings appear:

\begin{itemize}
\item  {\bf One can work on all DFSs in parallel:} Since we are able to
implement $SU(d_{n})$ on each DFS$_{n}$ ($n$=number of particles) {\em
independently}, we can in principle work on all DFSs in parallel. This means
that we can encode quantum information into each of the DFSs and perform
calculations (possibly different) on all of them {\em at once}.

\item  {\bf For the strong collective decoherence case the exchange gate is
asymptotically universal:} It is well known that the encoding
efficiency of the singlet space of the
strong collective decoherence-DFS for large $n$ approaches unity \cite{Zanardi:97c}. More precisely, let $k$ be the number of
{\em encoded} qubits in the singlet ($J=0$) sector of a Hilbert space of $n$
qubits, then
\begin{equation}
\lim_{n\rightarrow \infty }\frac{k}{n}=1-\frac{3}{2}\frac{\log _{2}n}{n}.
\end{equation}
We have established that the exchange gate alone (with an irrational phase)
implements universal computation on each DFS and on the singlet space in
particular. Thus, we find that, for large $n$, in order to achieve universal
computation with nearly perfect efficiency, all we need to be able to
perform is the exchange interaction. This result is very promising from
an experimental point of view, since the exchange interaction is
prevalent whenever there is a Heisenberg coupling between systems \cite
{Lidar:99c,Bacon:99b}. We emphasize that regardless of the decoherence
mechanism, this implies that universal quantum computation can be achieved
``asymptotically'' using a {\em single} gate \cite{DiVincenzo:99a}. We
conjecture that there are many more such two-body interactions which
similarly provide such ``asymptotic universality'' {\em on their own}.
\end{itemize}

\section*{Acknowledgments}

This material is based upon work supported by the U.S. Army Research Office
under contract/grant number DAAG55-98-1-0371 and NSF DMS-9971169. It
is a pleasure to acknowledge helpful discussions with Drs. Dorit
Aharonov and Alexei Kitaev.

\appendix

\section{The Partial Collective Angular Momentum Operators are a Set
of Commuting Observables}

\label{appD}

We prove here that the partial collective operators ${\bf S}_{\alpha
}^{k}\equiv {\bf S}_{\alpha }^{(1,2,\dots ,k)}=\sum_{i=1}^{k}\sigma _{\alpha
}^{i}$ form a commuting set and hence, a good operator basis. Note first
that
\begin{equation}
({\bf S}^{k})^{2}=\sum_{i,j=1}^{k}\sum_{\alpha =x,y,z}\sigma _{\alpha
}^{i}\sigma _{\alpha }^{j}.  \label{eq:jpauli}
\end{equation}
Thus
\begin{equation}
\lbrack ({\bf S}^{k})^{2},({\bf S}^{l})^{2} \rbrack =\left[ \sum_{i,j=1}^{k}\sum_{
\alpha =x,y,z}\sigma _{\alpha }^{i}\sigma _{\alpha
}^{j},\sum_{m,n=1}^{l}\sum_{\beta =x,y,z}\sigma _{\beta }^{m}\sigma _{\beta
}^{n}\right] .
\end{equation}
Terms with $\alpha =\beta $ obviously commute. Further, terms with
($m=i,n=j$), ($m=j,n=i$), or ($i\neq m,n, j\neq m,n$), commute, so we
need only consider ($i=m,j\neq n$), ($i=n,j\neq m$)
or ($i\neq m,j=n$), ($i\neq n,j=m$). In addition, assuming w.l.o.g. that $l\geq k$, terms
with $m,n>k$ also commute. Thus we are left with
\begin{equation}
\lbrack ({\bf S}^{k})^{2},({\bf S}^{l})^{2} \rbrack= 2 \sum_{i,j=1}^{k}\sum_{n\left(
\neq j\right) =1}^{k}\sum_{\beta \neq \alpha =x,y,z}\left[ \sigma _{\alpha
}^{i}\sigma _{\alpha }^{j},\sigma _{\beta }^{i}\sigma _{\beta }^{n}\right]
+2\sum_{i,j=1}^{k}\sum_{m\left( \neq i\right) =1}^{k}\sum_{\beta \neq \alpha
=x,y,z}\left[ \sigma _{\alpha }^{i}\sigma _{\alpha }^{j},\sigma _{\beta
}^{m}\sigma _{\beta }^{j}\right] .
\end{equation}
Using the fact that $[\sigma _{\alpha }^{i}\sigma _{\alpha }^{j},\sigma
_{\beta }^{i}\sigma _{\beta }^{n}]=i\sum_{\gamma }\varepsilon _{\alpha \beta
\gamma }\sigma _{\gamma }^{i}\sigma _{\alpha }^{j}\sigma _{\beta }^{n}$ and $
[\sigma _{\alpha }^{i}\sigma _{\alpha }^{j},\sigma _{\beta }^{m}\sigma
_{\beta }^{j}]=i\sum_{\gamma }\varepsilon _{\alpha \beta \gamma }\sigma
_{\alpha }^{i}\sigma _{\beta }^{m}\sigma _{\gamma }^{j}$:
\[
\lbrack ({\bf S}^{k})^{2},({\bf S}^{l})^{2} \rbrack=2\sum_{i,j=1}^{k}\sum_{n\left(
\neq j\right) =1}^{k}\sum_{\alpha ,\beta ,\gamma =\{x,y,z\}}\varepsilon
_{\alpha \beta \gamma }\sigma _{\gamma }^{i}\sigma _{\alpha }^{j}\sigma
_{\beta }^{n}+2\sum_{i,j=1}^{k}\sum_{m\left( \neq i\right)
=1}^{k}\sum_{\alpha ,\beta ,\gamma =\{x,y,z\}}\varepsilon _{\alpha \beta
\gamma }\sigma _{\alpha }^{i}\sigma _{\beta }^{m}\sigma _{\gamma }^{j},
\]
and both sums vanish due to the antisymmetric property of $\varepsilon
_{\alpha \beta \gamma }$.

\section{Maximal-$\lowercase{m}_{J}$ States of the Strong Collective Decoherence DFS}

\label{appA}

We show how to recursively express the $n$-particle total spin-$J$ states in
terms of ($n-1$)-particle states. Let us focus on DFS$_{n}(J)$ and in
particular on the maximal-$m_{J}$ state in it:
\begin{equation}
|\psi \rangle =|J_{1},\dots ,J_{n-1},J;m_{J}=J\rangle .
\end{equation}
In general ($J\neq 0,{n/2}$) there are two kinds of states: bottom ($|\psi
\rangle _{{\rm B}}$) and top ($|\psi \rangle _{{\rm T}}$) ones. The
angular momentum addition rule that must be satisfied for adding a single
spin-$\frac{1}{2}$ particle is that
\[
m_{J_{n-1}}\pm \frac{1}{2}=m_{J}.
\]
The B-state comes from adding a particle to the maximal $m_{J}$ state in DFS$
_{n-1}(J-1/2)$, which is:
\begin{equation}
|{\rm B}\rangle =|J_{1},\dots ,J_{n-2},J{-{\frac{1}{2}}};m_{J_{n-1}}=J-{{
\frac{1}{2}}}\rangle .
\label{eq:B}
\end{equation}
There is only one way to go from $|{\rm B}\rangle $ to $|\psi \rangle _{{\rm
B}}$, namely to add $1/2$ to $m_{J_{n-1}}=J-{{\frac{1}{2}}}$ in order to
obtain $m_{J}=J$. Thus
\begin{equation}
|\psi \rangle _{{\rm B}}=|{\rm B}\rangle |{\frac{1}{2}},{\frac{1}{2}}\rangle
\label{eq:psiB},
\end{equation}
where $|{\frac{1}{2}},{\frac{1}{2}}\rangle $ is the single-particle spin-up
state. The situation is different for the T-state, which is constructed by
adding a particle to
\begin{equation}
|{\rm T}_{\pm }\rangle =|J_{1},\dots ,J_{n-2},J+{{\frac{1}{2}}}
;m_{J_{n-1}}=J\pm {{\frac{1}{2}}}\rangle .
\label{eq:Tpm}
\end{equation}
These two possibilities give:
\begin{equation}
|\psi \rangle _{{\rm T}}=\alpha |{\rm T}_{+}\rangle |{\frac{1}{2}},-{\frac{1
}{2}}\rangle +\beta |{\rm T}_{-}\rangle |{\frac{1}{2}},{\frac{1}{2}}\rangle .
\label{eq:psiT}
\end{equation}
To find the coefficients $\alpha $ and $\beta $ we use the collective
raising operator ${\bf s}_{+}={\bf s}_{x}+i{\bf s}_{y}$, where we recall
that ${\bf s}_{\alpha }^{(k)}=\frac{1}{2}\sum_{i=1}^{k}\sigma _{\alpha }^{i}$. Since $|\psi \rangle $ is a maximal-$m_{J}$ state it is annihilated by $
{\bf s}_{+}\equiv {\bf s}_{\alpha }^{(n)}$. Similarly, $|{\rm T}_{+}\rangle $
is annihilated by ${\bf s}_{+}^{(n-1)}$. Therefore, since ${\bf s}_{+}={\bf s
}_{+}^{(n-1)}+\frac{1}{2}\sigma _{+}^{n}$:
\begin{eqnarray*}
{\bf s}_{+}|{\rm T}_{+}\rangle |{\frac{1}{2}},-{\frac{1}{2}}\rangle &=&|{\rm
T}_{+}\rangle |{\frac{1}{2}},{\frac{1}{2}}\rangle  \\
{\bf s}_{+}|{\rm T}_{-}\rangle |{\frac{1}{2}},{\frac{1}{2}}\rangle &=&\sqrt{
2J+1}|{\rm T}_{+}\rangle |{\frac{1}{2}},{\frac{1}{2}}\rangle ,
\end{eqnarray*}
where in the second line we used the elementary raising operator formula $
{\bf J}_{+}|j,m\rangle =\left[ j(j+1)-m(m+1)\right] ^{1/2}|j,m+1\rangle $
with $j=J+{{\frac{1}{2}}}$ and $m=J-{{\frac{1}{2}}}$. Application of ${\bf s}
_{+}$ to Eq.~(\ref{eq:psiT}) thus yields:
\begin{equation}
\alpha +\sqrt{2J+1}\beta =0
\end{equation}
Hence, up to an arbitrary phase choice, we find that
\begin{equation}
\alpha =-\sqrt{\frac{2J+1}{2J+2}}\quad \beta ={\frac{1}{\sqrt{
2J+2}}.}
\end{equation}
The special cases of $J=0,{n/2}$ differ only in that the corresponding DFSs
support just T- and B-states, respectively. The calculation of the
coefficients, therefore, remains the same.

In a similar manner one can carry the calculation one particle deeper. Doing
this we find for the maximal-$m_{J}$ states (provided they exist):
\begin{eqnarray}
|{\rm TT}\rangle \equiv |J_{1}, &&\dots ,J_{n-3},J+1,J+{\frac{1}{2}}
,J;m_{J}=J\rangle =\sqrt{\frac{2J+1}{2J+3}}|J_{1},\dots
,J_{n-3},J+1;m_{J_{n-2}}=J+1\rangle |{\frac{1}{2}},-{\frac{1}{2}}\rangle |{
\frac{1}{2}},-{\frac{1}{2}}\rangle   \nonumber \\
&&-\sqrt{\frac{2J+1}{(2J+2)(2J+3)}}|J_{1},\dots
,J_{n-3},J+1;m_{J_{n-2}}=J\rangle \left( |{\frac{1}{2}},{\frac{1}{2}}\rangle
|{\frac{1}{2}},-{\frac{1}{2}}\rangle +|{\frac{1}{2}},-{\frac{1}{2}}\rangle |{
\frac{1}{2}},{\frac{1}{2}}\rangle \right)   \nonumber \\
&&+\sqrt{\frac{2}{(2J+2)(2J+3)}}|J_{1},\dots
,J_{n-3},J+1;m_{J_{n-2}}=J-1\rangle |{\frac{1}{2}},{\frac{1}{2}}\rangle |{
\frac{1}{2}},{\frac{1}{2}}\rangle   \nonumber \\
|{\rm BT}\rangle \equiv |J_{1}, &&\dots ,J_{n-3},J,J+{\frac{1}{2}}
,J;m_{J}=J\rangle =-\sqrt{\frac{2J+1}{2J+2}}|J_{1},\dots
,J_{n-3},J;m_{J_{n-2}}=J\rangle |{\frac{1}{2}},{\frac{1}{2}}\rangle |{\frac{1
}{2}},-{\frac{1}{2}}\rangle   \nonumber \\
&&+{\frac{1}{\sqrt{(2J+2)(2J+1)}}}|J_{1},\dots
,J_{n-3},J;m_{J_{n-2}}=J\rangle |{\frac{1}{2}},-{\frac{1}{2}}\rangle |{\frac{
1}{2}},{\frac{1}{2}}\rangle   \nonumber \\
&&+\sqrt{\frac{2J}{(2J+1)(2J+2)}}|J_{1},\dots
,J_{n-3},J;m_{J_{n-2}}=J-1\rangle |{\frac{1}{2}},{\frac{1}{2}}\rangle |{
\frac{1}{2}},{\frac{1}{2}}\rangle   \nonumber \\
|{\rm TB}\rangle \equiv |J_{1}, &&\dots ,J_{n-3},J,J-{\frac{1}{2}}
,J;m_{J}=J\rangle =-\sqrt{\frac{2J}{2J+1}}|J_{1},\dots
,J_{n-3},J;m_{J_{n-2}}=J\rangle |{\frac{1}{2}},-{\frac{1}{2}}\rangle |{\frac{
1}{2}},{\frac{1}{2}}\rangle   \nonumber \\
&&+{\frac{1}{\sqrt{2J+1}}}|J_{1},\dots ,J_{n-3},J;m_{J_{n-2}}=J-1\rangle |{
\frac{1}{2}},{\frac{1}{2}}\rangle |{\frac{1}{2}},{\frac{1}{2}}\rangle
\nonumber \\
|{\rm BB}\rangle \equiv |J_{1}, &&\dots ,J_{n-3},J-1,J-{\frac{1}{2}}
,J;m_{J}=J\rangle =|J_{1},\dots ,J_{n-3},J-1;m_{J_{n-2}}=J-1\rangle |{\frac{1
}{2}},{\frac{1}{2}}\rangle |{\frac{1}{2}},{\frac{1}{2}}\rangle .
\label{eq:twodeep}
\end{eqnarray}
Caution must be exercised in using these expressions near the boundary
of Table~(\ref{tab1}), where
some of the states may not exist.

\section{Proofs of the Lemmas}

\label{appB}

{\it Enlarging Lemma---} Let ${\cal H}$ be a Hilbert space of
dimension $d$ and let $|i\rangle \in {\cal H}$. Assume we are given a set of
Hamiltonians ${\sf H}_{1}$ that generates $su(d-1)$ on the subspace of $
{\cal H}$ that does not contain $|i\rangle $, and another set ${\sf H}_{2}$
that generates $su(2)$ on the subspace of ${\cal H}$ spanned by $\left\{
|i\rangle ,|j\rangle \right\} $, where $|j\rangle $ is another state in $
{\cal H}$. Then $[{\sf H}_{1},{\sf H}_{2}]$ (all commutators)
generates $su(d)$ on ${\cal H}$
under closure as a Lie-algebra.

{\it Proof}. We explicitly construct the Lie-algebra $su(d)$ with the given
Hamiltonians.
Let ${\tilde{{\cal H}}}\subset
{\cal H}$ be the $d-1$ dimensional subspace ${\sf H}_{1}$ acts on. Let
us show that we can generate $su(2)$
between $|k\rangle \in {\tilde{{\cal H}}}$ and $|i\rangle $.

Let ${\bf X}_{ij}\equiv |i\rangle \langle j|+|j\rangle \langle i|\in {\sf H}
_{2}$ and ${\bf X}_{jk}\equiv |j\rangle \langle k|+|k\rangle \langle j|\in
{\sf H}_{1}$. Then
\begin{equation}
{\bf Y}_{ik}\equiv i[{\bf X}_{jk},{\bf X}_{ij}]=-i|i\rangle \langle
k|+i|k\rangle \langle i|
\end{equation}
acts as $\sigma _{y}$ on the states $|i\rangle ,|k\rangle $. Similarly
\begin{equation}
{\bf X}_{ik}\equiv i[{\bf Y}_{ij},{\bf X}_{jk}]=|i\rangle \langle
k|+|k\rangle \langle i|
\end{equation}
yields $\sigma _{x}$ on the space spanned by $|i\rangle ,|k\rangle $. These
two operations generate $su(2)$ on $|i\rangle ,|k\rangle $ for all
$|k\rangle$ in the subspace of ${\cal H}$ that does not contain
$|i\rangle$. Now the Mixing Lemma gives the desired 
result together with the observation that there we only use elements
in $[{\sf H}_{1},{\sf H}_{2}]$.

{\it Mixing Lemma}--- Consider the division of an $n$ dimensional Hilbert
space ${\cal H}$ into a direct sum of two subspaces ${\cal H}_{1}\oplus
{\cal H}_{2}$ of dimensions $n_{1}$ and $n_{2}$ respectively. Suppose that $
|i_{n}\rangle $ is an orthonormal basis for ${\cal H}_{n}$. Then the Lie
algebras generated by ${\bf X}_{i_{1},i_{2}}=|i_{1}\rangle \langle
i_{2}|+|i_{2}\rangle \langle i_{1}|$, ${\bf Y}_{i_{1},i_{2}}=i|i_{1}\rangle
\langle i_{2}|-i|i_{2}\rangle \langle i_{1}|$, and ${\bf Z}
_{i_{1},i_{2}}=|i_{1}\rangle \langle i_{1}|-|i_{2}\rangle \langle i_{2}|$
generate $su(n)$.

{\it Proof}. We explicitly construct the elements of $su(n)$. Consider $i[
{\bf X}_{i_{1},i_{2}},{\bf Y}_{j_{1},j_{2}}]$. Clearly, if $i_{1}\neq
i_{2}\neq j_{1}\neq j_{2}$ this equals zero and if $i_{1}=j_{1}$ and $
i_{2}=j_{2}$ then this commutator is $-{\bf Z}_{i_{1},i_{2}}$. If, however, $
i_{1}=j_{1}$ and $i_{2}\neq j_{2}$ this becomes
\begin{equation}
i[{\bf X}_{i_{1},i_{2}},{\bf Y}_{i_{1},j_{2}}]=-|i_{2}\rangle \langle
j_{2}|-|j_{2}\rangle \langle i_{2}|.
\end{equation}
Similarly:
\begin{equation}
i[{\bf X}_{i_{1},i_{2}},{\bf Y}_{j_{1},i_{2}}]=|i_{1}\rangle \langle
j_{1}|+|j_{1}\rangle \langle i_{1}|.
\end{equation}
Thus every $|i_{k}\rangle \langle j_{l}|+|j_{l}\rangle \langle i_{k}|$ is in
the Lie algebra. Similarly, $i[{\bf X}_{i_{1},i_{2}},{\bf X}_{j_{1},j_{2}}]$
yields
\begin{eqnarray}
i[{\bf X}_{i_{1},i_{2}},{\bf X}_{i_{1},j_{2}}] &=&i|i_{2}\rangle \langle
j_{2}|-i|j_{2}\rangle \langle i_{2}|  \nonumber \\
i[{\bf X}_{i_{1},i_{2}},{\bf X}_{j_{1},i_{2}}] &=&i|i_{1}\rangle \langle
j_{1}|-i|j_{1}\rangle \langle i_{1}|
\end{eqnarray}
Thus every $i|i_{k}\rangle \langle j_{l}|-i|j_{l}\rangle \langle i_{k}|$ is
in the Lie algebra. Taking the commutator of these with the $|i_{k}\rangle
\langle j_{l}|+|j_{l}\rangle \langle i_{k}|$ operators finally yields every $
|i_{k}\rangle \langle j_{l}|-|j_{l}\rangle \langle i_{k}|$. Since
$su(n)$ can be decomposed into a sum of overlapping $su(2)$'s
\cite{Cahn:book}, the Lie algebra is the entire $su(n)$, as claimed.

\newpage

\begin{thebibliography}{10}
\bibitem{Glossary}
Abbreviations frequently used in this paper are: QECC = quantum error
  correcting code, DFS = decoherence-free subsystem (or subspace, if no
  confusion can arise), WCD/SCD = weak/strong collective decoherence, irrep =
  irreducible representation.
\end{thebibliography}
\end{document}